\documentclass[prx,aps,twocolumn,article]{revtex4-2}

\usepackage[utf8]{inputenc}
\usepackage[english]{babel}

\usepackage{dcolumn}% Align table columns on decimal point
\usepackage{bm}% bold math
\usepackage{bbm}
\usepackage[colorlinks=true,citecolor=blue,linkcolor=blue,urlcolor=blue,pdfencoding=auto]{hyperref}

\usepackage{color} 
\usepackage{amssymb}
\usepackage{physics}
\usepackage{subeqnarray}
\usepackage{diagbox}
\usepackage{makecell}
\usepackage{amsmath}
\usepackage{mathtools}
\usepackage{empheq}
\usepackage[capitalise]{cleveref}
\usepackage{soul}
\usepackage[acronym]{glossaries}

\newcommand{\msf}[1]{\mathsf{#1}}
\newcommand{\bsb}[1]{\boldsymbol{#1}}
\newcommand{\mbf}[1]{\mathbf{#1}}

\newcommand{\hc}{\mathrm{h.c.}}
\newcommand{\bPhi}{\bsb{\Phi}}

\newcommand{\bQ}{\bsb{Q}}
\newcommand{\bO}{\bsb{O}}

\newcommand{\ba}{\bsb{a}}

\newcommand{\dbQ}{\dd\bQ}

\newcommand{\dbPhi}{\dd\bPhi}

\newcommand{\bV}{\bsb{V}\!}
\newcommand{\bI}{\bsb{I}\!}

\newcommand{\msI}{\msf{I}}

\newcommand{\msIX}{\msf{I}_\msf{X}}
\newcommand{\msT}{\msf{T}}
\newcommand{\msY}{\msf{Y}}
\newcommand{\msZ}{\msf{Z}}

\newcommand{\msZH}{\msZ_\mathrm{H}}

\newcommand{\msYH}{\msY_\mathrm{H}}

\newcommand{\msZTL}{\msZ_{\mathrm{TL}}}
\newcommand{\msG}{\msf{G}}
\newcommand{\msL}{\msf{L}}
\newcommand{\msC}{\msf{C}}
\newcommand{\msA}{\msf{A}}
\newcommand{\msM}{\msf{M}}
\newcommand{\msB}{\msf{B}}
\newcommand{\msD}{\msf{D}}
\newcommand{\msJ}{\msf{J}}
\newcommand{\msF}{\msf{F}}
\newcommand{\msE}{\msf{E}}
\newcommand{\msK}{\msf{K}}
\newcommand{\msU}{\msf{U}}

\newcommand{\msZD}{\msf{Z}_\mathrm{D}}

\newcommand{\msXD}{\msf{X}_\mathrm{D}}

\newcommand{\msKX}{\msf{K}_\msf{X}}
\newcommand{\msXA}{\msX_\mathrm{A}}
\newcommand{\msXH}{\msX_\mathrm{H}}
\newcommand{\msYnode}{\msY_{\rm node}}
\newcommand{\msP}{\msf{P}}
\newcommand{\msX}{\msf{X}}
\newcommand{\msR}{\msf{R}}

\newcommand{\msOmega}{\msf{\Omega}}
\newcommand{\msId}{\mathbbm{1}}
\newcommand{\wcut}{\omega_{\text{cut}}}

\glsdisablehyper
\newacronym{AE}{AE}{adiabatic elimination}
\newacronym{ME}{ME}{master equation}
\newacronym{DOF}{DOF}{degree of freedom}
\newacronym{JJ}{JJ}{Josephson junction}
\newacronym{TL}{TL}{transmission line}
\newacronym{TLs}{TLs}{transmission lines}
\newacronym{EM}{EM}{electromagnetic}

\newacronym{cQED}{cQED}{circuit quantum electrodynamics}
\crefname{section}{Sec.}{Secs.}

\definecolor{aprcolor}{rgb}{0.9,0.1,0.9}

\definecolor{pgcolor}{rgb}{0.4,0.6,0.2}

\begin{document}  
\title{Immittance formulas for exact blackbox quantization and \\ divergence-free effective models in circuit QED}

\author{Philippe Gigon}
\author{Peter Rabl}
\author{Adrian Parra-Rodriguez}
\email{adrian.parra.rodriguez@gmail.com}

\affiliation{Technical University of Munich, TUM School of Natural Sciences, Physics Department, 85748 Garching, Germany}
\affiliation{Walther-Meißner-Institut, Bayerische Akademie der Wissenschaften, 85748 Garching, Germany}
\affiliation{Munich Center for Quantum Science and Technology (MCQST), 80799 Munich, Germany}

\begin{abstract}
Building on the first-order circuit quantization method, we provide simple formulas to construct exact Hamiltonians for Josephson-junction-based superconducting qudits capacitively, inductively, or galvanically coupled to passive linear environments. These environments may be multiport, multimode, discrete or continuous, reciprocal or nonreciprocal, and are characterized directly by their impedance or admittance matrices. In the weak-coupling regime, we further derive \emph{divergence-free} dispersive Hamiltonians for mode-resolved environments and transition-resolved weak-coupling master equations for dissipative continua. Mode structure, frequency renormalizations, environment-mediated interactions, decay rates, and directional cross couplings then follow from the same causal immittance response, while spurious Lamb-shift divergences arising from uncontrolled approximations in previous treatments are made explicit and avoided. We apply the theory to a set of illustrative circuits comprising a discrete resonator filter, finite-band metamaterial environments, nonreciprocal waveguide-QED systems, and superconducting giant atoms, for which analytical response matrices can be obtained, although the method is particularly well suited to numerical responses from electromagnetic solvers or experimental characterization. We thereby extend the black-box quantization framework to multiport, dissipative, and nonreciprocal settings, establishing a simple and scalable route toward optimized and automated electromagnetic design of large-scale superconducting quantum hardware.
\end{abstract}

\maketitle
\section{Introduction}
Superconducting circuits~\cite{Devoret:2013}, composed of lumped capacitors, inductors, and nonlinear Josephson junctions, constitute a leading platform
for quantum information processing and engineered quantum dynamics, combining strong light-matter interactions~\cite{Devoret:2007,FornDiaz:2010,Niemczyk:2010} with precise control over dissipation, driving, and circuit topology. Within this setting, \gls{cQED}~\cite{Blais:2021} provides a powerful framework for controlling artificial atoms, i.e., superconducting qudits, and their direct
or photon-mediated interactions in \gls{EM} environments ranging from conventional~\cite{Blais:2004} and synthetic~\cite{Egger:2013,Mirhosseini:2018}
\gls{TLs} to 3D cavities~\cite{Paik:2011}.

Waveguide-QED architectures have more recently entered regimes in which propagation phases and spatial interference can be purposefully engineered. In analogy with cavity- and waveguide-QED experiments in atomic physics~\cite{Berman:1994,Raimond:2001,Sheremet:2023}, superconducting artificial atoms have conventionally been treated as localized objects coupled through lumped capacitive, inductive, or galvanic connections,
amounting in circuit language to local interconnections constrained by current conservation~\cite{Chua:1974}. Circuit architectures can instead couple the same nonlinear \gls{DOF} to the same \gls{EM} field at several
spatially separated points, leading to giant artificial
atoms~\cite{Gustafsson:2014,Kockum:2014,kannan:2020} and
molecules~\cite{Gheeraert:2020,Yin:2022,TLevy-Yeyati:2026}. More generally, \gls{cQED} also enables synthetic and exotic environmental band structures through complex interconnectivity~\cite{Kim:2021} and nonreciprocal
elements~\cite{Koch:2010,Kamal:2011,Kerckhoff:2015,Chapman:2017,Barzanjeh:2017,Caloz:2018}.

A defining advantage of \gls{cQED} is that, within the quasi-lumped approximation~\cite{Pozar:2009,Nigg:2012,Solgun:2015,MinevQL-Models:2021,ParraRodriguez:2025}, lumped nonlinear elements interface with an otherwise distributed or three-dimensional linear \gls{EM} environment through a finite set of circuit \emph{ports}, whose complete electromagnetic response naturally avoids artificial ultraviolet (UV) divergences~\cite{Houck:2008,Filipp:2011,Lalumiere:2013,Kockum:2014}. The linear sector is characterized by multiport immittance responses, i.e., impedance and admittance matrices (circuit Green's functions), directly accessible from simulation and measurement~\cite{Pozar:2009}. Such descriptions have proven powerful for circuit quantization and for deriving effective interactions in closed or weakly open systems~\cite{Nigg:2012,Bergenfeldt:2012,Solgun:2014,Solgun:2015,Mortensen:2016,Hassler:2019,Solgun:2019,Cattaneo:2019,MinevQL-Models:2021,MinevEPR:2021,Solgun:2022,Labarca:2024}, while enabling formulations free from artificial divergences~\cite{Paladino:2003,Bamba:2014,Gely:2017,Malekakhlagh:2017,Solgun:2014,Solgun:2015,ParraRodriguez:2018,ParraRodriguez:2019,ParraRodriguez:2022,ParraRodriguez:2025}.

For one-port environments, foundational constructions were developed by Caldeira and 
Leggett~\cite{CaldeiraLeggett_QT:1983} for admittance environments in the flux gauge, subsequently refined by Esteve et al.~\cite{Esteve:1986}, and complementarily by Paladino et al.~\cite{Paladino:2003} for impedance environments in the charge gauge. These works established that infinite-dimensional environments inductively or capacitively coupled to nonlinear inductors are naturally described in the corresponding gauges, with the capacitive construction yielding a regular high-frequency response and a circuit-determined cutoff~\cite{Paladino:2003}. Discrete multiport constructions were subsequently developed for reciprocal environments by Solgun et al.~\cite{Solgun:2015,Solgun:2019}, using canonical Cauer synthesis of the impedance response, with related extensions to nonreciprocal settings developed by Labarca, Benhayoune-Khadraoui et al.~\cite{Labarca:2024}.

\begin{figure}[t]
    \centering
    \includegraphics[width=\linewidth]{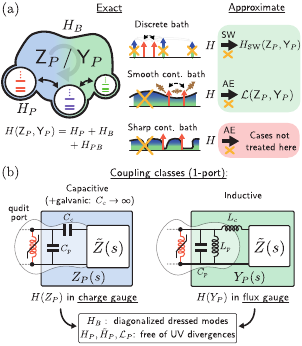}
    \caption{(a) From the immittance response $\msZ_P(s)$ or $\msY_P(s)$, we construct exact Hamiltonians for superconducting qudit ports linearly coupled to possibly nonreciprocal environments. We also derive effective Hamiltonian or Lindbladian dynamics when the environment is either discrete and dispersive or a smooth continuum remaining close to its ground state. Sharply varying baths or additional external driving may be treated using appropriate adiabatic-elimination schemes. (b) General one-port coupling classes considered. Qudits based on nonlinear inductive elements are coupled capacitively (inductively) to generic passive sub-environments characterized by $\tilde{Z}$. Combined with the charge (flux) gauge and first-order quantization, the $Z_P$ ($Y_P$) representation yields diagonal dressed environmental modes and UV-finite exact and effective models, including for the convergent infinite-mode environments considered here. The impedance class includes the galvanic limit $C_c\rightarrow\infty$, where the coupling capacitor becomes a short circuit~\cite{Gely:2017,Malekakhlagh:2017,ParraRodriguez:2018,Mehta:2022,Hyyppa:2022}.
    }
    \label{fig:Intro}
\end{figure}

In this work, we combine seminal results from circuit theory~\cite{Foster:1924,Cauer:1926,Brune:1931,Belevitch:1950,Newcomb:1966,Solgun:2015}
with the first-order circuit quantization method~\cite{ParraRodriguez:2024,Osborne:2024,ParraRodriguez:2025}, which bypasses the Legendre transformation and is particularly suited to nonreciprocal circuits. We use them to (i) derive closed-form immittance expressions for exact Hamiltonians of Josephson-junction-based superconducting qudits coupled to linear, possibly nonreciprocal, discrete, continuous, or mixed environments and (ii) derive reduced qudit dynamics when the environments remain close to their ground states and are either discrete and off-resonantly coupled or continuous with smooth spectra.
These reductions eliminate the environmental modes to obtain effective Hamiltonians or Lindbladian master equations, see Fig.~\ref{fig:Intro}(a).

The formalism covers capacitive, inductive, and galvanic couplings to mode-resolved and continuum electromagnetic environments, as illustrated by the one-port examples of Fig.~\ref{fig:Intro}(b). In the corresponding charge and flux gauges, the dressed environmental sector is naturally diagonal. For smooth dissipative immittance responses, constructing the exact Hamiltonian then reduces to extracting the poles at zero and infinity and diagonalizing a port-dimensional hermitian matrix.

Spectrally resolved environments yield dispersive effective Hamiltonians through a Schrieffer-Wolff transformation~\cite{Schrieffer:1966,Bravyi:2011}, whereas smooth continua lead to Gorini-Kossakowski-Sudarshan-Lindblad (GKSL) generators under weak-coupling and (partial-)secular approximations~\cite{Davies:1974,BreuerPetruccione:2002}, capturing decay, frequency renormalization, spatial propagation, and directional transport. We restrict these reductions to the two extreme gauges, although intermediate or mode-dependent gauges may be advantageous in specific regimes, with their optimal choice remaining open~\cite{Manucharyan:2017,DeBernardis:2018,Roth:2019,Stokes:2019,DiStefano:2019,Mehta:2022,Arwas:2023}.

We demonstrate the versatility of the approach through resonator-mediated networks, finite-band metamaterial environments~\cite{Snyman:2015,Goswami:2026}, including a case with coexisting discrete and continuous environmental modes, nonreciprocal waveguides, and giant artificial emitters~\cite{Kockum:2014,Kockum:2018}. For the waveguide examples, we evaluate the required immittances analytically assuming ideal TEM propagation, while the formalism itself applies to general linear environments. We explicitly show how commonly used cutoff-dependent models emerge as approximations to the first-order immittance formulas and assess their validity
for current experimental parameters. More generally, the exact Hamiltonians derived here provide a natural starting point for effective treatments beyond the Born--Markov approximation~\cite{Shi:2015,AGT:2017,Gu:2024}.

The manuscript is organized as follows. In \cref{Sec:ExactH_ZY}, we provide exact Hamiltonians for superconducting qudits coupled to discrete and continuous multiport environments directly from their impedance or admittance responses. In \cref{Sec:SW-Models}, we use a Schrieffer--Wolff transformation to obtain dispersive Hamiltonians for spectrally resolved environments and illustrate the results with a charge qudit and a fluxonium capacitively coupled through an LC mode, extending results from Refs.~\cite{Solgun:2019,Labarca:2024,Khan:2024}. In \cref{Sec:ME-models}, we derive master equations for smooth dissipative environments, including coherent frequency shifts, mediated interactions, and local and correlated decay rates. Examples with continuous environments are presented in \cref{Sec:DissipativeExamples}. Finally, \cref{Sec:ConclusionsOutlook} summarizes our results and discusses future directions.

\section{Exact Immittance Hamiltonian Construction}
\label{Sec:ExactH_ZY}

The class of circuits considered in this work is summarized in
\cref{fig:Intro} and shown schematically in \cref{fig:Fig_Cauer_MT}. A finite set of ports separates a nonlinear sector from an otherwise passive linear electromagnetic environment. The nonlinear sector is a finite set of explicitly retained port degrees of freedom associated with nonlinear inductive elements, together with any lumped linear elements and external biases assigned to their loops. The linear sector behind the ports may be multiport, multimode, discrete or continuous, reciprocal or nonreciprocal, and is characterized only through its impedance or admittance matrix.

The central goal of this section is to translate the linear environment's port response directly into an exact Hamiltonian
$H^{\msX}=H_P^{\msX}+H_B^{\msX}+H_{PB}^{\msX}$, where $H_P^{\msX}$ describes the dressed nonlinear port degrees of freedom, $H_B^{\msX}$ a diagonal basis of dressed environmental modes, and $H_{PB}^{\msX}$ their linear coupling. Here $\msX\in\{\msZ,\msY\}$ is the immittance response of the linear sector as seen by the ports. We decompose the immittance response into its zero-, finite-, and infinite-frequency contributions, map them onto a canonical multiport circuit representation~\cite{Newcomb:1966}, and quantize the resulting circuit using the first-order method~\cite{ParraRodriguez:2024,ParraRodriguez:2025}.

The resulting exact Hamiltonians provide the starting point for the dispersive and dissipative reductions derived in Secs.~\ref{Sec:SW-Models} and~\ref{Sec:ME-models}. Technical details of the port reduction are given in Appendix~\ref{AppSec:FirstOrderReview}.

\subsection{Brief review of Cauer expansions for nonreciprocal multiport environments}

\begin{figure}[t]
    \centering
    \includegraphics[width=\linewidth]{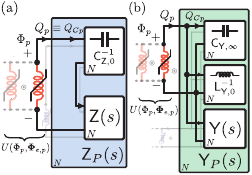}
    \caption{$N$ nonlinear port degrees of freedom coupled to linear dissipative environments
    described by (a) impedance or (b) admittance matrices, corresponding to
    generic (a) capacitive and (b) inductive coupling.  The inductive branches retained at port $p$ define $U(\Phi_p,\bPhi_{e,p})$, including their static flux offsets, and any parallel capacitances are included in the linear response. Multiport connectivity is shown faintly for simplicity (see \cref{AppSec:Z_construction_and_quantization,AppSec:Y_construction_and_quantization}). Time-dependent flux allocation with nonlinear inductors shunted by parasitic capacitors is treated in Appendix~\ref{AppSec:DrivenNonlinearPorts}.}
    \label{fig:Fig_Cauer_MT}
\end{figure}

To streamline the Hamiltonian construction below, we first review the aspects of multiport Cauer decompositions~\cite{Newcomb:1966} needed here, in particular the matrix-fraction expansions of impedance and admittance responses. We begin with purely discrete $N$-port environments whose ports are connected to nonlinear inductors. For capacitively coupled systems, we use an impedance representation and restrict ourselves to Laplace-domain impedance matrices of the form
\begin{align}
    \msZ_P(s) = \frac{\msC_{\msZ,0}^{-1}}{s}+\msZ(s),
    \label{eq:Z_P_class}
\end{align}
where $\msZ(s)$ contains no further poles at zero or infinity. That is, from the point of view of the nonlinear inductors, the environment responds with a set of voltages $\bV_P=(V_1,\dots, V_N)^T$ to injected currents $\bI_P=(I_1,\dots, I_N)^T$ via $\bV_P(s)=\msZ_P(s)\bI_P(s)$, see Fig.~\ref{fig:Fig_Cauer_MT}(a).

We further assume that the high-frequency dressing matrix $\msC_{\msZ,\infty}^{-1}\equiv \lim_{\eta \rightarrow \infty}\eta \msZ(\eta)$ (with $\eta\in \mathbbm{R}$) is finite. Interestingly, our analysis of $\msZ$-characterized environments remains valid as long as $(\msC_{\msZ,0}^{-1}+\msC_{\msZ,\infty}^{-1})$ is a full-rank positive matrix, including galvanic cases~\cite{CaldeiraLeggett_QT:1983,Esteve:1986} in which $\msC_{\msZ,0}^{-1}=0$~\footnote{This allows us, for example, to obtain analytically the continuum limit of Hamiltonians previously constructed numerically~\cite{Mehta:2022,Hyyppa:2022}.}. 

The remaining finite-frequency poles of $\msZ(s)$ can be written as a Cauer decomposition (matrix partial-fraction expansion), $\msZ(s)=\sum_{\alpha=1}^{M}\msZ_{\alpha}(s)$, with
\begin{align}
\msZ_{\alpha}(s)
=\frac{\msR_{\alpha}}{s+i\omega_{\alpha}}
+
\frac{\msR_{\alpha}^{T}}{s-i\omega_{\alpha}}\equiv
\frac{\msA_{\alpha}s+\msB_{\alpha}}
{\omega_{\alpha}^2+s^2},
\end{align}
where $\msR_{\alpha}=\underset{s=-i\omega_{\alpha}}{\operatorname{Res}}\msZ(s)$ is the residue (matrix), with rank
$r_\alpha\equiv\operatorname{rank}\msR_\alpha\leq N$, $\msA_{\alpha}=\msR_{\alpha}+\msR_{\alpha}^T$ and $\msB_{\alpha}=-i\omega_\alpha(\msR_{\alpha}-\msR_{\alpha}^T)$ are symmetric and antisymmetric matrices, respectively. The total impedance is thus realized as the series connection of a zero-frequency response (a capacitive network) and discrete finite-frequency poles represented by $r_\alpha\leq N$ possibly nonreciprocal oscillators coupled to the external ports. Details of this canonical structure are given in \cref{AppSec:Z_construction_and_quantization}. Minimal realizations require ideal transformers, also associated with energy-participation ratios in the superconducting-circuit literature~\cite{Ciani:2020}. These transformers encode sets of Kirchhoff constraints~\cite{Solgun:2015,ParraRodriguez:2024} and interface internally diagonalized modes with the external ports.

The decomposition given above naturally includes purely reciprocal poles, for which $\msB_\alpha=0$. Indeed, circuit equivalences also allow purely reciprocal responses to be represented using nonreciprocal elements. For example, a gyrator terminated at its second port by an inductor or capacitor appears from its first port as the corresponding dual element~\cite{Newcomb:1966,Carlin:1964}. More general impedance matrices containing an additional pole at infinity, $\msL_{\msZ,\infty}s$, or a constant nonreciprocal term, $\msB_\infty$, require additional steps (see Refs.~\cite{ParraRodriguez:2018,ParraRodriguezPhD:2021,Labarca:2024,ParraRodriguez:2025,Yang:2026} for some particular cases) that are left for future work.

For inductively coupled systems, we consider environments described by the admittance matrix 
\begin{align}
    \msY_P(s)
    =s\msC_{\msY,\infty}+
    \frac{\msL_{\msY,0}^{-1}}{s}
    +\msY(s),
    \label{eq:Y_P_class}
\end{align}
where $\msY(s)$ contains no poles at zero or infinity, see Fig.~\ref{fig:Fig_Cauer_MT}(b). We have also discarded a constant gyration term, which is not expected in realistic scenarios. We similarly assume that the high-frequency coefficient $\msL_{\msY,\infty}^{-1}\equiv\lim_{\eta\rightarrow\infty}\eta\msY(\eta)$ is finite. The discrete poles have the analogous Cauer decomposition
\begin{align}
    \msY(s)=\sum_{\alpha=1}^{M}{\msY}_{\alpha}(s),\quad\msY_{\alpha}(s)=\frac{\msD_{\alpha} s + \msE_{\alpha}}{\omega_{\alpha}^2 + s^2}.
\end{align}
The matrices $\msC_{\msY,\infty}$, $\msL_{\msY,0}^{-1}$, $\msL_{\msY,\infty}^{-1}$, and $\msD_{\alpha}$ are symmetric, while $\msE_{\alpha}$ is antisymmetric. The latter two encode, respectively, the transpose-symmetric and transpose-antisymmetric parts of the residue matrix.
The zero-frequency pole is synthesized by an inductive network, the infinite-frequency pole by a capacitive network, and the finite-frequency poles by possibly nonreciprocal oscillators. Details are given in \cref{AppSec:Y_construction_and_quantization}. To ensure well-defined port
degrees of freedom, we further restrict ourselves to cases in which $\msC_{\msY,\infty}$ has full rank. In what follows, we use $\msX(s)$ as a placeholder for $\msZ(s)$ or $\msY(s)$.

Although the Cauer expansions above were introduced for finite-dimensional discrete environments, the same description naturally extends to infinite discrete, continuous, and mixed spectra. In particular, continuous spectral components are obtained as limits of increasingly dense Cauer decompositions,
\begin{align}
    \msX(s)
    =
    \lim_{\substack{\Delta\omega\to0,\\[2pt] M\to\infty}}
    \sum_{\alpha=1}^{M}\msX_\alpha(s),
\end{align}
for which the finite-frequency sums become Riemann sums converging to a matrix-valued spectral measure~\cite{Senitzky:1960,Senitzky:1961,Feynman:1963,CaldeiraLeggett_QT:1983}. We denote by $\mathcal B_{\msX}\subseteq(0,\infty)$ the positive-frequency support of this finite-frequency spectral measure. For the class considered here, $\mathcal B_{\msX}$ may contain both isolated frequencies and continuous intervals, corresponding respectively to atomic and continuous components of the spectral measure. More general singular-continuous spectra, as can arise in quasiperiodic waveguides~\cite{Bonsel:2026}, are expected to be accessible through suitable limits of discrete Cauer networks, although we do not consider this extension explicitly.

\begin{figure}[t]
    \centering
    \includegraphics[width=\linewidth]{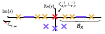}
    \caption{Schematic singularity structure of the general immittance responses $\msX_P(s)$ considered. The positive-frequency spectral support $\mathcal B_{\msX}$, represented by $s=-i\omega$ with $\omega>0$, may contain discrete atomic poles (yellow crosses) and dissipative continuous intervals (purple thick lines), the latter corresponding to branch cuts. Additional dissipative poles with $\Re(s)<0$ are shown in purple outside $\mathcal B_{\msX}$, while red denotes the zero- and infinite-frequency
contributions.}
\label{fig:Fig_Structure_Envs}
\end{figure}

In the corresponding spectral representation, the atomic and continuous components appear as isolated poles at $s=-i\omega_\alpha$ and branch cuts along $s=-i\omega$, respectively, over the continuous components of $\mathcal B_{\msX}$, together with their complex-conjugate counterparts, as illustrated in \cref{fig:Fig_Structure_Envs}. Particular responses may nevertheless admit simpler analytic continuations without explicit branch cuts, e.g., \cref{Sec:SingleQubitSemiInfiniteTL}.

We further define the boundary distribution $\msX[\omega]\equiv \msX(-i\omega+0^+)$, and denote the symmetric, antisymmetric, hermitian, and anti-hermitian parts of the response matrices by
\begin{align}
    \msX^\pm\equiv\frac{\msX\pm \msX^T}{2}, \quad  \msX_{\rm H/A}\equiv \frac{\msX\pm \msX^\dag}{2}.
\label{Eq:HermitianAntiHermitian}
\end{align}
The complex-conjugate symmetry $\msX(s^*)=\msX(s)^*$ follows from the reality of the time-domain response. Causality and passivity imply that the immittance is matrix positive real and hence analytic in the right half-plane~\cite{Newcomb:1966}. Finally, after the polynomial high-frequency contributions have been extracted, the remainder $\msX(s)$ vanishes along the positive real axis, $\lim_{\eta\to+\infty}\msX(\eta)=0$.

We stress that our $\msZ$- (\ref{eq:Z_P_class}) and $\msY$-based
(\ref{eq:Y_P_class}) constructions define two generally distinct environment classes. Their domains overlap for broad classes of finite-dimensional discrete environments, whereas inversion does not, in general, map one class into the other. For infinite discrete, continuous, or mixed environments, the overlap also requires that the corresponding dressing terms converge. Examples with and without such an overlap are discussed below.

\subsection{Exactly quantized circuit Hamiltonians}
\label{Sec:DiscreteQuantizedHamiltonian}

After obtaining an explicit canonical decomposition of the immittance response $\msX_P(s)$, we quantize the exact classical Hamiltonian using the first-order quantization algorithm of Refs.~\cite{ParraRodriguez:2024,ParraRodriguez:2025}, based on the Faddeev-Jackiw reduction method~\cite{Faddeev:1988,Jackiw:1993}. Details of the quantization procedure are provided in \cref{Sec:CanonicalHamiltonian,Sec:CanonicalHamiltonianY}. 
Throughout the manuscript, we use the standard definitions of branch fluxes [$\Phi_b(t)=\int_0^t d\tau\, V_b(\tau)$] and charges [$Q_b(t)=\int_0^t d\tau\, I_b(\tau)$] as the primary variables used to describe the circuits~\cite{Devoret:1997}.

We formulate the impedance-based $\msZ$ class (\ref{eq:Z_P_class}) in the charge gauge and the admittance-based $\msY$ class (\ref{eq:Y_P_class}) in the flux gauge to avoid cumbersome divergences~\cite{CaldeiraLeggett_QT:1983,Esteve:1986,Mehta:2022}. Omitting operator hats whenever no ambiguity can arise, the corresponding exact Hamiltonians can be decomposed as
\begin{align}
H^{\msX}
=
H_P^{\msX}
+
H_B^{\msX}
+
H_{PB}^{\msX},
\label{Eq:ExactHamiltonianDecomposition}
\end{align}
with $\msX\in\{\msZ,\msY\}$. For finite-mode realizations with rank-deficient $\msC_{\msZ,0}^{-1}$, the $\msZ$-Hamiltonian retains redundant variables. The residual Kirchhoff
constraints must therefore be reduced before quantization
in terms of independent canonical pairs, as detailed in
Appendix~\ref{AppSec:Z_construction_and_quantization}.

As previously mentioned, these two constructions apply, in general, to distinct classes of environments and should not be regarded as interchangeable gauge descriptions of an arbitrary circuit in infinite-dimensional cases. When a given circuit belongs to both classes, the resulting complete Hamiltonians describe the same physical system and are related by the corresponding canonical transformation, provided that this transformation remains unitarily implementable in the infinite-mode limit. The separate port, bath, and port-bath terms, as well as the associated normal modes and their frequencies, depend on the chosen construction and chosen gauge. 

We note that mixed gauges have previously been proposed as potentially advantageous starting points for deriving reduced effective  models~\cite{Roth:2019,Mehta:2022,Arwas:2023}. When a circuit admits both endpoint descriptions, continuous gauge interpolations can, in principle, be constructed; however, away from the two endpoints, the direct connection to the original immittance response and to a diagonal dressed environmental basis is no longer manifest. Here, we restrict ourselves to the two extreme gauge choices naturally associated with the two classes because they yield diagonal dressed environmental modes, closed-form Hamiltonians and analytical expressions for the exact and effective parameters, including for continuous environments, while providing divergence-free starting points for the effective models derived in \cref{Sec:SW-Models,Sec:ME-models}.

To avoid unnecessarily heavy notation, we suppress the superscript $\msX$ on $H_B^{\msX}$, $H_{PB}^{\msX}$, and their associated mode variables whenever the relevant class is clear from the context.

\subsubsection{Quantization of dressed-port degrees of freedom}

For the class of $\msZ$-environments (\ref{eq:Z_P_class}), the classical Hamiltonian for
the dressed external degrees of freedom is described, in the charge gauge~\cite{Roth:2019,Mehta:2022}, by
\begin{align}
    H_P^\msZ
    =
    U(\bPhi_P,\bPhi_{e,P})
    +
    \frac{1}{2}
    \bQ_C^T
    \left(
        \msC_{\msZ,0}^{-1}
        +
        \msC_{\msZ,\infty}^{-1}
    \right)
    \bQ_C,
    \label{Eq:ExternalHamiltonianZ}
\end{align}
where the flux and charge vectors are, respectively, $\bPhi_P=(\Phi_1,\ldots,\Phi_N)^T$ and $\bQ_C=(Q_{C1},\ldots,Q_{CN})^T\equiv\bQ_P$, with $\{\Phi_p,Q_{Cp'}\}=\delta_{pp'}$~\footnote{In the impedance representation, the internal charge vector associated with the capacitor network $\msC_{\msZ,0}^{-1}$ coincides with the port-charge vector, $\bQ_C\equiv\bQ_P$, because the corresponding elements are connected in series. This identification does not hold in the admittance representation, where several circuit blocks are connected in parallel and the port charges are distributed among them.}.

External fluxes are static in the main text and Appendices~\ref{AppSec:FirstOrderReview}--\ref{AppSec:Y_construction_and_quantization}. We assign all retained flux offsets to the explicit port branches, so they enter through $U(\bPhi_P,\bPhi_{e,P})$, as detailed in \cref{AppSubsec:ParallelNonlinearInductors}. This sector may include explicitly retained linear inductors. The passive immittance is defined for the remaining linear environment with no separate applied-flux offsets in its Hamiltonian. Time-dependent fluxes~\cite{You:2019,Riwar:2022,ParraRodriguez:2025} at nonlinear ports with parallel capacitances are treated in Appendix~\ref{AppSec:DrivenNonlinearPorts}.

We now specify the phase-space structure and quantization of the dressed port degrees of freedom. Since the immittance formulation is intended to be used directly at the level of the total Hamiltonian, without selecting or reconstructing a particular lumped-element realization, candidate compact directions are identified from the discrete translation symmetries of the full classical Hamiltonian~\cite{Devoret:2021,ParraRodriguez:2024,ParraRodriguez:2025}.

To characterize the compactness of the circuit variables, it is convenient to define the period lattice
\begin{align}
\Lambda_U = \left\{\boldsymbol{\lambda}\,:\,
U(\bPhi_P+\boldsymbol{\lambda},\bPhi_{e,P})
=U(\bPhi_P,\bPhi_{e,P}) \right\}.
\label{Eq:PortPotentialPeriodLattice}
\end{align}

In the charge gauge used for the $\msZ$-class, the environment coupling
depends only on the charges. Thus, each
$\boldsymbol{\lambda}\in\Lambda_U$ defines a discrete translation symmetry
of the full Hamiltonian. A static inductive shunt hidden in $\msZ_P$,
however, makes the corresponding port-flux direction extended. Whenever the limit exists, we define the dc inductive-shunt matrix inferred
from the impedance response,
\begin{align}
    \msL_{\msZ,\mathrm{dc}}^{-1}
    \equiv
    \lim_{\eta\to0^+}\eta\,\msZ_P^{-1}(\eta).
    \label{Eq:ZDCInductiveMatrix}
\end{align}
A sufficient condition for a periodic direction to be extended is then
\begin{align}
    \boldsymbol{\lambda}
    \notin
    \ker\{\msL_{\msZ,\mathrm{dc}}^{-1}\}.
    \label{Eq:ZExtendedDirectionCriterion}
\end{align}
If instead
$\boldsymbol{\lambda}\in\ker\{\msL_{\msZ,\mathrm{dc}}^{-1}\}$, the pole
structure does not obstruct compactness. A complete immittance-only classification of general galvanic and multiport cases is left for future
work.

For $k$ independent compact directions, the configuration and phase spaces may be taken as $(S^1)^k\times\mathbb{R}^{N-k}$ and $(S^1)^k\times\mathbb{R}^{2N-k}$, respectively, or equivalently treated on the fully extended covering space by resolving the corresponding translation sectors~\cite{ParraRodriguez:2024}. We use the former mixed compact--extended description throughout. See
Refs.~\cite{Averin:1985,Apenko:1989,Schoen:1990,Koch:2009,
Devoret:2021,ThanhLe:2020,ParraRodriguez:2024,Giacomelli:2026}
for further discussion.

For an extended canonical pair $(\Phi_p,Q_{Cp})\in\mathbbm{R}^2$,
quantization proceeds by promoting the classical variables to quantum
operators, i.e.,
$Q_{Cp}\overset{q.}{\rightarrow}\hat{Q}_{Cp}$ and
$\Phi_p\overset{q.}{\rightarrow}\hat{\Phi}_p$, satisfying
\begin{align}
    \left[\hat{\Phi}_p,\hat{Q}_{Cp'}\right]
    =
    i\hbar\,\delta_{pp'}.
\end{align}
Such pairs describe, for example, inductively shunted qudits such as
fluxonium~\cite{Manucharyan:2009}, as well as harmonic oscillators.

For a compact--extended canonical pair
$(\Phi_p,Q_{Cp})\in
(\mathbbm{R}/\Phi_0\mathbbm{Z})\times\mathbb{R}\simeq T^*S^1$,
where $\Phi_0=h/(2e)$ is the superconducting flux quantum, the quantum theory
is formulated in terms of the discrete Cooper-pair number operator
$\hat{n}_{Cp}$ and the globally defined unitary operator
$e^{i\hat{\varphi}_p}$. The corresponding quantization prescription is
\begin{align}
    Q_{Cp}
    &\overset{q.}{\rightarrow}
    \hat{Q}_{Cp}-Q_{g,p}
    =
    2e\left(\hat{n}_{Cp}-n_{g,p}\right),
    \\
    e^{i(2\pi/\Phi_0)\Phi_p}
    &\overset{q.}{\rightarrow}
    e^{i\hat{\varphi}_p},
\end{align}
where $\hat{n}_{Cp}\equiv\hat{Q}_{Cp}/(2e)$ and
$n_{g,p}\equiv Q_{g,p}/(2e)\in\mathbb{R}/\mathbb{Z}$. These operators satisfy
\begin{align}
    \left[
        \hat{n}_{Cp},
        e^{i\hat{\varphi}_{p'}}
    \right]
    =
    \delta_{pp'}e^{i\hat{\varphi}_{p'}}.
\end{align}
Grouping the compact and extended directions, we write
$\bQ_C\overset{q.}{\rightarrow}\hat{\bQ}_C-\bQ_{g,P}$.
The static components of $\bQ_{g,P}$ associated with extended flux coordinates
are unitarily removable and may be set to zero.

For the $\msY$ class in the flux gauge, the exact dressed-port Hamiltonian is
\begin{align}
\begin{aligned}
H_P^\msY
={}&
U(\bPhi_P,\bPhi_{e,P})
+
\frac{1}{2}
\bPhi_P^T
\left(
\msL_{\msY,0}^{-1}
+
\msL_{\msY,\infty}^{-1}
\right)
\bPhi_P\\
&+
\frac{1}{2}
\bQ_C^T
\msC_{\msY,\infty}^{-1}
\bQ_C.
\end{aligned}
\label{Eq:ExternalHamiltonianY}
\end{align}
The canonical variables satisfy
$\{\Phi_p,Q_{Cp'}\}=\delta_{pp'}$. 

As in the impedance construction, all retained static flux offsets are included in $U(\bPhi_P,\bPhi_{e,P})$. In contrast with the $\msZ$ case, compactness cannot be identified from $H_P^\msY$ alone because the term proportional to $\msL_{\msY,\infty}^{-1}$ may break port-flux translations. A port-flux
displacement can extend to a collective translation only if it leaves the
nonlinear potential invariant and lies in the kernel of the zero-frequency
inductive response. We therefore define
\begin{align}
\Lambda_{\mathrm c}
=
\Lambda_U\cap\ker\msL_{\msY,0}^{-1}.
\label{Eq:YCompactTranslationLattice}
\end{align}
For a finite-mode environment, every
$\boldsymbol{\lambda}\in\Lambda_{\mathrm c}$ extends to an exact collective
translation symmetry of the full Hamiltonian. The corresponding translation
simultaneously displaces the finite-frequency environmental coordinates,
compensating for the apparent breaking of the port-flux translation by
$\msL_{\msY,\infty}^{-1}$. For reciprocal environments, each independent generator defines a collective compact coordinate whose conjugate charge generally involves both port and environmental variables~\footnote{The same conclusion holds when $\Lambda_{\mathrm c}$
has a single independent generator. For general nonreciprocal multiport environments, simultaneous compact identification requires checking the compatibility of the quantum translation operators.}.

\begin{table}[t]
    \centering
    \begin{tabular}{@{}lll@{}}
        \hline
        Construction & Condition & Conclusion \\
        \hline
        $\msZ_P$
        &
        $\displaystyle
        \boldsymbol{\lambda}\notin
        \ker\msL_{\msZ,\mathrm{dc}}^{-1}$
        &
        Extended
        \\[2mm]
        $\msY_P$
        &
        $\displaystyle
        \boldsymbol{\lambda}\notin
        \ker\msL_{\msY,0}^{-1}$
        &
        Extended
        \\[2mm]
        $\msY_P$
        &
        $\displaystyle
        \boldsymbol{\lambda}\in\Lambda_{\mathrm c}$
        &
        \makecell[l]{Collective compact}
        \\
        \hline
    \end{tabular}
    \small
    \caption{Conditions for extended periodic port-flux directions $\boldsymbol{\lambda}\in\Lambda_U$ and collective translation symmetries. For finite $\msY_P$ environments, the compact conclusion holds under the conditions stated in the text. In infinite-mode environments, the displacement must also remain well defined and unitarily implementable.}
    \label{tab:ImmittanceCompactness}
\end{table}
For an infinite-mode discrete, continuous, or mixed environment, the same
conclusion holds provided that the corresponding environmental displacement
remains well defined in the infinite-mode limit and is unitarily
implementable after quantization. This requirement is automatic for a finite
number of modes, but may fail for infinite-mode environments, in particular
because of infrared divergences in gapless spectra.

If $\Lambda_{\mathrm c}=\{\bsb{0}\}$, all port directions are extended.
Otherwise, under the conditions stated above, 
each generator for which the corresponding collective translation remains well defined specifies a collective compact coordinate. In this manuscript, we mainly restrict ourselves to the case where all port canonical
pairs are extended, with $\left[\hat{\Phi}_p,\hat{Q}_{Cp'}\right]
=i\hbar\delta_{pp'}$. The systematic treatment of nontrivial compact sectors, connected to recent work on dissipative quantum phase transitions~\cite{Kuzmin:2019,Kuzmin:2021,Hakonen:2021,Mehta:2022,
Giacomelli:2024,Kuzmin:2025,Paris:2025,Giacomelli:2026}, is left for future
work.    

In summary, the sufficient conditions discussed above for identifying the character of periodic port-flux directions in the $\msZ$ and $\msY$ constructions are stated in \cref{tab:ImmittanceCompactness}. A concrete comparison for a circuit belonging to both classes is given below in the fluxonium example of \cref{Sec:ExampleFluxoniumInductive}.

\subsubsection{Direct port-port coupling terms} \label{Sec:MainDirectCoupling}

Having discussed the quantization of the external port flux variables and their conjugate charges, we now separate the terms that are local in the port labels from the cross-port contributions. For compactness, we define
\begin{align}
    \msM_{\msX,\nu}
    \equiv
    \begin{cases}
        \msC_{\msZ,\nu}^{-1}, & \msX=\msZ,\\[1mm]
        \msL_{\msY,\nu}^{-1}, & \msX=\msY,
    \end{cases}
    \qquad
    \nu\in\{0,\infty\}.
\label{Eq:DiscreteA0AinfDefinitions}
\end{align}
This common notation is used only after the separate impedance and admittance constructions have been established. In the corresponding derivations, we retain the physically explicit capacitance and inductance matrices.

The exact port Hamiltonian can then be written as
\begin{align}
    H_P^{\msX}
    =
    \sum_p h_p+H_{\mathrm{int}},
    \label{Eq:PortHamiltonianSplitUp}
\end{align}
where $\sum_p h_p$ contains all the diagonal (local) terms and $H_{\mathrm{int}}$ all the off-diagonal terms in $H_P^{\msX}$. In the following, it is convenient to separate the direct coupling terms into two contributions $H_{\mathrm{int}}
=H_{\mathrm{int}}^{(0)}+H_{\mathrm{int}}^{\mathrm{CT}}$, where the first contribution comes from direct capacitive (inductive) coupling and is given by
\begin{align}
\begin{aligned}
    H_{\mathrm{int}}^{(0)}
    ={}&
    \sum_{p'>p}
    \left(\msM_{\msX,0}\right)_{p,p'}
    O_{p}^{\msX} O_{p'}^{\msX}
    \\
    &+
    \delta_{\msX,\msY}
    \sum_{p'>p}
    \left(
        \msC_{\msY,\infty}^{-1}
    \right)_{p,p'}
    Q_{Cp}Q_{Cp'},
\end{aligned}
\label{Eq:DiscreteDirectInteraction}
\end{align}
while the cross-port part of the high-frequency counterterm is
\begin{align}
    H_{\mathrm{int}}^{\mathrm{CT}}
    =
    \sum_{p'>p}
    \left(\msM_{\msX,\infty}\right)_{p,p'}
    O_{p}^{\msX} O_{p'}^{\msX}.
\label{Eq:DiscreteDirectInteractionCounterTerm}
\end{align}
Here, $\delta_{\msX,\msY}=1$ in the admittance representation and vanishes in the impedance representation, and
\begin{align}
    O_{p}^{\msZ}=\hat Q_{Cp}-Q_{g,p},
    \qquad
    O_{p}^{\msY}=\hat\Phi_p.
    \label{Eq:CouplingOperatorDefMain}
\end{align}
For extended flux coordinates, the static offset charges are removable and set to zero. The second line of \cref{Eq:DiscreteDirectInteraction} is the direct capacitive interaction that may accompany the inductive coupling in the admittance representation.

The decomposition \cref{Eq:PortHamiltonianSplitUp} is exact and does not require the cross-port terms to be weak. For concreteness, the reduced models developed below assume that the remaining terms in $H_{\mathrm{int}}^{(0)}$ and $H_{\mathrm{int}}^{\mathrm{CT}}$ can be treated perturbatively after the diagonal contributions of the dressed capacitance and inductance matrices have been included exactly in the local Hamiltonians $h_p$. If some direct cross-port couplings are strong (compared to the corresponding level splitting), as may occur in giant-molecule architectures~\cite{Gheeraert:2020}, they should instead be incorporated into the system Hamiltonian and diagonalized together. The subsequent bath-coupling formulas remain valid after expanding the physical port operators in this collective eigenbasis. For the reduced models derived below, we restrict to time-independent external biases.

To proceed in the weak-direct-coupling regime, we diagonalize the local Hamiltonians,
\begin{align}
    h_p
    =
    \sum_i
    \hbar\omega_i^{(p)}
    \ket{i}_p\bra{i}_p,
    \label{Eq:DiagonalizedPortHamiltonian}
\end{align}
and define the transition frequencies $\omega_{ij}^{(p)}\equiv\omega_i^{(p)}-\omega_j^{(p)}$. In this basis, the coupling operator is decomposed as
\begin{align}
    O_{p}^{\msX}
    =
    \sum_{i,j}
    O_{p,ij}^{\msX}
    \sigma_{ji}^{(p)},
\label{Eq:DiscreteOperatorExpansion}
\end{align}
with $\sigma_{ji}^{(p)}\equiv\ket{i}_p\bra{j}_p$ and $O_{p,ij}^{\msX}\equiv\bra{i}O_{p}^{\msX}\ket{j}_p$.

For the admittance representation, we similarly decompose
\begin{align}
    Q_{Cp}
    =
    \sum_{i,j}
    Q_{Cp,ij}
    \sigma_{ji}^{(p)}.
\end{align}
The direct interaction \cref{Eq:DiscreteDirectInteraction} can therefore be expressed in the eigenbasis of the local Hamiltonian as
\begin{align}
    H_{\mathrm{int}}^{(0)}/\hbar
    =\frac{1}{2}
    \sum_{p'>p}
    \sum_{i,j,k,l}
    \left(
        \overline{g}_{ij,kl}^{p,p'}
        \sigma_{ji}^{(p)}
        \sigma_{lk}^{(p')}
        +
        \hc
    \right),
\label{Eq:DirectInteraction}
\end{align}
with
\begin{align}
\begin{aligned}
    \overline{g}_{ij,kl}^{p,p'}
    \equiv{}&
    \frac{
        O_{p,ij}^{\msX}
        O_{p',kl}^{\msX}
    }{\hbar}
    \left(\msM_{\msX,0}\right)_{p,p'}
    \\
    &+
    \delta_{\msX,\msY}
    \frac{
        Q_{Cp,ij}
        Q_{Cp',kl}
    }{\hbar}
    \left(
        \msC_{\msY,\infty}^{-1}
    \right)_{p,p'}.
\end{aligned}
\end{align}

\subsubsection{Bath and port-bath Hamiltonians}

The use of the canonical immittances in the chosen gauges ensures that the dressed linear environmental sector is diagonal. Consequently, $H_B$ is expressed in terms of independent (dressed) harmonic modes, with no residual quadratic mode-mode couplings, including diamagnetic-type terms~\cite{Paladino:2003,ParraRodriguez:2018}.

We first consider the atomic components of the spectral support $\mathcal B_{\msX}$, corresponding to isolated imaginary-axis poles. The associated bath Hamiltonian is
\begin{align}
    H_B
    =
    \sum_{\alpha=1}^{M}
    \sum_{\epsilon=1}^{r_\alpha}
    \hbar\omega_\alpha
    \left(
        a_{\alpha,\epsilon}^\dagger a_{\alpha,\epsilon}
        +\frac{1}{2}
    \right),\label{Eq:DiscreteInternalModeHamiltonianMaintext}
\end{align}
where $r_\alpha=\operatorname{rank}\msR_\alpha$ is the modal degeneracy associated with the pole at $\omega_\alpha$, and $[a_{\alpha,\epsilon},a_{\beta,\epsilon'}^\dagger]=\delta_{\alpha\beta}\delta_{\epsilon\epsilon'}$. 

The coupling between the external ports and the internal modes depends on the physical coupling topology shown in \cref{fig:Fig_Cauer_MT}. In both representations, it can be written as
\begin{align}
H_{PB}=\sum_\alpha\left[(\bO^\msX)^T\msG_\alpha\ba_\alpha + \text{h.c.}\right],
\label{Eq:CouplingHamiltonianDiscreteMaintext}
\end{align}
where the generic vector of operators $\bO^\msX$ is defined in \cref{Eq:CouplingOperatorDefMain}, and $\ba_\alpha$ is a vector of bosonic operators $a_{\alpha,\epsilon}$.

A possible overall sign of the admittance-representation interaction, inherited from the internal normal-coordinate convention, is absorbed into
$\msG_\alpha$ and has no physical consequence. The coupling matrix is related to the residue at the pole through
\begin{align}
    \msG_\alpha
    =
    \sqrt{\hbar|\omega_\alpha|}\,
    \msU_\alpha^\dagger
    \msR_{\mathrm D,\alpha}^{1/2},
    \label{Eq:G_alpha_discreteCoupling}
\end{align}
where
\begin{align}
    \msR_\alpha \equiv\underset{s=-i\omega_\alpha}{\operatorname{Res}}\msX(s)
    = \msU_\alpha^\dagger \msR_{\mathrm D,\alpha}\msU_\alpha
    \label{Eq:ResidueDiscreteMaintext}
\end{align}
is the hermitian positive-semidefinite residue matrix, with
$\msU_\alpha$ and $\msR_{\mathrm D,\alpha}$ denoting its eigendecomposition.

\subsection{Continuous and mixed environments}
\label{Sec:DissipativeEnvs}
We now turn to the continuous components of the spectral support $\mathcal B_{\msX}$ introduced above. For mixed environments, the isolated imaginary-axis poles, and their conjugate counterparts, are first separated from $\msX(s)$ and quantized according to \cref{Eq:DiscreteInternalModeHamiltonianMaintext,Eq:CouplingHamiltonianDiscreteMaintext}. After also extracting the possible zero- and infinite-frequency contributions already accounted for in the dressed port Hamiltonian, the remaining finite-frequency response contains only the continuous spectral component. For notational simplicity, we again denote this remaining response by $\msX(s)$, and in the continuum formulas below $\mathcal B_{\msX}$ denotes its continuous positive-frequency support, which may be bounded or disconnected. The resulting continuum of diagonal dressed environmental modes provides the multiport Caldeira--Leggett representation~\cite{CaldeiraLeggett_DQT:1981,CaldeiraLeggett_QT:1983,Devoret:1997}, building on the initial steps of Ref.~\cite{ParraRodriguez:2025}. For simplicity, throughout most of this work, we consider finite-frequency spectra that are either purely discrete or purely continuous. For a mixed environment, the complete bath and port-bath Hamiltonians are obtained by combining the discrete contributions above with the continuous contributions below, as illustrated for the left-handed metamaterial in \cref{Sec:LHTLMetamaterial}. The dressed environmental sector remains diagonal, with no residual quadratic mode-mode coupling terms.

In \cref{Sec:CauerDecomposition,Sec:CauerDecompositionY}, we derive the
continuum parameters directly from the immittance boundary values.
Here, we directly state the resulting bath and port-bath Hamiltonians. The dressed port
Hamiltonian $H_P^\msX$ retains the same form as in the discrete case, with the
capacitance and inductance matrices obtained from the same limiting
prescription. Dropping the irrelevant continuum zero-point constant, the internal modes are described by
\begin{align}
    H_B   =  \int_{\mathcal B_{\msX}}\!\mathrm d\omega\,
    \hbar\omega\,
    \ba^\dagger(\omega)\ba(\omega),
    \label{Eq:BathHamiltonianMaintext}
\end{align}
with $[a_\epsilon(\omega),a_{\epsilon'}^\dagger(\omega')]=\delta_{\epsilon\epsilon'}\delta(\omega-\omega')$. The modes couple to the nonlinear port degrees of freedom through
\begin{align}
    H_{PB}
    =
    \int_{\mathcal B_{\msX}}\!\mathrm d\omega\,
    \left[(\bO^{\msX})^T
        \msG(\omega)\ba(\omega)
        + 
        \text{h.c.}\right],
    \label{Eq:BathCouplingMaintext}
\end{align}
where the coupling matrix is related to the spectral decomposition of the immittance
$\msXH[\omega]=\msU^\dagger(\omega)\msXD(\omega)\msU(\omega)$ through
\begin{align}
    \msG(\omega) = \sqrt{\frac{\hbar|\omega|}{\pi}}\,
    \msU^\dagger(\omega)
    \msXD^{1/2}(\omega).
    \label{Eq:G_omega_continuousCoupling}
\end{align}
The matrix $\msG(\omega)$ need not have full rank. Indeed, internal modes associated with zero eigenvalues of $\msXH[\omega]$ are decoupled from the nonlinear port variables.

For the admittance representation, the continuum Hamiltonian is understood for extended port-flux directions, or more generally whenever the symmetry-adapted transformation associated with a compact collective sector is unitarily implementable. Compact sectors in gapless environments require the additional global and infrared analysis discussed after \cref{Eq:YCompactTranslationLattice}.

Before passing to effective models and applications of the formalism, we highlight that the exact Hamiltonian constructions derived above, in particular, \cref{Eq:ExternalHamiltonianZ,Eq:ExternalHamiltonianY,Eq:G_alpha_discreteCoupling,Eq:G_omega_continuousCoupling}, generalize seminal one-port constructions presented in Refs.~\cite{CaldeiraLeggett_QT:1983,Esteve:1986,Devoret:1997,Paladino:2003}, as well as multiport exact constructions in Refs.~\cite{Nigg:2012,Solgun:2015,ParraRodriguez:2019,Solgun:2019,ParraRodriguez:2022,Labarca:2024,ParraRodriguez:2025} and approximate reciprocal constructions in Refs.~\cite{Hassler:2019,Cattaneo:2019,MinevEPR:2021}. These simple immittance formulas constitute the fundamental new results of this manuscript and provide exact Hamiltonian descriptions for a broad class of circuit architectures and electromagnetic environments within a fully black-box framework. As such, they are directly compatible with standard electromagnetic simulation workflows.
\section{Dispersive models for discrete environments}
\label{Sec:SW-Models}
The exact Hamiltonians derived above provide a starting point for dispersive models of finite or countably infinite discrete environments, and allow us to generalize previous results~\cite{Solgun:2019,Labarca:2024} beyond the weakly anharmonic regime, or the approximate reciprocal cases~\cite{Khan:2024}. Here, we derive the effective model only in the two canonical charge and flux
gauges, leaving the systematic optimization over intermediate or mode-dependent gauges for future work~\cite{Roth:2019,Mehta:2022,Arwas:2023}.

For concreteness, we use the port-local basis of \cref{Eq:DiagonalizedPortHamiltonian} and assume that the residual cross-port terms satisfy the perturbative ordering associated with \cref{Eq:PortHamiltonianSplitUp}. If a subset of ports has strong direct coupling, the same construction is applied after diagonalizing that collective system Hamiltonian and expanding the physical operators $O_p^{\msX}$ in its eigenbasis.

Assuming that every retained finite-frequency transition is dispersive~\cite{Blais:2021} with respect to the internal modes, with the combined coupling to all modes remaining perturbative, and with convergent mode sums for countably infinite environments, we perform a standard Schrieffer--Wolff transformation $\tilde{H}=e^{S}He^{-S}$, with $S^\dagger=-S$ chosen such that $[S,H_0]=-H_{PB}$, where $H_0=\sum_p h_p+H_B$. The corresponding generator is
\begin{align}
\begin{aligned}
    S
    ={}&
    \sum_{\alpha,\epsilon,p}
    \sum_{i,j}
    \frac{
        (\msG_\alpha)_{p,\epsilon}
        O_{p,ij}^{\msX}
    }{
        \hbar
        \left(
            \omega_{ij}^{(p)}-\omega_\alpha
        \right)
    }
    a_{\alpha,\epsilon}
    \sigma_{ji}^{(p)}
    \\
    &+
    \sum_{\alpha,\epsilon,p}
    \sum_{i,j}
    \frac{
        (\msG_\alpha)_{p,\epsilon}^*
        O_{p,ij}^{\msX}
    }{
        \hbar
        \left(
            \omega_{ij}^{(p)}+\omega_\alpha
        \right)
    }
    a_{\alpha,\epsilon}^\dagger
    \sigma_{ji}^{(p)}.
\end{aligned}
\label{Eq:DiscreteSWGeneratorMain}
\end{align}
Projecting the transformed Hamiltonian onto the environmental vacuum and retaining terms up to second order yields
\begin{align}
    {H}_{\mathrm {SW}} = \sum_p h_p + H_{\mathrm{int}}^{(0)}  + H_{\rm LS}^{(2)} + H_{\rm int}^{(2)}.
\label{Eq:DiscreteEffectiveHamiltonianMain}
\end{align}
Here, the first two terms are the diagonalized port Hamiltonian and the direct interaction $H_{\mathrm{int}}^{(0)}$ of \cref{Eq:DirectInteraction}, respectively. 
The local second-order contribution is
\begin{align}
    H_{\rm LS}^{(2)}
    =
    \hbar
    \sum_p
    \sum_{i,l}
    \overline{\Lambda}_{il}^{(p)}
    \ket{i}_p\bra{l}_p,
\label{Eq:DiscreteLambShiftMain}
\end{align}
where $\overline{\Lambda}_{il}^{(p)}=\Lambda_{il}^{(p)}+\Lambda_{il,\infty}^{(p)}$ splits up into a frequency-independent contribution
\begin{align}
\begin{aligned}
    \Lambda_{il,\infty}^{(p)}
    ={}&
    -\frac{1}{2\hbar}
    \sum_j
    O_{p,ij}^{\msX}
    O_{p,jl}^{\msX}(\msM_{\msX,\infty})_{p,p},
\end{aligned}
\label{Eq:DiscreteStaticLambShiftCoefficientMain}
\end{align}
as well as a frequency-dependent contribution
\begin{align}
\begin{aligned}
    \Lambda_{il}^{(p)}
    ={}&
    \frac{1}{2\hbar}
    \sum_j
    O_{p,ij}^{\msX}
    O_{p,jl}^{\msX}
    \\
    &\times \sum_\alpha
    \left[
        \frac{\omega_{ji}^{(p)}(\msR_\alpha)_{p,p}}{\omega_\alpha+ \omega_{ji}^{(p)}}+ \frac{\omega_{jl}^{(p)}(\msR_\alpha)_{p,p}}{\omega_\alpha+ \omega_{jl}^{(p)}}
    \right].
\end{aligned}
\label{Eq:DiscreteLambShiftCoefficientMain}
\end{align}
Equation~\eqref{Eq:DiscreteLambShiftMain} retains the complete local second-order correction, including the matrix elements with $i\neq l$; no local secular approximation has been made.

The cross-port dispersive interaction is defined as
\begin{align}
    H_{\rm int}^{(2)}
    \equiv
    \tilde{H}_{\rm int}^{(2)}
    +
    H_{\mathrm{int}}^{\mathrm{CT}},
\end{align}
where $\tilde{H}_{\rm int}^{(2)}$ is generated by the SW transformation. Combining both contributions as derived in \cref{Sec:AppendixDiscreteSW}, it can be written as
\begin{align}
    H_{\rm int}^{(2)}/\hbar
    =
    \frac{1}{2}
    \sum_{p'>p}
    \sum_{i,j,k,l}
    \left(
        g_{ij,kl}^{p,p'}
        \sigma_{ji}^{(p)}
        \sigma_{lk}^{(p')}
        +
        \hc
    \right),
\label{Eq:DiscreteInteractionMain}
\end{align}
with the coupling coefficient
\begin{align}
\begin{aligned}
&g_{ij,kl}^{p,p'} = -i\frac{ O_{p,ij}^{\msX} O_{p',kl}^{\msX}}{2\hbar}
    \\
    &\times
    \left[
        \omega_{ij}^{(p)}
        \left(
            \msXA
            \left[
                \omega_{ij}^{(p)}
            \right]
        \right)_{p,p'}
        +
        \omega_{kl}^{(p')}
        \left(
            \msXA
            \left[
                \omega_{kl}^{(p')}
            \right]
        \right)_{p',p}
    \right].
\end{aligned}
\label{Eq:DiscreteGeneralCouplingMain}
\end{align}

The different frequency ordering obtained in Refs.~\cite{Solgun:2019,Labarca:2024} reflects a different gauge choice~\cite{Manucharyan:2017,DeBernardis:2018,Roth:2019,Stokes:2019,DiStefano:2019,Mehta:2022,Arwas:2023}, where an additional triangular canonical transformation is performed before the SW elimination. The resulting formulas (\ref{Eq:DiscreteGeneralCouplingMain}) agree with Refs.~\cite{Solgun:2019,Labarca:2024} at leading order in the near-resonant dispersive regime and have the same port-frequency ordering~\footnote{For the exchange coupling, the formulas in
Refs.~\cite{Solgun:2019,Labarca:2024} have the frequency
structure of \cref{Eq:DiscreteGeneralCouplingMain}, but with
$\omega_p(\msXA[\omega_p])_{p,p'}+\omega_{p'}(\msXA[\omega_{p'}])_{p,p'}\rightarrow
\omega_{p'}(\msXA[\omega_p])_{p,p'}+\omega_p(\msXA[\omega_{p'}])_{p,p'}$, where $\omega_p=\omega_{10}^{(p)}>0$ and
$\omega_{p'}=\omega_{10}^{(p')}>0$.
} as the reciprocal case developed in Ref.~\cite{Khan:2024}. 
In Ref.~\cite{Khan:2024}, the charge gauge frame was moreover found to provide a more accurate truncation for specific reciprocal examples. Determining which effective frame provides the more accurate truncation more generally remains an open question.

\subsection*{Example: CPB and fluxonium capacitively coupled to a common LC oscillator}
\label{Sec:DiscreteMixedQuditExample}
\begin{figure}[t]
    \centering
    \includegraphics[width=\linewidth]
    {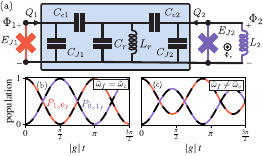}
    \caption{
    Resonator-mediated population exchange between $\ket{1_c,0_f}$ and
    $\ket{0_c,1_f}$ for a CPB-like charge qudit and a fluxonium
    capacitively coupled to the same LC mode (a). The lower panels show
    (b) the dressed-resonant case,
    $\tilde{\omega}_c=\tilde{\omega}_f$, and
    (c) an off-resonant case obtained by shifting the charge-qudit
    Josephson energy by
    $\Delta E_{J1}/h=1~\mathrm{MHz}$ from the dressed-resonant working
    point, thereby retaining a finite effective detuning between the
    transitions. Solid curves show the exact tripartite dynamics,
    including the resonator, whereas dashed curves are obtained from the
    complete effective qudit--qudit Hamiltonian in
    \cref{Eq:DiscreteEffectiveHamiltonianMain}, constructed using the
    immittance formulas. The circuit parameters are
    $\omega_r/(2\pi)=7.5~\mathrm{GHz}$,
    $C_{J1}=10~\mathrm{fF}$,
    $C_{c1}=0.4~\mathrm{fF}$,
    $C_{J2}=20~\mathrm{fF}$,
    $C_{c2}=0.8~\mathrm{fF}$,
    $C_r=90~\mathrm{fF}$,
    $E_{J1}^{\mathrm{res}}/h=5.55841~\mathrm{GHz}$,
    $E_{J1}^{\mathrm{off}}/h=5.55941~\mathrm{GHz}$,
    $E_{J2}/h=4.5~\mathrm{GHz}$,
    $E_{L2}/h=0.8~\mathrm{GHz}$,
    $n_g=0.5$, and
    $\Phi_{\rm ext}/\Phi_0=0$.
    The charge qudit, fluxonium, and resonator are truncated to
    $d_c=d_f=d_r=5$ states. The local charge-qudit and fluxonium
    Hamiltonians are numerically diagonalized in converged charge and
    phase bases, respectively.
}
\label{fig:Discrete_environment_example}
\end{figure}

To illustrate the applicability of the present dispersive formulas beyond the weakly anharmonic regime considered in previous impedance-based
approaches~\cite{Solgun:2019,Labarca:2024}, and complement related strongly anharmonic and multilevel treatments~\cite{Roth:2019,Khan:2024}, we consider
a CPB-like charge qudit and a fluxonium capacitively coupled to the same discrete LC mode, which may also represent the single-mode approximation of a finite transmission-line resonator
(see \cref{fig:Discrete_environment_example}(a)).
The multimode generalization follows directly by retaining the remaining finite-frequency impedance poles~\cite{ParraRodriguez:2018,Borrell:2026}. The total impedance of the two-port network can be readily obtained and is given by
\begin{align}
\msZ_P(s)
={}&
\frac{1}{s}
\begin{pmatrix}
C_{\Sigma1}^{-1} & 0 \\
0 & C_{\Sigma2}^{-1}
\end{pmatrix}
+
\frac{1}{C_{\mathrm{eff}}}
\frac{s}{s^2+\omega_r^2}
\bsb{\eta}_r\bsb{\eta}_r^{T},
\label{Eq:DiscreteExampleImpedance}
\end{align}
with $C_{\Sigma p}\equiv C_{Jp}+C_{cp}$, $\eta_p\equiv C_{cp}/C_{\Sigma p}$,
$\bsb{\eta}_r\equiv(\eta_1,\eta_2)^T$, $C_{\mathrm{eff}}\equiv C_r+C_{J1}\eta_1+C_{J2}\eta_2$, and $\omega_r^2\equiv1/(L_rC_{\mathrm{eff}})$.
The rank-one finite-frequency contribution makes explicit that it
corresponds to a single internal mode.

From the total impedance, one can directly extract
$\msC_{\msZ,0}^{-1}$ and $\msC_{\msZ,\infty}^{-1}$. Following the split in \cref{Eq:PortHamiltonianSplitUp}, we thus obtain the dressed port Hamiltonians for the charge and fluxonium
qudits,
\begin{align}
\begin{aligned}
    h_c
    &=
    4E_{C1}
    \left(
        \hat n_1-n_g
    \right)^2
    -
    E_{J1}\cos(\hat\varphi_1),
    \\
    h_f
    &=
    4E_{C2}\hat n_2^2
    +
    \frac{E_{L2}}{2}(\hat{\varphi}_2-\varphi_{\rm ext})^2
    -
    E_{J2}
    \cos
    \left(
        \hat\varphi_2
    \right),
    \end{aligned}
    \label{Eq:DispersiveExampleLocalHamiltonian}
\end{align}
where $E_{Cp}=e^2/(2\tilde{C}_{Jp})$ and the effective dressed
capacitances are
$\tilde{C}_{Jp}^{-1}\equiv
C_{\Sigma p}^{-1}+\eta_p^2 C_{\mathrm{eff}}^{-1}$.
The coupling operators for the two qudits are given by
$O_c=2e(\hat n_1-n_g)$ and $O_f=2e\hat n_2$.

For this circuit, $H_{\mathrm{int}}^{(0)}=0$, while the high-frequency
cross-port contribution reads
\begin{align}
H_{\mathrm{int}}^{\mathrm{CT}}
=
\frac{4e^2\eta_1\eta_2}{C_{\mathrm{eff}}}
\left(
\hat n_1-n_g
\right)
\hat n_2.
\label{Eq:DiscreteExampleDirectInteraction}
\end{align}
To obtain the coupling to the internal resonator mode, we compute the residue at the resonator frequency
[\cref{Eq:ResidueDiscreteMaintext}] and find $\msR_r=\bsb{\eta}_r\bsb{\eta}_r^{T}/(2C_{\mathrm{eff}}) =\msC_{\msZ,\infty}^{-1}/2$. Its rank-one decomposition directly gives the coupling vector in
\cref{Eq:G_alpha_discreteCoupling},
\begin{align}
    \msG_r
    \equiv
    \sqrt{\frac{\hbar\omega_r}{2C_{\mathrm{eff}}}}
    \bsb{\eta}_r,
\end{align}
such that the coupling Hamiltonian in
\cref{Eq:CouplingHamiltonianDiscreteMaintext} is given by
\begin{align}
H_{PB}
=
2e\sqrt{\frac{\hbar\omega_r}{2C_{\mathrm{eff}}}}
\left[
\,\eta_1\left(\hat n_1-n_g\right)
+
\,\eta_2\hat n_2
\right]
\hat a
+
\mathrm{h.c.}
\label{Eq:ExactCouplingInternalModesDiscrete}
\end{align}

We stress that the combination of
\cref{Eq:DispersiveExampleLocalHamiltonian},
\cref{Eq:DiscreteExampleDirectInteraction}, and
\cref{Eq:ExactCouplingInternalModesDiscrete}, together with
$H_B=\hbar\omega_r\hat a^\dagger\hat a$, yields the exact,
non-singular Hamiltonian of the circuit shown in
\cref{fig:Discrete_environment_example}(a) directly from its port
impedance, without the need to explicitly construct a Lagrangian,
perform the Legendre transformation, and quantize the full set of
circuit degrees of freedom using standard circuit-quantization
procedures~\cite{Devoret:1997,Burkard:2004,ParraRodriguez:2024,Osborne:2024,ParraRodriguez:2025}.

We now test the dispersive formulas derived above. Eliminating the LC
mode gives the effective multilevel qudit--qudit Hamiltonian
\cref{Eq:DiscreteEffectiveHamiltonianMain}, including the complete
local second-order corrections and transition-resolved mediated
interactions. To interpret the dominant population exchange shown in
\cref{fig:Discrete_environment_example}, we consider its block restricted to the
single-excitation subspace
$\{\ket{1_c,0_f},\ket{0_c,1_f}\}$. After subtracting the effective
ground-state energy, this block reads
\begin{align}
    \frac{H_{10,01}^{c,f}}{\hbar}
    =
    \begin{pmatrix}
        \tilde{\omega}_c & g  \\
        g^* & \tilde{\omega}_f
    \end{pmatrix},
\label{Eq:EffectiveExchangeBlock}
\end{align}
where, consistently to second order,
\begin{align}
\begin{aligned}
    \tilde{\omega}_c
    &=
    \omega_{10}^{(1)}
    +
    \overline{\Lambda}_{11}^{(1)}
    -
    \overline{\Lambda}_{00}^{(1)},
    \\
    \tilde{\omega}_f
    &=
    \omega_{10}^{(2)}
    +
     \overline{\Lambda}_{11}^{(2)}
    -
     \overline{\Lambda}_{00}^{(2)}.
\end{aligned}
\label{Eq:DressedTransitionFrequenciesExample}
\end{align}
Here, $\overline{\Lambda}_{ii}^{(p)}$ denotes the diagonal matrix
element of the complete local correction in
\cref{Eq:DiscreteLambShiftMain}, and
$g\equiv g_{10,01}^{c,f}
=\bra{1_c,0_f}H_{\rm int}^{(2)}\ket{0_c,1_f}/\hbar$
is the effective exchange coupling between the two transitions.

In \cref{fig:Discrete_environment_example}, we compare the complete multilevel effective model with the dynamics obtained from the full tripartite Hamiltonian, including the LC resonator. Panels (b) and (c) show the population exchange for the dressed-resonant $\delta_{\rm eff}\equiv\tilde{\omega}_c-\tilde{\omega}_f=0$
and off-resonant $\delta_{\rm eff}\neq0$ cases, respectively. The dispersive Hamiltonian accurately reproduces the full dynamics in this strongly anharmonic regime. The two-state block in
\cref{Eq:EffectiveExchangeBlock} is used only to characterize the dominant exchange process, while the complete multilevel effective Hamiltonian is retained in the simulations. Further few-level corrections, such as $ZZ$ interactions~\cite{Solgun:2022}, can be obtained straightforwardly by perturbatively eliminating higher qudit levels.

\section{Master equations for dissipative environments}
\label{Sec:ME-models}
We now focus on dissipative environments of the form described in
\cref{Sec:DissipativeEnvs}, in the weak-coupling regime in which the continuous bath modes can be eliminated using the standard Born--Markov adiabatic elimination~\cite{BreuerPetruccione:2002}. We retain the same organization used in \cref{Sec:SW-Models}: the diagonal port dressings are included in the local Hamiltonians $h_p$, while residual direct cross-port interactions are treated perturbatively. Strongly coupled port clusters should instead be incorporated into the system Hamiltonian prior to adiabatic elimination. As in the dispersive construction, we restrict the derivation to the two canonical charge and flux gauges.

\subsection{Elimination of bath modes in the continuum limit}

We use the diagonalized local Hamiltonians in
\cref{Eq:DiagonalizedPortHamiltonian} together with the continuum
Hamiltonian of \cref{Sec:DissipativeEnvs}. Within the Born approximation,
$\rho(t)\simeq\rho_P(t)\otimes\rho_B$, and we take the bath to be at zero
temperature, such that
\begin{align}
\begin{aligned}
    \langle
        a^\dag_\epsilon(\omega)
        a_{\epsilon'}(\omega')
    \rangle_B
    &=0,
    \\
    \langle
        a_\epsilon(\omega)
        a^\dag_{\epsilon'}(\omega')
    \rangle_B
    &=
    \delta_{\epsilon,\epsilon'}
    \delta(\omega-\omega').
\end{aligned}
\label{Eq:BathCorrelator}
\end{align}
To second order in $H_{PB}$, and after performing the secular or partial-secular
approximation described below, the reduced generator takes the form
\begin{align}
\begin{aligned}
    \mathcal L\rho_P
    ={}&
    -\frac{i}{\hbar}
    \left[
        \sum_p h_p
        +
        H_{\mathrm{int}}^{(0)}+H_{\rm LS}^{(2)}
        +
        H_{\rm int}^{(2)},
        \rho_P
    \right]
    \\
    & +
    \mathcal L_P^{(2)}\rho_P
    +
    \mathcal L_{\rm int}^{(2)}\rho_P,
\end{aligned}
\label{Eq:EffectiveME}
\end{align}
where
$H_{\mathrm{int}}^{(0)}$ is the direct capacitive/inductive interaction (\ref{Eq:DirectInteraction}), and the superscript $(2)$ denotes contributions retained at second order in the weak-coupling expansion. The complete second-order coherent cross-port interaction is
\begin{align}
    H_{\rm int}^{(2)}
    \equiv
    \tilde{H}_{\rm int}^{(2)}
    +
    H_{\mathrm{int}}^{\mathrm{CT}},
\label{Eq:MEInteractionSplit}
\end{align}
where the high-frequency counterterm $H_{\mathrm{int}}^{\mathrm{CT}}$ is treated perturbatively and combined with the environment-mediated interaction $\tilde{H}_{\rm int}^{(2)}$ generated at second order. Details of the nonsecular Born--Markov equation and its reduction to \cref{Eq:EffectiveME} are given in \cref{Sec:AdiabaticElimination}.

\subsection{Effective coefficients under secular and partial-secular approximations}
The secular approximation acts on Bohr-frequency blocks rather than on
individual transitions. For each positive Bohr frequency $\Omega$, we
define
\begin{align}
    \mathcal T_p(\Omega)
    \equiv
    \left\{
        (i,j):
        i>j,\;
        \omega_{ij}^{(p)}=\Omega
    \right\},
\end{align}
and the corresponding lowering component of the physical port operator,
\begin{align}
    O_p^{\msX}(\Omega)
    \equiv
    \sum_{(i,j)\in\mathcal T_p(\Omega)}
    O_{p,ji}^{\msX}
    \sigma_{ij}^{(p)}.
\label{Eq:MainTransitionBlockOperator}
\end{align}
Thus, exactly degenerate transitions are retained in the same block. Nearly degenerate transitions may similarly be grouped within a partial-secular frequency cluster centered at a common frequency
$\bar\Omega$, provided that the bath response is sufficiently smooth across that cluster and the intra-cluster splittings are not resolved on the dissipative timescale; see \cref{Sec:AdiabaticElimination}.

The complete second-order dissipator can then be written compactly as
\begin{align}
\begin{aligned}
    &
    \left(
        \mathcal L_P^{(2)}
        +
        \mathcal L_{\rm int}^{(2)}
    \right)\rho_P
    =
    \frac{2}{\hbar}
    \sum_{\Omega>0}
    \sum_{p,p'}
    \left(
        \Omega\msXH[\Omega]
    \right)_{p,p'}
    \\
    &\hspace{20mm}\times
    \mathcal D
    \left[
        O_{p'}^{\msX}(\Omega),
        {O_p^{\msX}(\Omega)}^\dagger
    \right]
    \rho_P,
\end{aligned}
\label{Eq:MainSecularDissipator}
\end{align}
where $\mathcal D[A,B]\rho \equiv A\rho B  -\frac{1}{2}
\left\{BA,\rho \right\}$ and $\mathcal D[A]\equiv\mathcal D[A,A^\dagger]$. The $p=p'$ terms yield local decay and are grouped in $\mathcal L_P^{(2)}$, whereas the $p\neq p'$
terms correspond to collective decay and are contained in $\mathcal L_{\rm int}^{(2)}$. Note that passivity implies $\msXH[\Omega]\succeq0$, and therefore guarantees that each exactly degenerate Bohr-frequency block has a positive-semidefinite Kossakowski matrix.

For a one-dimensional local transition block,
\cref{Eq:MainSecularDissipator} reduces to the familiar decay rate
\begin{align}
    \gamma_{ij}^{p}
    =
    \frac{
        2|O_{p,ij}^{\msX}|^2
    }{\hbar}
    \left(
        |\omega_{ij}^{(p)}|
        \msXH[
            \omega_{ij}^{(p)}
        ]
    \right)_{p,p},
\label{Eq:LocalDecay}
\end{align}
with the corresponding contribution
$\gamma_{ij}^{p}\mathcal D[\sigma_{ij}^{(p)}]\rho_P$.
More generally, two transitions
$(i,j)\in\mathcal T_p(\Omega)$ and
$(k,l)\in\mathcal T_{p'}(\Omega)$ have the correlated coefficient
\begin{align}
    \gamma_{ij,lk}^{p,p'}
    =
    \frac{
        2
        O_{p,ij}^{\msX}
        O_{p',lk}^{\msX}
    }{\hbar}
    \left(
        \Omega
        \msXH[\Omega]
    \right)_{p,p'}.
\label{Eq:CorrelatedDecay}
\end{align}
These expressions are special cases of the block form in
\cref{Eq:MainSecularDissipator}; in particular, degenerate transitions
within the same port are not treated as independent dissipative channels.

For well-resolved local energy eigenspaces, the secular approximation can be applied to the coherent second-order local correction, yielding the Lamb shift
\begin{align}
    H_{\rm LS}^{(2)}
    =
    \sum_p\sum_i
    \hbar\overline{\Lambda}_{i}^{(p)}
    \ket{i}_p\bra{i}_p,
\end{align}
with
$\overline{\Lambda}_{i}^{(p)}
=\Lambda_i^{(p)}+\Lambda_{i,\infty}^{(p)}$, where the static contribution is given by 
\begin{align}
\Lambda_{i,\infty}^{(p)}
=
-\frac{1}{2}
\frac{
    \langle i|
    (O_p^{\msX})^2
    |i\rangle_p
}{\hbar}
\left(
    \msM_{\msX,\infty}
\right)_{p,p},
\label{Eq:LambRenormalization}
\end{align}
and the frequency-dependent contribution reads
\begin{align}
\Lambda_i^{(p)}
=
\sum_j
\frac{|O_{p,ij}^{\msX}|^2}{\hbar}
\frac{\omega_{ji}^{(p)}}{\pi}
\mathcal P\!\int_0^\infty
\frac{
    \left(
        \msXH[\omega']
    \right)_{p,p}
}{
    \omega'+\omega_{ji}^{(p)}
}
\,\mathrm d\omega'.
\label{Eq:LambShift}
\end{align}
If instead some of the local energy eigenspaces are nearly degenerate, off-diagonal contributions in $H_{\rm LS}^{(2)}$ must be retained, leading to coherent mixing between local states (refer to \cref{Sec:AdiabaticElimination} for the full expression before performing the secular approximation).

For the coherent interaction between different ports, we define the exactly resonant transition set
\begin{align}
    \mathcal R_{ij}^{p,p'}
    \equiv
    \left\{
        (k,l):
        \omega_{kl}^{(p')}
        =
        -\omega_{ij}^{(p)}
    \right\}.
\end{align}
After combining the second-order principal-value contribution
$\tilde{H}_{\rm int}^{(2)}$ with
$H_{\mathrm{int}}^{\mathrm{CT}}$ according to
\cref{Eq:MEInteractionSplit}, the interaction takes the compact form
\begin{align}
    H_{\rm int}^{(2)}/\hbar
    =
    \sum_{p'>p}
    \sum_{i>j}
    \sum_{(k,l)\in\mathcal R_{ij}^{p,p'}}
    \left(
        g_{ij,kl}^{p,p'}
        \sigma_{ji}^{(p)}
        \sigma_{lk}^{(p')}
        +
        \hc
    \right),
\end{align}
with
\begin{align}
    g_{ij,kl}^{p,p'}
    =
    -i
    \frac{
        O_{p,ij}^{\msX}
        O_{p',kl}^{\msX}
    }{\hbar}
    \left(
        \omega_{ij}^{(p)}
        \msXA[
            \omega_{ij}^{(p)}
        ]
    \right)_{p,p'}.
\label{Eq:CoherentCoupling}
\end{align}
For a partial-secular cluster, the corresponding bath kernels are evaluated
at its common representative frequency $\bar\Omega$, as described in \cref{Sec:AdiabaticElimination}. The independent direct interaction
$H_{\mathrm{int}}^{(0)}$ remains explicitly separated in
\cref{Eq:EffectiveME}.

\subsection{Validity of the effective master equation}
\label{Sec:ValidityAE}

Besides weak system--bath coupling, the Markov approximation requires the
environmental response to vary weakly over the frequency scales induced by
the reduced dynamics~\cite{BreuerPetruccione:2002,Hassler:2019}. In the
spirit of the self-consistency condition of Ref.~\cite{Hassler:2019}, a
convenient local criterion for a transition block centered at $\Omega$ is
\begin{align}
    \epsilon_{\mathrm{AE}}
    \equiv
    \frac{O_{\Omega}^{2}}{\hbar}
    \left\|
        \partial_\omega
        \left(
            \omega\msX[\omega]
        \right)_{\omega=\Omega}
    \right\|
    \ll 1,
    \label{Eq:ValidityAE}
\end{align}
where $O_{\Omega}^{2}$ denotes the largest relevant product
$|O_{p,ij}^{\msX}O_{p',lk}^{\msX}|$ among the transitions retained in the block.

This local condition should be understood as a heuristic criterion for the validity of the Markov approximation. If the first derivative vanishes, or the response varies nonlinearly over the dynamically resolved frequency window, higher-order variations must also be considered. Independently, the secular approximation requires well-separated Bohr-frequency blocks, while unresolved transitions must be retained within the same partial-secular
cluster~\cite{Schaller:2008,Cattaneo:2019,Trushechkin:2021}. More refined validity criteria remain an active topic in open-system theory.

\section{Circuit examples with
continuous environments}
\label{Sec:DissipativeExamples}

We now apply the exact continuum construction and, where appropriate, the dissipative formalism to representative continuous environments. We first consider one-port reciprocal systems, including conventional \gls{TLs} and metamaterial environments, and then turn to waveguide-mediated  and giant-atom configurations, including nonreciprocal settings. Focusing mainly on capacitively coupled qudits, including the galvanic limit, we explicitly show exact Hamiltonian constructions and controlled effective models obtained from their immittance responses.

\subsection{Single nonlinear element coupled to reciprocal continuous environments}

\label{Sec:ExamplesSingleQubits}

We begin with one-port reciprocal environments. These examples illustrate
the exact impedance- and admittance-based Hamiltonian constructions and the
validity of the Markovian reduction for smooth and structured spectra.

\subsubsection{Charge qudit capacitively coupled to semi-infinite \gls{TL}}
\label{Sec:SingleQubitSemiInfiniteTL}

\begin{figure}[t]
    \centering
\includegraphics[width=.95\linewidth]{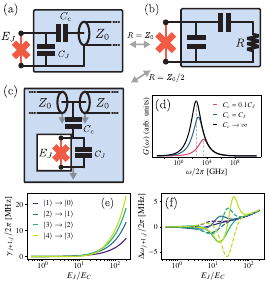}
    \caption{The charge qudit capacitively coupled to (a) a semi-infinite and (c) an infinite TL, both of characteristic impedance $Z_0$, can be represented by the equivalent circuit in panel (b), with $R=Z_0$ and $R=Z_0/2$ for the semi-infinite and infinite lines, respectively. (d) Coupling parameter $G(\omega)$ for $R=Z_0$ and $C_J=10~\mathrm{fF}$ as a function of frequency for different coupling capacitances, including the galvanic limit ($C_c\rightarrow\infty$). Panels (e) and (f) show the decay rates and corrections to the transition frequencies, as defined in \cref{Eq:FrequencyCorrection}, for the semi-infinite line in panel (a). In both panels, $Z_0=50~\Omega$, $C_J=10~\mathrm{fF}$, $C_c=C_J/50$, and $E_J$ is varied to explore the transition from the CPB to the transmon regime. The solid lines in panel (f) correspond to $n_g=0$, and the dashed lines to $n_g=0.5$.}
    \label{fig:1CQ_Cc_TLs}
\end{figure}

As a first example, we study the system depicted in Fig.~\ref{fig:1CQ_Cc_TLs}(a) comprising a charge qudit capacitively coupled to a semi-infinite \gls{TL} with characteristic impedance $Z_0$. Previous derivations of the exact Hamiltonian for this case can be found in Refs.~\cite{Paladino:2003,ParraRodriguez:2018}. Using the input-impedance equivalence between a semi-infinite \gls{TL} and a resistor of resistance $Z_0$~\cite{Pozar:2009} (see \cref{Sec:ImpedanceSemiInf} for details), the total impedance seen from the \gls{JJ} port in the Laplace domain is, in canonical form, 
\begin{align}
    Z_{P}(s)=\frac{1}{sC_\Sigma}+\frac{C_c^2}{C_\Sigma^2}\frac{Z_0}{1+s/\wcut},
\end{align}
where we introduced $C_\Sigma\equiv C_J+C_c$ and the cutoff frequency $\wcut\equiv(Z_0C_cC_J/C_\Sigma)^{-1}$. We directly identify the residue of the zero-frequency pole as $C_{Z,0}^{-1}\equiv C_\Sigma^{-1}$, and the remaining regularized impedance $Z(s)$ has a single pole on the negative real axis. Furthermore, the high-frequency effective capacitance is readily found to be $C_{Z,\infty}=C_JC_\Sigma/C_c$, allowing us to obtain the local port Hamiltonian from \cref{Eq:ExternalHamiltonianZ},
\begin{align}
    H_P^\msZ=4E_C(\hat{n}-n_g)^2-E_J\cos(\hat\varphi),
    \label{Eq:BareChargeHamiltonian}
\end{align}
with $E_C\equiv E_{C_J}=e^2/(2C_J)$. Importantly, the capacitance entering the local Hamiltonian in the charge gauge is the bare capacitance $C_J$ rather than the capacitance $C_\Sigma$ commonly used in perturbative treatments. 

Using the expression for the coupling strength in \cref{Eq:G_omega_continuousCoupling}, we obtain the full Hamiltonian
\begin{align}
    H=H_P^\msZ+H_B +\hat Q \int_{\mathbbm{R}^+}\!\mathrm d\omega\,
     G(\omega)[a(\omega)+a^\dag(\omega)],\label{eq:H_1CQ_Cc_Semi-Inf}
\end{align}
with $\hat Q=2e(\hat n-n_g)$, $H_B$ given in \cref{Eq:BathHamiltonianMaintext} and 
\begin{align}
    G(\omega)=\frac{C_c}{C_\Sigma}\sqrt{\frac{\hbar\omega}{\pi}\frac{Z_0 }{1+(\omega/\wcut)^2}}.
\end{align}
This example illustrates a regular galvanic limit $C_c\rightarrow\infty$. In this case, the coupling ratio saturates (i.e., $C_c/C_\Sigma\rightarrow1$) and the cutoff frequency shifts to lower frequencies
[$\wcut\rightarrow\tilde{\omega}_{\rm cut}\equiv(Z_0C_J)^{-1}$], such that the coupling parameter [see Fig.~\ref{fig:1CQ_Cc_TLs}(d)] becomes 
\begin{align}
 G_{\rm gal}(\omega)=\sqrt{\frac{\hbar\omega}{\pi}\frac{Z_0}{1+(\omega/\tilde{\omega}_{\rm cut})^2}},
\end{align}
whereas the dressed inverse capacitance $C_{Z,0}^{-1}+C_{Z,\infty}^{-1}=1/C_J$ remains invariant~\cite{Mehta:2022}.

The above Hamiltonian is an exact representation of the circuit given in \cref{fig:1CQ_Cc_TLs}(a) and can be used as a starting point to model the full system without invoking further approximations, in particular, the Markov approximation. Here, however, we focus on the weak-coupling effective description developed in \cref{Sec:ME-models}. We first extract the decay rate from \cref{Eq:LocalDecay} by taking the real part of the regular impedance. For two states $i>j$, we obtain
\begin{align}
    \gamma_{ij}=\frac{2|Q_{ij}|^2}{\hbar}\frac{C_c^2}{C_\Sigma^2}\frac{|\omega_{ij}|Z_0}{1+\omega_{ij}^2/\wcut^2}.
\end{align}    
In \cref{fig:1CQ_Cc_TLs}(e), we show the decay rates between pairs of neighboring states for a charge qudit with different ratios of $E_J/E_C$. We first observe that the decay between neighboring eigenstates decreases as $E_J/E_C$ is reduced. This observation can be understood by considering the limiting case $E_J=0$, for which the qudit is diagonal in the charge basis, and hence the capacitive coupling to the \gls{TL} cannot induce transitions between different states. Note that the crossover occurs earlier for higher-excited states, as they approach the charge eigenstates even for finite $E_J$. In the opposite limit of large $E_J/E_C$, the system approaches a harmonic oscillator, for which the decay rate between states $j+1$ and $j$ scales with the bosonic enhancement $|Q_{j+1,j}|^2\sim j+1$.

Next, we examine the correction to the charge-qudit energy spectrum. As mentioned above, we use the bare charging energy [$E_C=E_{C_J}$ in \cref{Eq:BareChargeHamiltonian}], whereas the literature has historically (e.g., Refs.~\cite{CaldeiraLeggett_QT:1983,Esteve:1986,Kockum:2014,Hassler:2019,Gheeraert:2020} among others) used the dressed charging energy $E_C=E_{C_\Sigma}=e^2/(2C_\Sigma)$ and omitted or regularized ultraviolet-divergent Lamb-shift contributions from the coupling to the \gls{TL}. We therefore compare the transition energies between two neighboring states obtained from the dressed Hamiltonian (\cref{Eq:BareChargeHamiltonian} with $E_C=E_{C_\Sigma}$) with those obtained from the bare Hamiltonian ($E_C=E_{C_J}$) after including the \textit{finite} Lamb-shift corrections [\cref{Eq:LambShift,Eq:LambRenormalization}] from our effective model, that is, we consider 
\begin{align}
\begin{aligned}
    \Delta \omega_{j+1,j}\equiv&\, \omega_{j+1,j}(E_{C_J})+(\overline{\Lambda}^{(j+1)}_{E_{C_J}}-\overline{\Lambda}^{(j)}_{E_{C_J}})\\
&-\omega_{j+1,j}(E_{C_\Sigma}). 
\end{aligned}
\label{Eq:FrequencyCorrection}
\end{align} 
In \cref{fig:1CQ_Cc_TLs}(f), we show the discrepancies in the transition frequencies for different $E_J/E_C$ and for two values of $n_g$. Three regimes can be distinguished. For $E_J/E_C\ll1$, the discrepancy vanishes. Indeed, as mentioned above, at $E_J=0$, the Hamiltonian and charge operator are simultaneously diagonal in the charge basis, such that the frequency-dependent contribution in \cref{Eq:LambShift} vanishes, whereas \cref{Eq:LambRenormalization} exactly reproduces the capacitive dressing. In the opposite limit, $E_J/E_C\gg1$, the system approaches a harmonic oscillator, and the level shifts between adjacent levels become constant. Consequently, all neighboring transition frequencies acquire the same correction, and the dependence on $n_g$ disappears. In the crossover regime, Josephson tunneling mixes several charge states, resulting in state- and $n_g$-dependent transition frequencies and charge-operator matrix elements. This produces the pronounced nonmonotonic corrections observed between the two limiting regimes.

\subsubsection{Charge qudit capacitively coupled to infinite \gls{TL}}
\label{Sec:SingleQubitInfiniteTL}

We now consider the example shown in \cref{fig:1CQ_Cc_TLs}(c), where the semi-infinite line is extended to infinity in the opposite direction as well. The impedance seen by the junction can be computed in the same way as for the semi-infinite case studied above, with the crucial difference that the infinite \gls{TL} is equivalent, as seen from the injection point, to the parallel combination of two semi-infinite \gls{TL}s (see \cref{Sec:ImpedanceSemiInf}) and therefore presents an effective impedance $Z_0/2$. Following the same steps as in the semi-infinite case, we obtain the Hamiltonian in \cref{eq:H_1CQ_Cc_Semi-Inf} with the replacement $Z_0\rightarrow Z_0/2$, such that the corresponding cutoff becomes $2\wcut$. After the analogous elimination of the bath, the decay rate is readily obtained as
\begin{align}
    \gamma_{ij}=\frac{2|Q_{ij}|^2}{\hbar}\frac{C_c^2}{C_\Sigma^2}\frac{|\omega_{ij}|Z_0/2}{1+\omega_{ij}^2/(2\wcut)^2}.
\end{align}

Consequently, it is straightforward to show that the decay rate of the charge qudit coupled to the infinite \gls{TL} satisfies $\gamma_{ij}^{\text{inf}}\approx \gamma_{ij}^{\text{semi-inf}}/2$ when the transition frequencies are well below the cutoff, $|\omega_{ij}| \ll \wcut$, whereas $\gamma_{ij}^{\text{inf}}\approx 2\gamma_{ij}^{\text{semi-inf}}$ in the high-frequency regime $|\omega_{ij}| \gg \wcut$.

\subsubsection{Coupling to a low-$Q$ resonator}
\label{Subsubsec:Low-Q-resonator}

\begin{figure}[t]
    \centering
    \includegraphics[width=\linewidth]{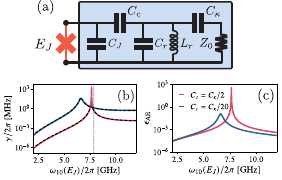}
    \caption{(a) Transmon qudit ($C_J=200~\mathrm{fF}$, variable $E_J$, and $n_g=0$) capacitively coupled ($C_c=C_J/50$) to a lossy resonator ($C_r=200~\mathrm{fF}$, $E_{L_r}/h=80~\mathrm{GHz}$). (b) Decay rate of the transmon for different ratios of $C_c/C_\kappa$ and (c) validity condition of adiabatic elimination. The result from the impedance formula is compared with the decay rate obtained from the poles of the linearized circuit (dashed lines). The bare resonator frequency is indicated by the dotted line; the actual pole is shifted due to the coupling capacitances.
    }
    \label{fig:Low-Q-resonator}
\end{figure}

In the previous examples, the artificial atom was directly coupled to a \gls{TL}. Here, we study a structured bath, in which the coupling of the charge qudit to the \gls{TL} is mediated by an LC resonator [see \cref{fig:Low-Q-resonator}(a)]. The impedance seen from the \gls{JJ} port can be readily found (see \cref{App:LowQEnvironment})
\begin{align}
    Z_P(s)
    =
    \frac{1}{sC_\Sigma}+
    \frac{C_c^2}{C_\Sigma^2}
    \frac{Z_r(s)}{1+sC_\ast Z_r(s)},
\end{align}
with $C_\ast=C_c C_J/C_\Sigma$, $C_\Sigma=C_J+C_c$, and $Z_r(s)$ the loaded resonator impedance given in \cref{App:LowQEnvironment}. 

The impedance $Z(s)= Z_P(s)-1/(sC_\Sigma)$ only contains dissipative poles in the left half-plane and thus describes a structured dissipative environment. Provided that its spectral response is sufficiently smooth around the relevant qudit transition frequencies, the weak-coupling master equation derived in \cref{Sec:ME-models} can be directly applied. Near resonance, this requires the qudit-resonator coupling to be much smaller than the cavity linewidth~\cite{Blais:2021}. For the parameters considered below, increasing $C_\kappa/C_c$ favors this bad-cavity regime.

In \cref{fig:Low-Q-resonator}(b), we show the decay rate of the first transmon transition as a function of the uncoupled transmon frequency ($C_c=0$). We fix $C_J=200~\mathrm{fF}$ and tune the junction energy $E_J$ within the transmon regime, $E_J/E_C\gtrsim 50$, corresponding to the frequency range shown in the figure, for two different ratios of $C_c/C_\kappa$. We compare our impedance formula with the decay rate extracted from the poles of the linearized version of the circuit shown in \cref{fig:Low-Q-resonator}(a)~\cite{Houck:2008,Hassler:2019}. As expected, for $C_\kappa\gg C_c$, the two decay rates agree almost perfectly, whereas for $C_c\approx C_\kappa$ the two formulas disagree around the Purcell peak. This behavior is consistent with the deterioration of the heuristic condition (\ref{Eq:ValidityAE}), as can be seen in \cref{fig:Low-Q-resonator}(c). Both the linearized model and the impedance formula capture physics beyond the standard Purcell approximation~\cite{purcell:1946,Blais:2021}, such as the asymmetry in the decay rate below and above the resonator frequency~\cite{Houck:2008}. While the comparison in \cref{fig:Low-Q-resonator}(b) is restricted to the transmon regime, our formulas also apply to strongly anharmonic qudits.

\subsubsection{Fluxonium inductively coupled to semi-infinite TL}
\label{Sec:ExampleFluxoniumInductive}

\begin{figure}[t]
    \centering
    \includegraphics[width=\linewidth]
        {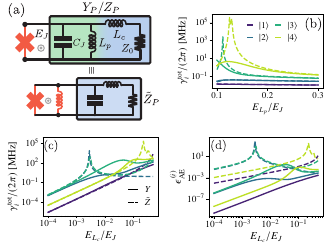}
    \caption{
    (a) Fluxonium qudit inductively coupled to a semi-infinite
    \gls{TL}, with the $Y_P$, $Z_P$, and $\tilde Z_P$
    descriptions considered in the text. The static external flux
    is set to $\varphi_{\mathrm{ext}}=0$. (b) Total decay rates from state $i$ to all lower-lying states as a function of $E_{L_p}/E_J$, for $E_J/h=6$~GHz,
    $E_C/h=1$~GHz, and $L_c=3.41~\mu\mathrm{H}$.
    (c) Total decay rates obtained from the $Y$ and $\tilde Z$
    constructions as a function of the inductive coupling, for
    fixed $(E_J,E_C,E_{L_p})/h=(6,1.5,0.4)$~GHz.
    (d) Corresponding adiabatic-elimination parameters
    $\epsilon_{\mathrm{AE}}^{(i)}$.
    Solid and dashed lines correspond to the $Y$ and $\tilde Z$
    constructions, respectively.
    }
    \label{fig:ExampleFluxonium}
\end{figure}

We now consider a fluxonium~\cite{Manucharyan:2009} inductively
coupled to a semi-infinite \gls{TL} with characteristic impedance
$Z_0$, see \cref{fig:ExampleFluxonium}(a). Since this circuit belongs
to both the $Y_P$ and $Z_P$ classes, it provides a simple setting
in which the two constructions can be compared explicitly.
We first analyze it using the admittance formulation.

The total admittance seen by the junction is
\begin{align}
    Y_P(s)
    =
    sC_J+\frac{1}{sL_p}+\frac{1}{Z_0+sL_c},
\end{align}
with cutoff frequency $\wcut=Z_0/L_c$. The coefficients in
\cref{eq:Y_P_class} are
$C_{Y,\infty}=C_J$, $L_{Y,0}=L_p$, and
$L_{Y,\infty}^{-1}=L_c^{-1}$.
Following Appendices~\ref{AppSec:FirstOrderReview} and \ref{AppSec:Y_construction_and_quantization}, we assign
the static external flux to the junction branch,
$\Phi_J=\Phi+\Phi_e$. Writing
$\varphi=2\pi\Phi/\Phi_0$ and
$\varphi_{\mathrm{ext}}=2\pi\Phi_e/\Phi_0$,
\cref{Eq:ExternalHamiltonianY} gives
\begin{align}
    H_P^Y
    =
    4E_C\hat n^2
    -E_J\cos(\hat\varphi+\varphi_{\mathrm{ext}})
    +\frac{E_{L_p}+E_{L_c}}{2}\hat\varphi^2,
    \label{eq:H_P_Y_fluxonium}
\end{align}
where $E_C=e^2/(2C_J)$ and
$E_{L_\nu}=(\Phi_0/2\pi)^2/L_\nu$, with $\nu=p,c$.

The regular admittance entering the system-bath coupling is
$Y(s)=1/(Z_0+sL_c)$, which gives the Lorentz--Drude coupling
\begin{align}
    G_Y(\omega)
    =
    \sqrt{\frac{\hbar\omega}{\pi Z_0}}\,
    \frac{1}{\sqrt{1+(\omega/\wcut)^2}}.
    \label{Eq:FluxoniumGY}
\end{align}

For this one-port circuit, the admittance construction is
well established~\cite{CaldeiraLeggett_QT:1983,Esteve:1986,
Devoret:1997}. The full junction impedance $Z_P=Y_P^{-1}$
satisfies $\lim_{\eta\to0^+}\eta Z_P^{-1}(\eta) =1/L_p$, so the criterion of \cref{Eq:ZExtendedDirectionCriterion} identifies the port-flux coordinate as extended.

For an alternative impedance partition, we retain $L_p$
explicitly in the local potential and characterize the remaining
passive environment by
\begin{align}
\begin{aligned}
    \tilde Z_P(s)
    &\equiv
    \left[
        Y_P(s)-\frac{1}{sL_p}
    \right]^{-1}
    \\
    &=
    \frac{Z_0+sL_c}
    {1+sC_J(Z_0+sL_c)}.
\end{aligned}
\label{Eq:FluxoniumZtilde}
\end{align}
We obtain $C_{\tilde Z,0}^{-1}=0$ and
$C_{\tilde Z,\infty}^{-1}=C_J^{-1}$.
Keeping the external flux in the junction potential as prescribed
in Appendix~\ref{AppSec:FirstOrderReview}, the port Hamiltonian is
\begin{align}
    H_P^{\tilde Z}
    =
    4E_C\hat n^2
    -E_J\cos(\hat\varphi+\varphi_{\mathrm{ext}})
    +\frac{E_{L_p}}{2}\hat\varphi^2,
\end{align}
and the coupling function reads
\begin{align}
    G_{\tilde Z}(\omega)
    =
    \sqrt{\frac{\hbar\omega Z_0}{\pi}}\,
    \frac{1}{
        \sqrt{
            (1-C_JL_c\omega^2)^2
            +(Z_0C_J\omega)^2
        }
    }.
    \label{Eq:FluxoniumGZtilde}
\end{align}
The explicit inductive term in the local potential keeps
the port-flux coordinate extended in this partition as well.
The $Y$ and $\tilde Z$ constructions thus provide different
system-bath partitions of the same circuit while using the
same external-flux convention.

In \cref{fig:ExampleFluxonium}(b), we plot the total decay rate
$\gamma_i^{\mathrm{tot}}\equiv\sum_{j<i}\gamma_{ij}$ of the first
four excited states as a function of $E_{L_p}/E_J$, comparing
the $Y_P$ and $\tilde Z_P$ constructions. For
$E_{L_p}/E_J\gtrsim0.2$, the low-energy states are confined to
an approximately quadratic minimum, the two gauges converge,
and the flux operator follows approximately harmonic-oscillator
selection rules, with
$\gamma_i^{\mathrm{tot}}\simeq\gamma_{i,i-1}$ and
$|\varphi_{i,i-1}|^2\propto i$.
At smaller $E_{L_p}/E_J$, additional Josephson minima become
relevant and the increasingly fluxon-like states produce
nonmonotonic, state-dependent decay rates, still dominated
by neighboring-level transitions.

In \cref{fig:ExampleFluxonium}(c), we compare instead the $Y$
and $\tilde Z$ constructions as the inductive coupling is varied.
Their decay rates converge in the weak-coupling limit and differ
increasingly as the adiabatic elimination becomes less controlled,
as quantified by $\epsilon_{\mathrm{AE}}^{(i)}$ in
\cref{fig:ExampleFluxonium}(d). For the parameters shown, the two reductions agree closely in the effectively harmonic regime and converge in the vanishing-coupling limit. Away from these limits, equivalent exact Hamiltonians may yield different finite-order reduced models because their system-bath
partitions differ. The optimal gauge or partition may therefore depend on the approximation and parameter regime and remains an open question in  general~\cite{Manucharyan:2017,DeBernardis:2018,Roth:2019,Mehta:2022,Arwas:2023}.

\subsection{Metamaterial environments}
\label{Sec:MetamaterialExamples}

We now consider one-port metamaterial environments, where engineered
propagation bands and potentially large impedances can enhance
electromagnetic dressing and make weak-loading approximations increasingly
restrictive~\cite{Zueco:2012,Egger:2013,PuertasMartinez:2019,
Leger:2019,Indrajeet:2020}. We illustrate the immittance construction with
the two examples shown in \cref{fig:Metamaterials}.

\subsubsection{Josephson-junction-array metamaterial}
\label{Sec:JJAMetamaterial}

\begin{figure}[t]
\centering
\includegraphics[width=\linewidth]{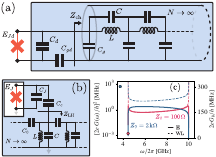}
\caption{(a) A Josephson junction coupled to a semi-infinite linearized Josephson-junction array as in Ref.~\cite{Snyman:2015}, with array junctions described by $L\parallel C$ and each node shunted to ground by $C_g$. Constant voltage sources present in the original circuit are omitted for clarity. (b) A Josephson junction capacitively coupled to an infinite left-handed \gls{TL} as in Ref.~\cite{Goswami:2026}, formed by series capacitances $C_0$ and shunt $L\parallel C$ branches. (c) Continuum coupling parameter $[2eG(\omega)/h]^2$ obtained from
the exact loaded impedance (solid, E) and from the weak-loading approximation (dashed, WL), for the left-handed circuit in (b). We use $C_J=200~\mathrm{fF}$, $C_c=C_J/6$, $\omega_{\rm IR}/(2\pi)=4.5~\mathrm{GHz}$,
$\omega_{\rm UV}/(2\pi)=10~\mathrm{GHz}$, corresponding to     $\omega_0/(2\pi)\simeq10.08~\mathrm{GHz}$ with
$\omega_0=(LC_0)^{-1/2}$, and compare
$Z_0=\sqrt{L/C_0}=100~\Omega$ and $2~\mathrm{k}\Omega$.
The colored circles indicate the additional finite-frequency pole below the propagation band, with its coupling $2eG_b/h$ shown on the right axis. For $Z_0=100~\Omega$ ($2~\mathrm{k}\Omega$), 
$\omega_b/(2\pi)\simeq4.496~\mathrm{GHz}$
($3.769~\mathrm{GHz}$) and    $2eG_b/h\simeq28.4~\mathrm{MHz}$  ($304~\mathrm{MHz}$).
}
\label{fig:Metamaterials}
\end{figure}

We first consider a nonlinear Josephson junction coupled to a semi-infinite Josephson-junction array as in Ref.~\cite{Snyman:2015}, shown in \cref{fig:Metamaterials}(a), and recover its Hamiltonian construction directly from the port impedance
(see \cref{AppSec:JJAMetamaterial} for the derivation). All junctions in the array are linearized as $L\parallel C$ [with admittance $Y(s)=sC+1/(sL)$], while each array node has capacitance $C_g$ to ground. In the thermodynamic limit, the driving-point impedance of the linearized array is
\begin{align}
Z_{\rm ch}(s)
=
\frac{1}{2Y(s)}
\left[
\sqrt{1+\frac{4Y(s)}{sC_g}}-1
\right],
\label{Eq:JJAChainImpedance}
\end{align}
with spectral domain $\mathcal B_Z=(0,\omega_{\rm UV})$ and
$\omega_{\rm UV}=2/\sqrt{L(C_g+4C)}$.

The \gls{JJ}, with energy $E_{Jd}$ and parallel capacitance $C_d$, defines the port between the island shunted by $C_{gd}$ and the first array node. The total impedance seen from the terminals of the \gls{JJ} is given by
\begin{align}
Z_P(s)
=&
\left[
sC_d+
\left(
\frac{1}{sC_{gd}}+Z_{\rm ch}(s)
\right)^{-1}
\right]^{-1}
\\=&
\frac{1}{sC_\Sigma}
+
\eta^2
\frac{Z_{\rm ch}(s)}
{1+sC_*Z_{\rm ch}(s)},
\label{Eq:JJARegularZ}
\end{align}
where
$C_\Sigma=C_d+C_{gd}$,
$C_*=C_dC_{gd}/C_\Sigma$, and
$\eta=C_{gd}/C_\Sigma$.
The low- and high-frequency limits give
$C_{Z,0}^{-1}=C_\Sigma^{-1}$ and
$C_{Z,\infty}^{-1}=\eta^2/(C_b+C_*)$, with
$C_b=[C_g+(C_g^2+4CC_g)^{1/2}]/2$. 

The exact Hamiltonian then follows as
\begin{align}
\hat H={}&
4E_C(\hat n-n_g)^2-E_{Jd}\cos(\hat\varphi)
+\int_{\mathcal B_Z}
\mathrm d\omega\,
\hbar\omega\,\hat a^\dagger(\omega)\hat a(\omega)
\nonumber\\
&+
2e(\hat n-n_g)
\int_{\mathcal B_Z}
\mathrm d\omega\,
G(\omega)
\left[
\hat a(\omega)+\hat a^\dagger(\omega)
\right],
\label{Eq:JJAExactHamiltonian}
\end{align}
where $E_C=e^2/(2C_{\rm eff})$, with the resulting local charging capacitance
$C_{\rm eff}\equiv(C_{Z,0}^{-1}+C_{Z,\infty}^{-1})^{-1}
=C_d+C_{gd}C_b/(C_{gd}+C_b)$.
From the pole-subtracted impedance, we obtain the coupling
\begin{align}
G^2(\omega)
=
\frac{\hbar\eta^2\omega}{\pi}
\sqrt{\frac{L}{C_g}}\,
\frac{
\sqrt{1-(\omega/\omega_{\rm UV})^2}
}{
1-\omega^2LC+r(1+r)\omega^2LC_g
},
\label{Eq:JJAG}
\end{align}
for $\omega\in\mathcal B_Z$, with $r=C_*/C_g$. Thus, we recover the continuum Hamiltonian of Ref.~\cite{Snyman:2015} directly from the one-port impedance, without explicitly constructing the array of modes.

\subsubsection{Left-handed TL metamaterial}
\label{Sec:LHTLMetamaterial}
As a complementary metamaterial example, we consider the transmon capacitively coupled
to the infinite left-handed transmission line of
Ref.~\cite{Goswami:2026}, shown in \cref{fig:Metamaterials}(b).
Each unit cell contains a series capacitance $C_0$ and a shunt
$L\parallel C$. With $Z_s(s)=1/(sC_0)$ and
$Y_p(s)=sC+1/(sL)$, the impedance seen from an internal node of the infinite line is
\begin{align}
Z_{\rm LH}(s)
=
\frac{1}{
\sqrt{Y_p(s)\left[Y_p(s)+4/Z_s(s)\right]}
}.
\label{Eq:LHTLImpedance}
\end{align}
Its propagation band is
$\mathcal B_Z=(\omega_{\rm IR},\omega_{\rm UV})$, with
$\omega_{\rm IR}=1/\sqrt{L(C+4C_0)}$ and
$\omega_{\rm UV}=1/\sqrt{LC}$, and the corresponding dispersion is
\begin{align}
\omega_k
=
\frac{1}{
\sqrt{L[C+2C_0(1-\cos(kd))]}},
\label{Eq:LHTLDispersion}
\end{align}
see also Ref.~\cite{Goswami:2026}.

Defining $C_\Sigma=C_J+C_c$, $C_*=C_JC_c/C_\Sigma$, and
$\eta=C_c/C_\Sigma$, we find the impedance at the port of the \gls{JJ} 
\begin{align}
Z_P(s)
=
\frac{1}{sC_\Sigma}
+
\eta^2
\frac{Z_{\rm LH}(s)}
{1+sC_*Z_{\rm LH}(s)}.
\label{Eq:LHTLRegularZ}
\end{align}
Since $Z_{\rm LH}(s)\sim1/(sC_b)$ at high frequency, with
$C_b=\sqrt{C(C+4C_0)}$, the resulting local charging capacitance is 
\begin{align}
C_{\rm eff}
=
C_J+\frac{C_cC_b}{C_c+C_b}.
\label{Eq:LHTLCeff}
\end{align}
Besides the continuum supported on $\mathcal B_Z$, the regular part of
\cref{Eq:LHTLRegularZ} contains one isolated pole at $\omega_b$ below the
propagation band. Using
$Z_{\rm LH}(s)
=sL/\sqrt{(1+s^2/\omega_{\rm UV}^2)
(1+s^2/\omega_{\rm IR}^2)}$, its position follows directly from
\begin{align}
\sqrt{
\left(1-\frac{\omega_b^2}{\omega_{\rm UV}^2}\right)
\left(1-\frac{\omega_b^2}{\omega_{\rm IR}^2}\right)
}
=
\omega_b^2LC_*,
\qquad
0<\omega_b<\omega_{\rm IR}.
\label{Eq:LHTLBoundPole}
\end{align}
For every finite $C_*>0$ this equation has a unique solution strictly
below $\omega_{\rm IR}$. The pole reaches the continuum edge only in the zero-loading limit $C_*\rightarrow0$. Its Foster/Cauer contribution can be written as
\begin{align}
Z_b(s)
=
\frac{2R_b s}{s^2+\omega_b^2},
\label{Eq:LHTLBoundPoleFoster}
\end{align}
with the residue $R_b\equiv\operatorname*{Res}_{s=-i\omega_b}Z_P(s)$ given by
\begin{align}
R_b
=
\frac{\eta^2}{C_*}
\left[
2
+
\frac{
(\omega_b/\omega_{\rm UV})^2
}{
1-(\omega_b/\omega_{\rm UV})^2
}
+
\frac{
(\omega_b/\omega_{\rm IR})^2
}{
1-(\omega_b/\omega_{\rm IR})^2
}
\right]^{-1},
\label{Eq:LHTLBoundCoupling}
\end{align}
and $G_b^2=\hbar\omega_bR_b$. Thus, both the frequency and coupling of the additional discrete mode
follow directly from the position and residue of the finite pole of the
port response.

The exact Hamiltonian is
$\hat H=\hat H_P^Z+\hat H_{PB}+\hat H_B$, with
\begin{align}
\begin{aligned}
\hat H_P^Z
={}&
4E_C(\hat n-n_g)^2-E_J\cos(\hat\varphi),
\\
\hat H_{PB}
={}&
(\hat n-n_g)
g_b(\hat b+\hat b^\dagger)
\\
&+
(\hat n-n_g)
\int_{\mathcal B_Z}
\mathrm d\omega\,
g(\omega)
[\hat a(\omega)+\hat a^\dagger(\omega)],
\\
\hat H_B
={}&
\hbar\omega_b\hat b^\dagger\hat b
+
\int_{\mathcal B_Z}
\mathrm d\omega\,
\hbar\omega\,
\hat a^\dagger(\omega)\hat a(\omega),
\end{aligned}
\label{Eq:LHTLExactHamiltonian}
\end{align}
where $E_C=e^2/(2C_{\rm eff})$, $g_b\equiv2eG_b$, and
$g(\omega)\equiv2eG(\omega)$. The exact continuum coupling follows
directly from the loaded port response as
\begin{align}
G^2(\omega)
=
\frac{\hbar\eta^2\omega^2L}{\pi}
\frac{
\sqrt{
\left(1-\omega^2/\omega_{\rm UV}^2\right)
\left(\omega^2/\omega_{\rm IR}^2-1\right)
}
}{
\left(1-\omega^2/\omega_{\rm UV}^2\right)
\left(\omega^2/\omega_{\rm IR}^2-1\right)
+\omega^4L^2C_*^2
},
\label{Eq:LHTLG}
\end{align}
with $\omega\in\mathcal B_Z$. In the weak-loading limit,
$|sC_*Z_{\rm LH}(s)|\ll1$, the finite-$C_c$ dressing of the line is neglected, yielding instead
\begin{align}
G_{\rm WL}^2(\omega)
=
\frac{\hbar\eta^2\omega^2L}{
\pi
\sqrt{
\left(1-\omega^2/\omega_{\rm UV}^2\right)
\left(\omega^2/\omega_{\rm IR}^2-1\right)
}
}.
\label{Eq:LHTLGWeakLoading}
\end{align}
%This approximation corresponds to the one made in Ref.~\cite{Goswami:2026}, where the qubit and LHTL are quantized independently after neglecting the capacitive loading of the line. 
Together with the two-level and rotating-wave approximations, this recovers the left-handed coupling of Ref.~\cite{Goswami:2026}, $g_k\propto\omega_k^{3/2}$. Their further approximation $g_k=g/\sqrt N$ gives $J(\omega)\simeq2g^2\omega_0/\omega^2$, with $\omega_0=(LC_0)^{-1/2}$, away from the band edges.

The difference is illustrated in \cref{fig:Metamaterials}(c), where we deliberately choose a parameter regime different from that of Ref.~\cite{Goswami:2026} to highlight the possible importance of the additional discrete bound mode. The weak-loading approximation produces a strongly enhanced coupling near the bare band edges, whereas the exact capacitive loading regularizes the continuum coupling and transfers part of the spectral weight to the discrete pole below $\omega_{\rm IR}$. This effect is small for $Z_0=100~\Omega$ but becomes pronounced for $Z_0=2~\mathrm{k}\Omega$, where the pole moves further below the band and acquires a substantially larger coupling. Thus, increasing the metamaterial impedance makes the independent-quantization approximation progressively more restrictive.

We do not pursue weak-coupling master-equation reductions for these metamaterial examples. The Josephson-array circuit is of particular interest in the high-impedance regime, where strong dressing and nonperturbative approaches, such as polaron transformations, become relevant~\cite{Bera:2014,Snyman:2015}, while the left-handed line provides a complementary, strongly structured finite-band environment. More generally, the same immittance construction may provide a systematic
starting point for improved quantum models of superconducting metamaterial circuits explored in \gls{cQED}~\cite{Zueco:2012,PuertasMartinez:2019,
Indrajeet:2020}.

\subsection{Waveguide-mediated coupling between distant qudits}
\label{Sec:TwoPortExamples}

\begin{figure}[t]
    \centering
    \includegraphics[width=\linewidth]
    {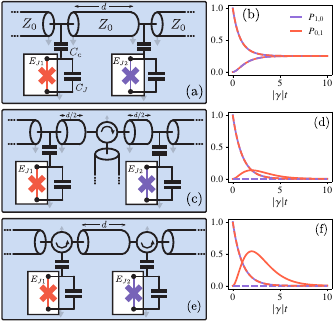}
    \caption{Two transmon qubits coupled through an infinite \gls{TL} at injection points separated by a distance $d$. We consider the cases (a) without a circulator between the injection points, (c) with a circulator between the two injection points, and (e) with a circulator at each injection point. Panels (b,d,f) show the corresponding effective dynamics when one of the two qubits is initialized in its excited state and then allowed to decay. The solid lines correspond to an initial excitation in the first qubit, whereas the dashed lines are obtained from starting with an excitation in the second qubit. In panel (b), the qubits are spaced at $d=\lambda/2$. }
\label{fig:Examples_waveguide_mediated_qudits}
\end{figure}

Having considered several one-port environments, we now turn to waveguide-mediated interactions between two charge qudits, including chiral configurations~\cite{lodahl:2017} generated by circulators~\cite{Viola:2014,Kerckhoff:2015,Chapman:2017,Barzanjeh:2017}. The three cases studied are coupling (i) directly to a \gls{TL}, (ii) to a \gls{TL} with a circulator placed symmetrically between the injection points and its third port terminated by an impedance-matched semi-infinite \gls{TL}, and (iii) through a circulator at each injection point, see
\cref{fig:Examples_waveguide_mediated_qudits}.

For simplicity, we consider charge qudits in the transmon regime and directly project the multilevel master equation onto their ground and first excited states, evaluating all coefficients from the multilevel formulas at the $0\leftrightarrow1$ transitions. As for the discrete example, no perturbative elimination of higher levels is performed. Such an elimination at the Liouvillian level~\cite{Kessler:2012} could yield improved few-level models. We refer to the resulting systems as transmon qubits and assume them to be identical, so that \cref{Eq:EffectiveME} reduces to
\begin{align}
\begin{aligned}
    \mathcal{L}\rho_P
    ={}&
    -i\left[
        \sum_{p}\frac{\tilde{\omega}}{2}\sigma_z^{(p)}
        +
        \left(
            g\sigma_+^{(1)}\sigma_-^{(2)}
            +
            \mathrm{h.c.}
        \right),
        \rho_P
    \right]
    \\
    &+
    \left(
        \gamma_{12}^*
        \mathcal{D}[\sigma_-^{(1)},\sigma_+^{(2)}]
        +
        \gamma_{12}
        \mathcal{D}[\sigma_-^{(2)},\sigma_+^{(1)}]
    \right)\rho_P
    \\
    &+
    \gamma
    \left(
        \mathcal{D}[\sigma_-^{(1)}]
        +
        \mathcal{D}[\sigma_-^{(2)}]
    \right)\rho_P,
\end{aligned}
\label{Eq:METwoQubits}
\end{align}
where the transition operators are defined as
$\sigma_+^{(p)}\equiv\sigma_{01}^{(p)}$,
$\sigma_-^{(p)}\equiv\sigma_{10}^{(p)}$, and
$\sigma_z^{(p)}=2\sigma_+^{(p)}\sigma_-^{(p)}-1$.
We use the notation
$\tilde{\omega}=\omega_{10}+(\overline{\Lambda}_1-\overline{\Lambda}_0)$,
$g=g_{10,01}^{1,2}$, $\gamma=\gamma_{10}^1=\gamma_{10}^2$, and
$\gamma_{12}=\gamma_{10,01}^{1,2}$.

\subsubsection{Qubits coupled to a bidirectional TL}
\label{Sec:WaveguideReciprocal}
Let us first consider the case shown in
\cref{fig:Examples_waveguide_mediated_qudits}(a), where two \gls{JJ}s couple capacitively to a bidirectional \gls{TL} at injection points separated by a distance $d$. The $N=2$ specialization of the reciprocal \gls{TL} impedance derived in
\cref{AppSec:TransmissionLineImpedances} is
\begin{align}
   \msZTL^{\rm R}(s)
   =
   \frac{Z_0}{2}
   \begin{pmatrix}
        1 & e^{-s \tau} \\
        e^{-s \tau} & 1
   \end{pmatrix},
   \label{Eq:TwoPortTL}
\end{align}
where $\tau=d/v$ is the propagation time between the two injection points. The total impedance seen from the \gls{JJ}s is obtained by including the coupling in the two-port network.

For simplicity, we assume identical capacitances and no direct capacitive coupling between the \gls{JJ}s, hence $\msC_c=C_c\msId_2$ and
$\msC_J=C_J\msId_2$, with $\msId_n$ the $n$-dimensional identity matrix. Therefore, the exact impedance matrix seen by the \gls{JJ}s is
\begin{align}
    \msZ_P(s)
    =
    \left[
        sC_J\msId_2
        +
        \left(
            \msId_2/sC_c
            +
            \msZTL^{\rm R}(s)
        \right)^{-1}
    \right]^{-1}.
\end{align}

We can extract the residue at zero
$\msC_{\msZ,0}^{-1}=C_\Sigma^{-1}\msId_2$ and denote the remaining
impedance by $\msZ(s)\equiv\msZ_P(s)-\msId_2/(sC_\Sigma)$. For capacitively coupled ports separated by a finite propagation delay, the high-frequency dressing capacitance is local, $\msC_{\msZ,\infty}^{-1}=C_c/(C_J C_\Sigma)\msId_2$.

The dissipative and coherent couplings follow directly from
\cref{Eq:LocalDecay,Eq:CorrelatedDecay,Eq:CoherentCoupling} and depend
strongly on the qubit separation. For qubits separated by $d=\lambda/4$,
where $\lambda=2\pi v/\omega_{10}$, we find
$\gamma_{12}\propto\omega_{10}/\wcut\approx0$ and $\gamma\approx2g$.
In contrast, for $d=\lambda/2$ we obtain
$\gamma=-\gamma_{12}$ and $g\approx0$. In the latter case, the
dissipators in \cref{Eq:METwoQubits} can be recast in purely collective
form 
$\gamma\mathcal{D}[\sigma_-^{(1)}-\sigma_-^{(2)}]$, yielding the standard
collective waveguide-QED relaxation~\cite{lodahl:2017}, as shown in
\cref{fig:Examples_waveguide_mediated_qudits}(b).

\subsubsection{Circulator between injection points}
\label{Subsubsec:CirculatorInbetween}

We next consider the scenario shown in
\cref{fig:Examples_waveguide_mediated_qudits}(c), with a circulator placed
between the two qubits, each at a distance $d/2$ from it. We assume an
ideal, zero-delay circulator that routes signals from
$1\rightarrow2\rightarrow3\rightarrow1$, with its third port terminated
by a semi-infinite, impedance-matched \gls{TL}. The corresponding $N=2$ directional response derived in
\cref{AppSec:TransmissionLineImpedances} is
\begin{align}
    \msZTL^{\rm IC}(s)
    =
    \frac{Z_0}{2}
    \begin{pmatrix}
        1 & 0\\
        e^{-s \tau} & 1
    \end{pmatrix},
\end{align}
where the asymmetry reflects the chirality of the system as seen from the
qubit ports. Replacing $\msZTL^{\rm R}(s)$ by $\msZTL^{\rm IC}(s)$ in the exact construction above and subtracting the pole at $s=0$, the effective
coefficients in \cref{Eq:METwoQubits} can be evaluated analytically as
\begin{align}
\gamma
&=
\frac{|Q_{10}|^2}{\hbar}
\frac{C_c^2}{C_\Sigma^2}
\frac{
    Z_0\omega_{10}
}{
    1+\omega_{10}^2/(2\wcut)^2
},
\\
\gamma_{12}
&=
\frac{|Q_{10}|^2}{2\hbar}
\frac{C_c^2}{C_\Sigma^2}
\frac{
    Z_0\omega_{10}
}{
    \left(
        1+i\omega_{10}/(2\wcut)
    \right)^2
}
e^{-i\omega_{10}\tau}.
\end{align}
The coherent coupling fulfills $\gamma_{12}=-2ig$, corresponding to fully
chiral waveguide-mediated coupling, i.e., an initial excitation in the
qubit at $x=-d/2$ excites the qubit at $x=d/2$, whereas the reverse
process is absent [see 
\cref{fig:Examples_waveguide_mediated_qudits}(d)]. Here
$\gamma=\gamma^{\mathrm{inf}}$, where $\gamma^{\mathrm{inf}}$ is the local decay of a qubit into an infinite \gls{TL} (as discussed in  \cref{Sec:SingleQubitInfiniteTL}), since both
qubits locally see an infinite \gls{TL}, and $|\gamma_{12}|=\gamma/2$.

\subsubsection{Circulators at injection points}

If the circulators are instead placed at the qubit injection points, the
$N=2$ specialization of the directional response in
\cref{AppSec:TransmissionLineImpedances} is
\begin{align}
    \msZTL^{\rightarrow}(s)
    =
    Z_0
    \begin{pmatrix}
        1 & 0\\
        2e^{-s \tau} & 1
    \end{pmatrix}.
\end{align}
In contrast to the case discussed in the previous subsubsection, each qubit
now locally sees a semi-infinite \gls{TL} due to the circulator at the
coupling point. Consequently,
$\gamma=\gamma^{\text{semi-inf}}$ (refer to \cref{Sec:SingleQubitSemiInfiniteTL}) and $|\gamma_{12}|=\gamma$.
Together with $\gamma_{12}=-2ig$, this yields the standard cascaded master
equation~\cite{carmichael_quantum_1993,gardiner_driving_1993}, with the
characteristic dynamics shown in
\cref{fig:Examples_waveguide_mediated_qudits}(f).

These three examples therefore connect the standard bidirectional and
cascaded waveguide-QED limits directly to the corresponding exact circuit
responses. In particular, placing the circulators at the injection points
changes the local response from that of an infinite to a semi-infinite
\gls{TL}, increasing both the local and directional couplings relative to
the configuration with circulators between the injection points.

\subsection{Giant artificial atoms}
\label{Sec:ExamplesGiantAtoms}

In the previous subsection, we studied waveguide-mediated interactions
between charge qudits operated in the transmon regime and coupled at single
injection points to a \gls{TL}. We now apply our method to the same class of
qudits coupled at two different positions of the same \gls{TL}, with
extensions to multiple connection points being straightforward. As above, we hard-project their multilevel dynamics onto the two lowest levels and use
the standard giant-atom terminology for the resulting two-level models.
These so-called giant atoms~\cite{Kockum:2014} have attracted considerable
interest in recent years~\cite{Kockum:2018,andersson:2019,kannan:2020,
Soro:2022,joshi:2023,chen:2025,jouanny:2025,TLevy-Yeyati:2026,
almanakly:2026,diekmann:2026}, as interference between the injection points allows the tailoring of distinct effective coherent and dissipative interactions among the individual port qudits.

Our effective master equation, derived in \cref{Sec:ME-models}, provides a direct and systematic treatment of these systems. In the
weak-capacitive-loading regime, we recover standard theoretical results from the literature~\cite{Kockum:2014,Kockum:2018,Soro:2022}, while the exact impedance captures corrections beyond this approximation within the validity regime of the master equation. Furthermore, we naturally obtain finite Lamb-shift contributions without the need to introduce ad hoc high-frequency cutoffs~\cite{Kockum:2014}.

The two specific examples considered in this section are two braided giant atoms coupled either (i) directly to the \gls{TL}
[\cref{fig:Examples_TLs_Giant_atoms}(a)], or (ii) via circulators at the injection points [\cref{fig:Examples_TLs_Giant_atoms}(b)]. Generalizations to different braiding configurations are straightforward and briefly
discussed in \cref{App:BraidedGiantAtomsImpedance}.

\subsubsection{Two braided giant atoms in a standard TL}
\begin{figure}[t]
    \centering
    \includegraphics[width=\linewidth]{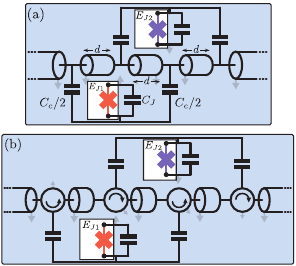}
    \caption{
    Giant artificial atoms (charge qudits) coupled to a standard \gls{TL}
    (a) directly, or (b) via ideal circulators (forming an effective chiral
    system). The circuit parameters and distances in panel (b) are identical
    to those in panel (a), with the addition of ideal circulators at the
    injection points.
    }
    \label{fig:Examples_TLs_Giant_atoms}
\end{figure}

We consider two identical giant atoms coupled capacitively at four equally  spaced points of the same infinite transmission line in the braided
ordering $(a,b,a,b)$, see \cref{fig:Examples_TLs_Giant_atoms}(a).
Each atom has a total coupling capacitance $C_c$, distributed equally
between its two connection points. The total impedance seen from the \gls{JJ}s is obtained by first mapping the
four-port impedance seen after the coupling capacitors onto a two-port
admittance at the two junction nodes,
\begin{align}
    \msYnode(s)
    =
    \msP^T
    \left[
        \left(\frac{2}{sC_c}\right)\msId_4
        +
        \msZTL^{\rm R}(s)
    \right]^{-1}
    \msP.
    \label{Eq:NodeAdmittance}
\end{align}

Here, $\msP$ maps the junction voltages onto the four coupling-leg inputs, while $\msP^T$ sums their currents at the junction nodes. The matrix $\msZTL^{\rm R}(s)$ is the four-port reciprocal \gls{TL} impedance defined in
\cref{AppSec:TransmissionLineImpedances,App:BraidedGiantAtomsImpedance}.

Including the junction capacitances, the exact port impedance seen from
the two \gls{JJ}s is
\begin{align}
    \msZ_P(s)
    =
    \left[
        sC_J\msId_2+\msYnode(s)
    \right]^{-1}
    =
    \frac{\msId_2}{sC_\Sigma}
    +
    \msZ(s).
    \label{Eq:Z_P_braidedGAExact}
\end{align}
From this exact impedance, the coefficients of the effective
dissipative model~(\ref{Eq:METwoQubits}) can be numerically evaluated without further approximating the circuit response.

To connect with previous analytical results~\cite{Kockum:2018,Soro:2022}, we now consider the weak-capacitive-loading regime
\begin{align}
    \epsilon_Z
    =
    \left\|
        \frac{sC_c}{2}\msZTL(s)
    \right\|
    \ll 1.
    \label{Eq:ValidityPerturbativeImpedance}
\end{align}
Under this condition, the matrix inversion in \cref{Eq:Z_P_braidedGAExact} can be performed perturbatively, while the residue of the zero-frequency pole remains $\msC_{\msZ,0}^{-1}=\msId_2/C_\Sigma$. Evaluating the resulting regular
impedance on the boundary ($s=-i\omega+0^+$) gives
\begin{align}
    \msZ[\omega]
    \simeq
    Z_0
    \left(
        \frac{C_c}{2C_\Sigma}
    \right)^2
    \begin{pmatrix}
        1+r^2 & \dfrac{3r+r^3}{2} \\
        \dfrac{3r+r^3}{2} & 1+r^2
    \end{pmatrix},
    \label{Eq:ApproxGABidirectional}
\end{align}
with $r=e^{i\omega\tau}$.

Inserting this approximate response into the immittance formulas, we
recover the standard expressions~\cite{Kockum:2014,Kockum:2018,Soro:2022}
\begin{align}
\begin{aligned}
    \gamma &= 2\gamma_0 \left(1+\cos(2\theta)\right),
    \\
    \gamma_{12} &= \gamma_0 \left(3\cos(\theta)+\cos(3\theta)\right),
    \\
    g &= \frac{\gamma_0}{2}
    \left(3\sin(\theta)+\sin(3\theta)\right),
\end{aligned}
\label{Eq:GAStandardResults}
\end{align}
with $\theta=\omega_{10}\tau$, and
\begin{align}
    \gamma_0
    =
    \frac{|Q_{10}|^2}{4\hbar}
    \omega_{10}Z_0
    \left(
        \frac{C_c}{C_\Sigma}
    \right)^2.
\end{align}
At the braided-interaction point $\theta=\pi/2$, the waveguide-mediated
dissipator vanishes within the two-level model,
$\gamma=\gamma_{12}=0$, whereas the coherent exchange interaction remains
finite, $g=\gamma_0$.

\begin{figure}[t]
    \centering
    \includegraphics[width=\linewidth]
    {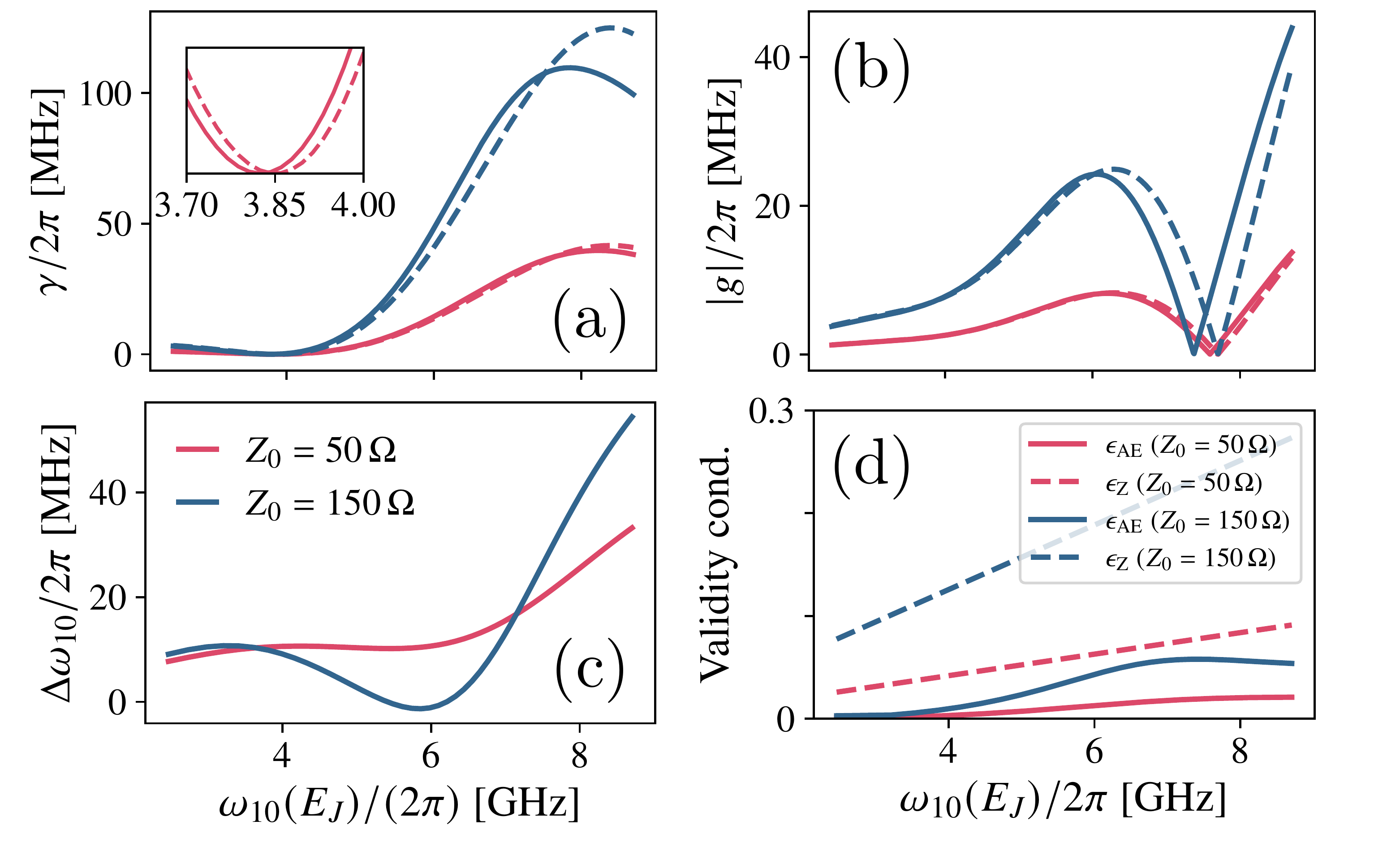}
    \caption{
    (a) Local decay rate $\gamma$, (b) coherent coupling $g$,
    (c) frequency correction [\cref{Eq:FrequencyCorrection}], and
    (d) heuristic master-equation validity condition and
    weak-capacitive-loading parameter for the braided giant-atom
    configuration shown in \cref{fig:Examples_TLs_Giant_atoms}(a).
    For each case, two different values of the characteristic impedance
    $Z_0$ are plotted. The solid lines in (a) and (b) correspond to the exact
    evaluation of the impedance, whereas the dashed lines are obtained for
    the approximate impedance in the weak-capacitive-loading regime
    \cref{Eq:ValidityPerturbativeImpedance}. The frequency correction in
    (c) is shown only for the exact impedance, since extending the
    approximate response to high frequencies yields a divergent Lamb-shift
    contribution. In all panels, we vary $E_J$ at fixed
    $C_J=200~\mathrm{fF}$, $n_g=0$, and $C_c=C_J/6$. The separation is chosen such that there is a decoherence-free interaction point near $\omega_{10}^*/(2\pi)=3.85$~GHz.
    }
    \label{fig:GA_Approx_vs_Full}
\end{figure}

The exact impedance given in \cref{Eq:Z_P_braidedGAExact} extends this result
beyond the weak-capacitive-loading approximation in
\cref{Eq:ValidityPerturbativeImpedance}. In
\cref{fig:GA_Approx_vs_Full} we compare the local decay rate (a) and the
coherent coupling strength (b) for two different values of the \gls{TL}
impedance $Z_0\in\{50,150\}\Omega$ as functions of the qubit frequency (when tuning $E_J$).
The qubit spacing is chosen such that the two identical transmon-regime
charge qudits ($C_J=200~\mathrm{fF}$, $C_c=C_J/6$) reach a decoherence-free interaction point near
$\omega_{10}/(2\pi)=3.85$~GHz. By tuning the effective Josephson energy $E_J$, we probe the frequency dependence of the effective coefficients \cref{Eq:GAStandardResults}. As expected, the dissipation rate vanishes
close to $\omega_{10}/(2\pi)=3.85$~GHz
[see \cref{fig:GA_Approx_vs_Full}(a)] for both the approximate
(dashed line) and exact (solid line) impedance. As the qubit frequency
increases, discrepancies between the approximate and exact impedance
become visible. These differences become more pronounced for larger
values of the characteristic impedance $Z_0$. Current giant-atom  experiments~\cite{kannan:2020,almanakly:2026} mostly use a
characteristic \gls{TL} impedance $Z_0=50~\Omega$, for which the
perturbative expansion of the impedance is well controlled over the
relevant frequency range [see \cref{fig:GA_Approx_vs_Full}(d)].
For larger characteristic impedances, corrections beyond the
weak-capacitive-loading approximation become more relevant.

Importantly, the approximate impedance \cref{Eq:ApproxGABidirectional} does not retain the high-frequency cutoff produced by the coupling capacitors.
Consequently, extending this low-frequency approximation to the entire
frequency range in the Lamb-shift integral produces a
divergence~\cite{Kockum:2014}. By contrast, the exact impedance
\cref{Eq:Z_P_braidedGAExact} includes the full capacitive response and yields a convergent Lamb-shift contribution when inserted into
\cref{Eq:LambShift}. The resulting correction to the qubit transition
frequency, $\Delta\omega_{10}$ as defined in
\cref{Eq:FrequencyCorrection}, is shown in
\cref{fig:GA_Approx_vs_Full}(c).

\subsubsection{Giant atoms coupled to chiral TL}
We finally consider the modified setup shown in
\cref{fig:Examples_TLs_Giant_atoms}(b), in which ideal circulators enforce
directional propagation at each coupling point. This configuration is
closely related to that studied in Ref.~\cite{Soro:2022}. 
The exact impedance is obtained using the same nodal construction as above,
with $\msZTL^{\rm R}(s)$ replaced by the directional response
$\msZTL^{\rightarrow}(s)$; details are given in
\cref{AppSec:TransmissionLineImpedances,App:BraidedGACirculators}.

In the weak-capacitive-loading regime of
\cref{Eq:ValidityPerturbativeImpedance}, evaluating the regular part of the
impedance on the boundary $s=-i\omega+0^+$ gives
\begin{align}
    \msZ[\omega]
    \simeq
    Z_0
    \left(
        \frac{C_c}{2C_\Sigma}
    \right)^2
    \begin{pmatrix}
        2(1+r^2) & 2r\\
        2r(r^2+2) & 2(1+r^2)
    \end{pmatrix}.
    \label{Eq:ApproxGAChiral}
\end{align}
The asymmetry of the off-diagonal elements,
$Z_{12}\neq Z_{21}$, directly encodes the nonreciprocal propagation.

\begin{figure}[t]
    \centering
    \includegraphics[width=\linewidth]
    {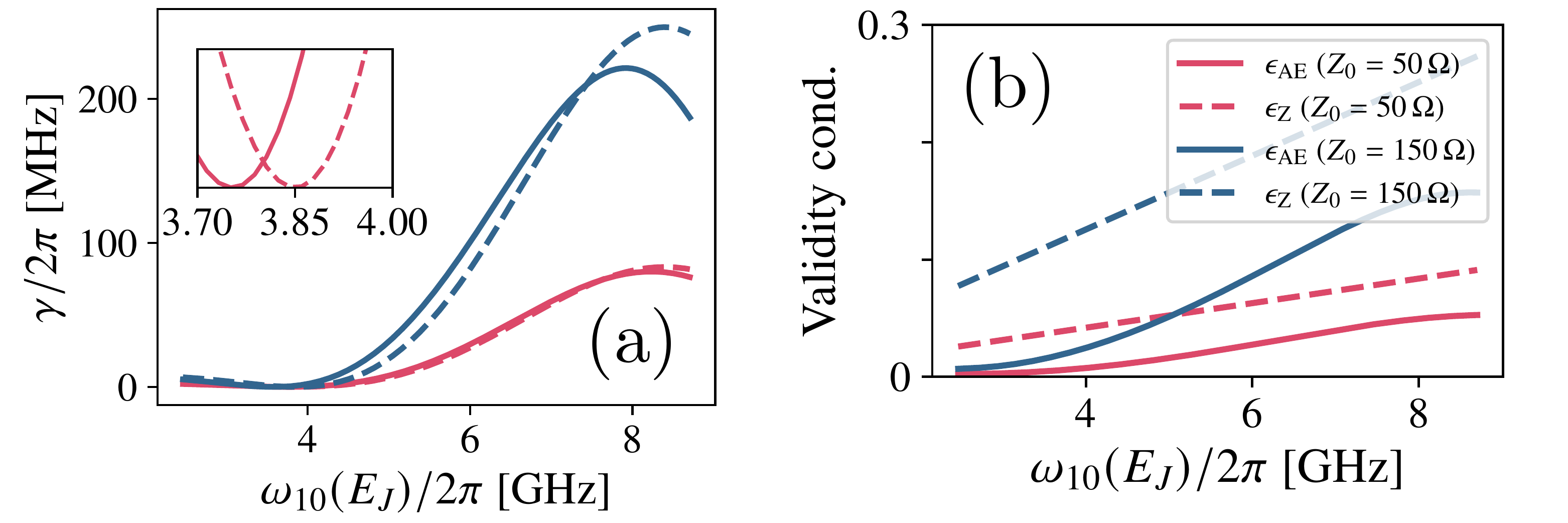}
    \caption{
    (a) Local decay rate and (b) validity parameters for the chiral
    configuration shown in \cref{fig:Examples_TLs_Giant_atoms}(b), with
    parameters identical to those in
\cref{fig:GA_Approx_vs_Full}.
    }
    \label{fig:Examples_TLs_Giant_atoms_chiral}
\end{figure}

Applying the immittance formulas gives
\begin{align}
\begin{aligned}
    \gamma
    &=
    4\gamma_0
    \left(
        1+\cos(2\theta)
    \right),
    \\
    \gamma_{12}
    &=
    2\gamma_0
    \left(
        e^{i\theta}
        +
        2e^{-i\theta}
        +
        e^{-3i\theta}
    \right),
    \\
    g
    &=
    i\gamma_0
    \left(
        2e^{-i\theta}
        +
        e^{-3i\theta}
        -
        e^{i\theta}
    \right).
\end{aligned}
\end{align}
At $\theta=\pi/2$, the waveguide-induced dissipation again vanishes,
$\gamma=\gamma_{12}=0$, whereas the coherent interaction remains finite,
$g=2\gamma_0$. Thus, chirality preserves the decoherence-free interaction
point while doubling the coherent coupling relative to the bidirectional
configuration.

Figure~\ref{fig:Examples_TLs_Giant_atoms_chiral}(a) shows the decay rate
for the chiral case with otherwise identical parameters to those in
\cref{fig:GA_Approx_vs_Full}. As already mentioned for a single transmon
coupled to a \gls{TL}, the presence of the circulator increases the local
decay rate because the semi-infinite \gls{TL} presents twice the impedance
of the infinite \gls{TL} in the low-frequency regime. Aside from this
quantitative difference, we still observe a decoherence-free interaction
point near the target frequency $\omega_{10}^*/(2\pi)=3.85$~GHz. As in the bidirectional case, the agreement between the exact and approximate impedance is best at low frequencies and small $Z_0$, where the capacitive-loading parameter remains small. Deviations become increasingly visible as either the frequency or the characteristic impedance is increased.

The weak-coupling Markovian reduction becomes less controlled at the upper
end of the parameter range, where the heuristic condition
$\epsilon_{\mathrm{AE}}\ll1$ is only marginally satisfied
[see \cref{fig:Examples_TLs_Giant_atoms_chiral}(b)]. For example,
$\epsilon_{\mathrm{AE}}>0.1$ for the $Z_0=150~\Omega$ \gls{TL} at
$\omega_{10}/(2\pi)\approx8$~GHz. For still larger $Z_0$ or qubit
frequencies, the full circuit Hamiltonian should instead be retained rather
than eliminating the continuum within the Markov approximation.

More generally, corrections beyond the weak-capacitive-loading approximation
are expected to become increasingly relevant in high-impedance or strongly
structured environments, such as metamaterial transmission lines. In such
regimes, retaining the exact immittance provides a direct route to controlled effective models whenever the weak-coupling Markovian reduction itself remains valid.

\section{Conclusions \& Outlook}
\label{Sec:ConclusionsOutlook}

In summary, we have extended the black-box framework~\cite{Nigg:2012,Solgun:2014,Solgun:2015,ParraRodriguez:2019,Solgun:2019,MinevEPR:2021,Labarca:2024,ParraRodriguez:2025} of circuit quantization to Josephson-junction-based superconducting qudits capacitively, inductively, or galvanically coupled to a broad class of passive linear environments, including multiport, multimode, discrete or continuous, reciprocal or nonreciprocal settings. This generalization retains simple closed-form immittance formulas that provide a direct route from analytical, numerically simulated, or experimentally characterized impedance and admittance matrices to exact Hamiltonian descriptions.

Starting from these exact dressed Hamiltonians, we further obtained divergence-free dispersive Hamiltonians for spectrally resolved environments and weak-coupling master equations for smooth dissipative continua. Mode structure, frequency renormalizations and Lamb shifts, environment-mediated coherent interactions, and local and correlated decay rates are thereby determined within the same framework from a single causal immittance response. We illustrated the formalism with a discrete environment and a range of continuous environments, including conventional and resonator-filtered transmission lines, finite-band metamaterial environments, nonreciprocal waveguide-QED systems, and giant artificial atoms. These examples show how widely used circuit- and waveguide-QED models emerge as controlled approximations to the exact construction, while also exposing spurious UV divergences and finite-coupling corrections that may be missed when approximate responses are extended beyond their regime of validity. Such exact Hamiltonians and systematic corrections are expected to become particularly important in high-impedance and metamaterial \gls{cQED}, where vacuum fluctuations, multimode hybridization, and environment-induced renormalizations are enhanced~\cite{PuertasMartinez:2019,Leger:2019,Pechenezhskiy:2020,Kuzmin:2025}.

Looking ahead, natural extensions include more general mixed capacitive-inductive coupling networks, including balanced-coupling configurations~\cite{ParraRodriguez:2018,Sank:2025,Yang:2026}, and effective models for sharply structured continuous environments beyond the weak-coupling Markovian regime. 

The incorporation of additional drive lines~\cite{Solgun:2019,Solgun:2022,Labarca:2024,Khan:2024} and flux drives threading unresolved loops of the linear environment~\cite{You:2019,Riwar:2022,Lu:2026} also remains to be addressed. Finally, the immittance constructions developed here may provide a systematic framework to investigate optimal gauges for reduced models of both discrete and continuous environments~\cite{Manucharyan:2017,DeBernardis:2018,Roth:2019,Mehta:2022,Arwas:2023}. Together, these results point toward a unified and scalable black-box description of superconducting quantum hardware and metamaterials, in which electromagnetic responses can be translated systematically into exact and controlled reduced quantum models without reconstructing and quantizing an explicit circuit representation of the linear environment.

\begin{acknowledgments}
We thank I. L. Egusquiza for proofreading the first version of the manuscript. P. G. thanks A. Misselwitz and L. Schamriß for stimulating comments and discussions. P. G. acknowledges support from the Swiss National Science Foundation through Project No. CRSII 222812/1. A. P.-R. acknowledges support from the European Union’s Marie Skłodowska-Curie Actions (MSCA) under grant agreement No.~101204967 (FTMcQED). This research is part of the Munich Quantum Valley, which is supported by the Bavarian state government with funds from the Hightech Agenda Bayern Plus.

This manuscript was prepared with the assistance of ChatGPT. ChatGPT was used to assist with code development, proofs, and language editing in accordance with the authors' intentions. All content, claims, and conclusions have been reviewed and verified by the authors to ensure accuracy and originality.

\end{acknowledgments}

\appendix

\section{First-order quantization method in a nutshell}
\label{AppSec:FirstOrderReview}
For completeness, we briefly summarize the ingredients of the first-order circuit-quantization procedure developed in Refs.~\cite{ParraRodriguez:2024,ParraRodriguez:2025} based on the Faddeev-Jackiw algorithm~\cite{Faddeev:1988,Jackiw:1993} that will be used in the following Appendices. We restrict the review to the energetic, source, and ideal constraint elements required for the $\msZ$- and $\msY$-environment realizations considered in this manuscript.

\begin{figure}[t]
    \centering
    \includegraphics[width=\linewidth]{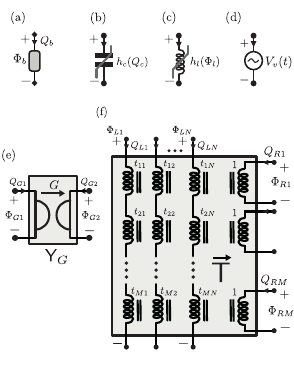}
    \caption{
Elementary lumped-element circuit components used in the first-order construction: (a) branch charge and flux variables, (b) a nonlinear capacitor, (c) a nonlinear inductor, (d) a voltage source, (e) a two-port gyrator, and (f) an ideal right transformer. Branch orientations fix the signs of the charge and flux relations. The voltage source in panel~(d)
is used in Appendix~\ref{AppSec:DrivenNonlinearPorts}.}   \label{fig:App_FirstOrderElements}
\end{figure}

Each two-terminal energetic element is assigned a branch charge and flux pair $(Q,\Phi)$. Its contribution to the construction consists of a term in the precanonical two-form and a term in the energy function. With the branch orientations of \cref{fig:App_FirstOrderElements}, a
capacitor and an inductor contribute, respectively,
\begin{align}
    \omega_{2B}^{C}
    &=
    \frac{1}{2}
    \mathrm{d}Q_c
    \wedge
    \mathrm{d}\Phi_c,\quad\text{and}\quad
    h_c(Q_c),
    \nonumber\\
    \omega_{2B}^{L}
    &=
    \frac{1}{2}
    \mathrm{d}\Phi_l
    \wedge
    \mathrm{d}Q_l,
    \quad\text{and}\quad
    h_l(\Phi_l).
    \label{Eq:AppEnergeticElementRules}
\end{align}
Thus, for a general collection of capacitive and inductive branches,
\begin{align}
    \omega_{2B}
    =
    \frac{1}{2}
    \left(
        \sum_{c}
        \mathrm{d}Q_c\wedge\mathrm{d}\Phi_c
        +
        \sum_{l}
        \mathrm{d}\Phi_l\wedge\mathrm{d}Q_l
    \right),
    \label{Eq:AppPrecanonicalTwoForm}
\end{align}
while
\begin{align}
    H
    =
    \sum_c h_c(Q_c)
    +
    \sum_l h_l(\Phi_l).
    \label{Eq:AppEnergyFunction}
\end{align}

The voltage-source rule used for driven fluxes is given in Appendix~\ref{AppSec:DrivenNonlinearPorts}.

These contributions constitute the precanonical  data prior to the imposition of circuit constraints. Kirchhoff's laws, together with the transformer and gyrator relations, restrict these variables to the physical circuit manifold.    

The corresponding constraint of a two-port gyrator~\cite{Tellegen:1948} [\cref{fig:App_FirstOrderElements}(e)] mixes charge and flux directions via
\begin{align}
    \mathrm{d}\bQ_G
    =\msY_G\mathrm{d}\bPhi_G\equiv
    G\msf{J}\,
    \mathrm{d}\bPhi_G,
    \label{Eq:AppGyratorConstraint}
\end{align}
where 
\begin{align}
    \msf{J}
    =
    \begin{pmatrix}
        0 & 1\\
        -1 & 0
    \end{pmatrix}
\end{align}
is the canonical symplectic matrix. Ideal Belevitch transformers~\cite{Belevitch:1950} also do not contribute to the energy function, but impose linear constraints between the branch variables on their two sides. For a right transformer [\cref{fig:App_FirstOrderElements}(f)], the constraints are
\begin{align}
    \mathrm{d}\bQ_R
    &=
    -\msT\,
    \mathrm{d}\bQ_L,
    &
    \mathrm{d}\bPhi_L
    &=
    \msT^T
    \mathrm{d}\bPhi_R.
    \label{Eq:AppRightTransformerConstraint}
\end{align}
For the mirrored left transformer used in the admittance decomposition (see \cref{AppSec:Y_construction_and_quantization}), the roles of the two sides are
interchanged,
\begin{align}
    \mathrm{d}\bQ_L
    &=
    -\msT\,
    \mathrm{d}\bQ_R,
    &
    \mathrm{d}\bPhi_R
    &=
    \msT^T
    \mathrm{d}\bPhi_L.
    \label{Eq:AppLeftTransformerConstraint}
\end{align}
The two orientations therefore represent the same ideal transformer
element, differing only by the side toward which the transformer is
oriented.

Together with Kirchhoff's current and voltage laws, these ideal-element relations form the complete set of kinematic constraints. Schematically, writing all branch variables as $\bsb{z}_{2B}$, they can be collected as
\begin{align}
    \msF\,
    \mathrm{d}\bsb{z}_{2B}
    =
    0,
    \qquad
    \msF
    =
    \begin{pmatrix}
        \msF_{\mathrm{Kir}}\\
        \msF_G\\
        \msF_T
    \end{pmatrix}.
    \label{Eq:AppTotalPfaffConstraints}
\end{align}
Consequently, transformers and gyrators can increase the number of
independent constraints beyond those imposed by Kirchhoff's laws
alone; the actual number is the rank of the combined constraint
system, since some of the relations may be linearly dependent.

A basis of allowed circuit directions can be collected in a matrix
$\msK$ satisfying
\begin{align}
    \msF\msK=0,
    \qquad
    \mathrm{d}\bsb{z}_{2B}
    =
    \msK\,\mathrm{d}\bsb{z},
    \label{Eq:AppConstraintKernel}
\end{align}
which defines the restriction to the constrained circuit manifold.
The precanonical two-form and energy are then pulled back to this
manifold,
\begin{align}
    \omega
    =
    \iota^*\omega_{2B},
    \qquad
    H
    =
    \iota^*H_{2B}.
    \label{Eq:AppPullback}
\end{align}
Equivalently, for a linear embedding and matrix representations of the two-forms ($\omega \leftrightarrow\msOmega$ and $\omega_{2B}\leftrightarrow\msOmega_{2B}$),
\begin{align}
   \msOmega
    =
    \msK^T
    \msOmega_{2B}
    \msK.
    \label{Eq:AppTwoFormPullback}
\end{align}

If $\omega$ is nondegenerate, one may directly identify Darboux
coordinates and proceed to canonical quantization. If it is
degenerate, each zero mode generates either a gauge redundancy or a dynamical constraint obtained by contraction with $\mathrm{d}H$. The Faddeev-Jackiw (a modern version of the Dirac-Bergmann algorithm~\cite{Dirac:1950,Bergmann:1949}) procedure consists of restricting to the resulting constraint manifold and repeating this step until a nondegenerate symplectic form is obtained~\cite{Faddeev:1988,Jackiw:1993,
ParraRodriguez:2024,Osborne:2024,ParraRodriguez:2025}. We note that, in the geometrical formulation of Refs.~\cite{ParraRodriguez:2024,ParraRodriguez:2025}, the integration of Kirchhoff's constraints, together with the topological axioms assigned to the branch manifolds, allows the discrete translation symmetries associated with compact directions to be identified algorithmically. Here, for simplicity, we start from the extended branch manifold $\mathbb{R}^{2B}$ and identify discrete symmetries after constructing the Hamiltonian; their interpretation as compact sectors requires the global conditions discussed in the main text.

\subsection*{Reduction of parallel nonlinear inductive elements to port variables}
\label{AppSubsec:ParallelNonlinearInductors}

\begin{figure}[t]
    \centering
    \includegraphics[width=\linewidth]
        {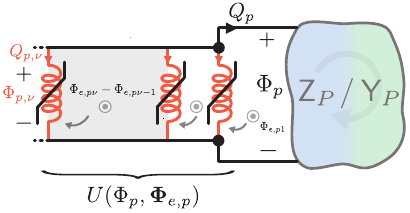}
    \caption{Reduction of parallel inductive elements at port $p$ to a
    single effective port, with the passive environment described
    by $\msZ_P$ or $\msY_P$.
    The integrated Kirchhoff voltage constraints identify a common
    port flux $\Phi_p$ and static branch offsets $\Phi_{e,p\nu}$.
    Counting branches outward from the box, the first-loop offset
    is $\Phi_{e,p1}$, while the oriented flux between neighboring
    branches $\nu$ and $\nu-1$ is $\Phi_{e,p\nu}-\Phi_{e,p,\nu-1}$. The branch currents combine into the port current.
    }
    \label{fig:App_ParallelInductivePort}
\end{figure}

The inductive elements at the ports used in the following Appendices
can be reduced directly using the first-order construction.
Consider a port $p$ composed of $m_p$ inductive elements connected
in parallel, as depicted in \cref{fig:App_ParallelInductivePort}.
These may include explicitly retained linear inductors. We denote their branch variables by $(Q_{p\nu},\Phi_{p\nu})$, with
$\nu=1,\ldots,m_p$, and choose the branch-current orientations
opposite to the port-current direction.

At fixed external fluxes, Kirchhoff's voltage constraints give
$\mathrm{d}\Phi_{p\nu}=\mathrm{d}\Phi_p$, which integrate to
\begin{align}
    \Phi_{p\nu} = \Phi_p + \Phi_{e,p\nu}.
    \label{Eq:ParallelInductorFluxConstraint}
\end{align}
Here, $\Phi_{e,p\nu}$ is the static offset of branch $\nu$
relative to the chosen flux reference of the passive environment.
The difference $\Phi_{e,p\nu}-\Phi_{e,p\nu'}$ gives the oriented
flux between branches $\nu$ and $\nu'$, with the sign convention
fixed by \cref{Eq:ParallelInductorFluxConstraint}.
Whenever the passive environment imposes no independent static
inductive constraint, as for a capacitively coupled $\msZ_P$ port,
we may choose the first branch as the flux reference and set
$\Phi_{e,p1}=0$ without changing these differences.
We retain $\Phi_{e,p1}$ in the general notation and figure to
include fluxes through loops with a passive inductive return path,
as in the $\msY_P$ construction.
All retained static flux offsets enter the external potential
through \cref{Eq:ParallelInductorFluxConstraint}.

With the orientations of \cref{fig:App_ParallelInductivePort},
Kirchhoff's current constraint reads
$\mathrm{d}Q_p=-\sum_{\nu=1}^{m_p}\mathrm{d}Q_{p\nu}$, and hence
$Q_p=-\sum_{\nu=1}^{m_p}Q_{p\nu}+Q_{g,p}$ upon integration. The integration constant $Q_{g,p}$ is irrelevant for the inductive
energy and drops out of the two-form.

Using \cref{Eq:AppEnergeticElementRules}, the combined
precanonical contribution of the inductive elements reduces to
\begin{align}
    \omega_p
    =
    \frac{1}{2}\sum_{\nu=1}^{m_p}
    \mathrm{d}\Phi_{p\nu}\wedge\mathrm{d}Q_{p\nu}
    =
    \frac{1}{2}
    \mathrm{d}Q_p\wedge\mathrm{d}\Phi_p.
    \label{Eq:ParallelInductorPortTwoForm}
\end{align}
Their energy contributions combine into the port potential
\begin{align}
    U_p
    \left(
        \Phi_p,\bPhi_{e,p}
    \right)
    =
    \sum_{\nu=1}^{m_p}
    h_{p\nu}
    \left(
        \Phi_p+\Phi_{e,p\nu}
    \right).
    \label{Eq:ParallelInductorPortPotential}
\end{align}
Repeating the construction independently for all $N$ ports gives
\begin{align}
    \omega_P
    &=
    \frac{1}{2}
    \dbQ_P^T\wedge\dbPhi_P,
    \nonumber\\
    U(\bPhi_P,\bPhi_{e,P})
    &=
    \sum_{p=1}^{N}
    U_p(\Phi_p,\bPhi_{e,p}).
    \label{Eq:ReducedNonlinearPorts}
\end{align}
The parallel inductive elements at each port can therefore be
replaced, for the subsequent first-order reduction, by a single
port pair $(\Phi_p,Q_p)$ and the corresponding potential
$U_p(\Phi_p,\bPhi_{e,p})$. The extension to time-dependent fluxes with parallel branch capacitances is given in Appendix~\ref{AppSec:DrivenNonlinearPorts}.

\section{Details on the quantization of the nonlinear port coupled to a $\msZ$-environment}
\label{AppSec:Z_construction_and_quantization}

In this Appendix, we provide details to construct the discrete and continuous Hamiltonians for the circuit in Fig.~\ref{fig:Fig_Cauer_MT}(a). We provide first the canonical matrix fraction expansion (Cauer decomposition)~\cite{Newcomb:1966} for discrete $\msZ$-environments and take the continuous limit to represent dissipative ones. We then obtain the exact Hamiltonian by applying the first-order
construction~\cite{ParraRodriguez:2024,ParraRodriguez:2025} summarized in \cref{AppSec:FirstOrderReview} to the
discrete realization, and subsequently take the continuum limit.
Both discrete and continuous coupling Hamiltonians are written in the
charge gauge.

\subsection{Cauer decompositions of discrete and continuous $\msZ$ environments}
\label{Sec:CauerDecomposition}
    In this subsection, we first consider a set of nonlinear flux elements that define the ports of a canonical impedance matrix with a discrete number of poles (see the canonical circuit representation in Fig.~\ref{fig:App_Zcircuit}). We then take the appropriate continuum limit to recover the pole structure of continuous environments. The discrete impedance can be decomposed as
\begin{align}
    \msZ_P(s)
    =
    \frac{\msC_{\msZ,0}^{-1}}{s}
    +
    \sum_{\alpha=1}^{M}\msZ_{\alpha}(s)
    =
    \frac{\msC_{\msZ,0}^{-1}}{s}
    +
    \msZ(s).
    \label{Eq:CauerDiscrete}
\end{align}
The finite-frequency poles are given by
\begin{align}
    \msZ_{\alpha}(s)
    =
    \frac{\msA_{\alpha}s+\msB_{\alpha}}
    {\omega_{\alpha}^2+s^2}.
\end{align}
    Here, $\msA_{\alpha}$ are symmetric positive-semidefinite matrices, whereas $\msB_{\alpha}$ are antisymmetric matrices. The pole at zero is extracted as $\msC_{\msZ,0}^{-1}=\lim_{s\to0}s\msZ_P(s)$. In the canonical realization used below, the physical matrix $\msC_{\msZ,0}^{-1}$ is encoded entirely in the ideal transformer, so the corresponding internal capacitive branch matrix may be chosen as the identity without loss of generality~\cite{Newcomb:1966}. In the remainder of this section, we discuss how the remaining finite frequency contribution $\msZ(s)$ can be synthesized with nonreciprocal oscillators and ideal transformers.

For a discrete number of poles, one can directly extract the matrices $\msA_\alpha$ and $\msB_\alpha$ by computing the symmetric and antisymmetric parts of the residues at the corresponding pole frequencies,
\begin{align}
    \msA_\alpha
    =
    2\msR_\alpha^+,
    \qquad
    \msB_\alpha
    =
    -2i\omega_\alpha\msR_\alpha^-,
    \label{Eq:DiscretePoleMapping}
\end{align}
with $ \msR_\alpha^\pm\equiv\underset{s=-i\omega_\alpha}{\operatorname{Res}}\msZ^\pm(s)$, where $\msZ^\pm=(\msZ\pm\msZ^T)/2$. Equivalently, the full residue is
\begin{align}
    \msR_\alpha
    &=
    \frac{1}{2}
    \left(
        \msA_\alpha
        +
        \frac{i}{\omega_\alpha}
        \msB_\alpha
    \right)
    \succeq0,
    \label{Eq:FullPoleResidue}
\end{align}
where positivity follows from passivity~\cite{Newcomb:1966}. We denote its rank by
\begin{align}
    r_\alpha
    \equiv
    \operatorname{rank}\msR_\alpha.
    \label{Eq:PoleRank}
\end{align}

To take the continuum limit, we first consider the causal boundary value of each finite-frequency pole,
\begin{align}\label{Eq:NonReciprocalBoundaryDist}
    &\msZ_{\alpha}[\omega]
    \equiv
    \lim_{\eta\to0^+}
    \msZ_{\alpha}(-i\omega+\eta)
    \\
    &=
    \frac{\pi\msA_{\alpha}}{2}
    \left[
        \delta(\omega-\omega_{\alpha})
        +
        \delta(\omega+\omega_{\alpha})
    \right]
    -
    i\omega\msA_{\alpha}
    \mathcal{P}
    \left(
        \frac{1}{\omega_{\alpha}^2-\omega^2}
    \right)\nonumber
    \\
    &+
    i\frac{\pi\msB_{\alpha}}{2|\omega_{\alpha}|}
    \left[
        \delta(\omega-\omega_{\alpha})
        -
        \delta(\omega+\omega_{\alpha})
    \right]
    +
    \msB_{\alpha}
    \mathcal{P}
    \left(
        \frac{1}{\omega_{\alpha}^2-\omega^2}
    \right),\nonumber
\end{align}
where $\mathcal{P}$ denotes the Cauchy principal value. We have used the standard Sokhotski--Plemelj identity
\begin{align}
\begin{aligned}
    \lim_{\eta\to0^+}
    &\frac{1}
    {\omega_{\alpha}^2+(-i\omega+\eta)^2}
    =
    \mathcal{P}
    \left(
        \frac{1}{\omega_{\alpha}^2-\omega^2}
    \right)
    \\
    &+
    i\frac{\pi}{2|\omega_{\alpha}|}
    \left[
        \delta(\omega-\omega_{\alpha})
        -
        \delta(\omega+\omega_{\alpha})
    \right].
\end{aligned}
\end{align}
For simplicity, we denote the boundary value
$\msX(s=-i\omega+0^+)\equiv\msX[\omega]$
in the remainder of the Appendix.

Equation~\eqref{Eq:NonReciprocalBoundaryDist} shows that the dissipative part of a discrete impedance is an atomic matrix-valued distribution supported at the pole frequencies. For a continuous impedance, the corresponding positive-frequency densities are
\begin{align}
    \msA(\omega)
    =
    \frac{2}{\pi}
    \Re\{\msZ^+[\omega]\},
    \qquad
    \msB(\omega)
    =
    \frac{2\omega}{\pi}
    \Im\{\msZ^-[\omega]\}.
    \label{Eq:ContinuousPoleMapping}
\end{align}
For bounded or disconnected continua, these spectral densities vanish outside
$\mathcal B_{\msZ}$, so the following integrals may still be written over
$0<\omega<\infty$. A uniform discretization $\omega_\alpha=\alpha\Delta\omega$ is therefore obtained by assigning the Riemann weights
\begin{align}
    \msA_\alpha
    =
    \msA(\omega_\alpha)\Delta\omega,
    \qquad
    \msB_\alpha
    =
    \msB(\omega_\alpha)\Delta\omega.
    \label{Eq:RiemannPoleMapping}
\end{align}

In the continuum limit, the symmetric and antisymmetric parts of the finite-frequency impedance combine to give the total impedance
\begin{align}
    \msZ(s)
    =
    \msZ^+(s)+\msZ^-(s)
    =
    \int_0^\infty
    \mathrm{d}\omega\,
    \frac{
        s\msA(\omega)+\msB(\omega)
    }{
        s^2+\omega^2
    }.
    \label{Eq:CauerContinuum}
\end{align}

This discrete-to-continuum construction is the circuit counterpart of the standard representation of quantum dissipation in terms of a system coupled to a continuum limit of discrete environmental modes~\cite{Senitzky:1960,Senitzky:1961,Feynman:1963,CaldeiraLeggett_QT:1983}.
The conjugation symmetry of $\msZ$ implies $\msA(-\omega)=\msA(\omega)$ and $\msB(-\omega)=\msB(\omega)$, while causality reconstructs the complementary parts of $\msZ^\pm$ through the Kramers--Kronig relations.

Each discrete pole can be synthesized using $r_\alpha$
nonreciprocal oscillators and ideal transformers as depicted in \cref{fig:App_Zcircuit}. We choose unit internal capacitances (i.e., $C_\alpha^{a,\epsilon}=C_\alpha^{b,\epsilon}=1$) and
\begin{align}
    G_\alpha
    =
    \omega_\alpha,
    \label{Eq:GaugeChoiceZ}
\end{align}
for all polarizations
$\epsilon\in\{1,\dots,r_\alpha\}$ associated with the pole.
For each polarization, we define the transformer
\begin{align}
    \msT_\alpha^\epsilon
    =
    \begin{pmatrix}
        \msT_{\alpha,a}^\epsilon\\
        \msT_{\alpha,b}^\epsilon
    \end{pmatrix},
    \qquad
    \msT_{\alpha,a}^\epsilon,
    \msT_{\alpha,b}^\epsilon
    \in
    \mathbb{R}^{1\times N}.
    \label{Eq:DiscreteTransformerDefinition}
\end{align}
For convenience, we collect the polarization-resolved transformer
rows into
\begin{align}
\begin{aligned}
    \msT_{\alpha,a}
    &=
    \begin{pmatrix}
        \msT_{\alpha,a}^{1}\\
        \vdots\\
        \msT_{\alpha,a}^{r_\alpha}
    \end{pmatrix},
    \qquad
    \msT_{\alpha,b}
    =
    \begin{pmatrix}
        \msT_{\alpha,b}^{1}\\
        \vdots\\
        \msT_{\alpha,b}^{r_\alpha}
    \end{pmatrix},\\
    \msT_\alpha
    &=
    \begin{pmatrix}
        \msT_{\alpha,a}\\
        \msT_{\alpha,b}
    \end{pmatrix},
    \qquad
    \msT_{\alpha,a},\msT_{\alpha,b}
    \in
    \mathbb{R}^{r_\alpha\times N}.
\end{aligned}
\end{align}

For this (non-unique) choice of parametrization, the pole matrices are related to the circuit parameters
\begin{align}
\begin{aligned}
    \msA_\alpha
    &=
    \msT_{\alpha,a}^T\msT_{\alpha,a}
    +
    \msT_{\alpha,b}^T\msT_{\alpha,b},
    \\
    \msB_\alpha
    &=
    \omega_\alpha
    \left(
        \msT_{\alpha,b}^T\msT_{\alpha,a}
        -
        \msT_{\alpha,a}^T\msT_{\alpha,b}
    \right).
\end{aligned}
\label{Eq:NonReciprocalTransformerImpedanceRelation}
\end{align}

Equivalently, each pole is realized as the sum of the $r_\alpha$ elementary two-port sections
\begin{align}
    \msZ_\alpha(s)
    =
    \sum_{\epsilon=1}^{r_\alpha}
    \left(\msT_\alpha^\epsilon\right)^T
    \frac{1}{s^2+\omega_\alpha^2}
    \begin{pmatrix}
        s & -\omega_\alpha\\
        \omega_\alpha & s
    \end{pmatrix}
    \msT_\alpha^\epsilon.
    \label{Eq:CanonicalPoleRealization}
\end{align}
For a given polarization, if either of the two transformer rows vanishes, the corresponding antisymmetric contribution is zero and the section reduces to a purely reciprocal oscillator. More generally, this occurs whenever $\msT_{\alpha,a}^\epsilon$ and $\msT_{\alpha,b}^\epsilon$ are linearly dependent. In the former case, the capacitor at the transformer-decoupled gyrator port is seen through the gyrator as an effective inductor at the coupled port, recovering the usual reciprocal LC realization~\cite{Newcomb:1966}. The two relations in \cref{Eq:NonReciprocalTransformerImpedanceRelation}
can be combined into the single residue factorization
\begin{align}
\begin{aligned}
    &
    \left(
        \msT_{\alpha,b}^T
        -
        i\msT_{\alpha,a}^T
    \right)
    \left(
        \msT_{\alpha,b}^T
        -
        i\msT_{\alpha,a}^T
    \right)^\dag
    \\
    &\qquad=
    \msA_\alpha
    +
    \frac{i}{\omega_\alpha}
    \msB_\alpha
    =
    2\msR_\alpha.
\end{aligned}
\label{Eq:TransformerResidueFactorization}
\end{align}
Therefore, an eigendecomposition
\begin{align}
    \msR_\alpha
    &=
    \msU_\alpha^\dag
    \msD_\alpha
    \msU_\alpha,
    \nonumber\\
    \msD_\alpha
    &=
    \operatorname{diag}
    \left(
        \rho_{\alpha,1},
        \ldots,
        \rho_{\alpha,r_\alpha}
    \right),
    \qquad
    \rho_{\alpha,\epsilon}>0,
    \label{Eq:ResidueEigendecomposition}
\end{align}
provides one convenient transformer gauge,
\begin{align}
    \msT_{\alpha,b}^T
    -
    i\msT_{\alpha,a}^T
    =
    \sqrt{2}\,
    \msU_\alpha^\dag
    \msD_\alpha^{1/2}.
    \label{Eq:TransformerEigenGauge}
\end{align}
Writing $\bsb{u}_{\alpha,\epsilon}$ for the corresponding normalized
eigenvectors, the two rows of the transformer associated with each
polarization are explicitly
\begin{align}
\begin{aligned}
    \left(
        \msT_{\alpha,b}^{\epsilon}
    \right)^T
    &=
    \sqrt{2\rho_{\alpha,\epsilon}}\,
    \Re\left\{
        \bsb{u}_{\alpha,\epsilon}
    \right\},
    \\
    \left(
        \msT_{\alpha,a}^{\epsilon}
    \right)^T
    &=
    -\sqrt{2\rho_{\alpha,\epsilon}}\,
    \Im\left\{
        \bsb{u}_{\alpha,\epsilon}
    \right\}.
\end{aligned}
\label{Eq:TransformerEigenPolarizations}
\end{align}
Thus, the real and imaginary parts of each residue eigenvector
directly determine the two-row transformer
$\msT_\alpha^\epsilon$ connecting the external ports to the
corresponding elementary nonreciprocal oscillator. Unitary rotations among the $r_\alpha$ polarizations give equivalent factorizations.

For the Riemann discretization in \cref{Eq:RiemannPoleMapping}, we can take
\begin{align}
    \msT_\alpha
    =
    \msT(\omega_\alpha)\sqrt{\Delta\omega},
    \qquad
    \msT(\omega)
    =
    \begin{pmatrix}
        \msT_a(\omega)\\
        \msT_b(\omega)
    \end{pmatrix}.
    \label{Eq:ContinuumTransformerDefinition}
\end{align}

\subsection{Derivation of the exact circuit Hamiltonian}
\label{Sec:CanonicalHamiltonian}

\begin{figure}[t]
    \centering
    \includegraphics[width=\linewidth]{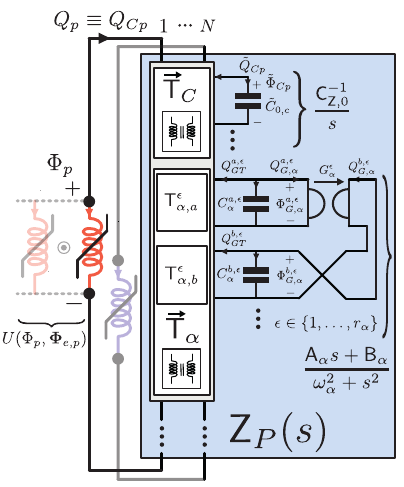}
    \caption{Schematic of nonlinear inductive ports coupled to a multiport linear black box described by a canonical impedance decomposition. Two parallel nonlinear inductive branches with static loop fluxes are shown at one port to illustrate the reduction in \cref{AppSubsec:ParallelNonlinearInductors}. Their parallel capacitances are included in the passive impedance. More general qudit ports can be treated using the tools of Refs.~\cite{ParraRodriguez:2024,ParraRodriguez:2025}.}
    \label{fig:App_Zcircuit}
\end{figure}

After having identified the canonical circuit ingredients to synthesize the pole structure of the impedance $\msZ(s)$, we now proceed to quantize the full circuit Hamiltonian (for notation refer to \cref{fig:App_Zcircuit}). Each external port (possibly containing multiple nonlinear inductive elements) is represented by a single port pair $(\Phi_p,Q_p)$ (following the discussion in \cref{AppSubsec:ParallelNonlinearInductors}) and contributes the energy term $U(\bPhi_P,\bPhi_{e,P})$. The combined $N$ external port variables are conveniently grouped into the vectors $\bPhi_P$ and $\bQ_P$, with their compact or extended character understood as in the main text. Note that in the impedance realization, the series constraints imply
$\bQ_C\equiv\bQ_P$. All external fluxes $\bPhi_{e,P}$ in this
Appendix are static and are already included in
$U(\bPhi_P,\bPhi_{e,P})$ through the branch offsets of
\cref{AppSubsec:ParallelNonlinearInductors}.

The internal degrees of freedom are split up into a zero-frequency capacitive pole with the internal variables $\tilde{\bQ}_C,\tilde{\bPhi}_C$, and a set of finite-frequency nonreciprocal oscillators. In the following, we combine the pole and polarization indices into a single index
\begin{align}
    k
    \equiv
    (\alpha,\epsilon),
    \qquad
    \sum_k
    \equiv
    \sum_{\alpha=1}^{M}
    \sum_{\epsilon=1}^{r_\alpha},
    \qquad
    \omega_k
    \equiv
    \omega_\alpha.
\end{align}
Using the polarization-resolved transformers introduced in
\cref{Eq:DiscreteTransformerDefinition}, we similarly write
\begin{align}
    \msT_k
    \equiv
    \msT_\alpha^\epsilon
    =
    \begin{pmatrix}
        \msT_{k,a}\\
        \msT_{k,b}
    \end{pmatrix}
    \in
    \mathbb{R}^{2\times N}.
\end{align}
For each $k$, the variables
\begin{align}
    \bPhi_{G,k}
    &=
    \begin{pmatrix}
        \Phi_{G,k}^{a}\\
        \Phi_{G,k}^{b}
    \end{pmatrix},
    \,
    \bQ_{G,k}
    =
    \begin{pmatrix}
        Q_{G,k}^{a}\\
        Q_{G,k}^{b}
    \end{pmatrix},
    \,
    \bQ_{GT,k}
    =
    \begin{pmatrix}
        Q_{GT,k}^{a}\\
        Q_{GT,k}^{b}
    \end{pmatrix}\nonumber
\end{align}
are two-component vectors in the $a,b$ space of the corresponding
nonreciprocal oscillator. We now apply the first-order construction summarized in
\cref{AppSec:FirstOrderReview} to the canonical circuit.

\subsubsection{Canonical two-form}
\label{Sec:TwoForm}

Following \cref{AppSec:FirstOrderReview}, our first goal is to  construct the canonical two-form of the circuit depicted in \cref{fig:App_Zcircuit}.

We start with the zero-frequency capacitive pole, whose contribution
to the precanonical  two-form is
\begin{align}
    \omega^C_{2B}
    =
    \frac{1}{2}
    \mathrm{d}\tilde{\bQ}_C^T
    \wedge
    \mathrm{d}\tilde{\bPhi}_C.
\end{align}
Using the right-transformer constraint in
\cref{Eq:AppRightTransformerConstraint}~\footnote{When $\msC_{\msZ,0}^{-1}$ is full rank, it can be naturally decomposed into right or left transformers plus a diagonal matrix via matrix inversion.}, $\dbQ_{CT}=-\msT_C\dbQ_P=-\dd\tilde{\bQ}_C$ (observe that in Fig.~\ref{fig:App_Zcircuit}, the arrow for an element of $\dbQ_{CT}$ is not named),
this becomes
\begin{align}
    \omega^C
    =
    \frac{1}{2}
    \dbQ_P^T
    \wedge
    \left(
        \msT_C^T
        \mathrm{d}\tilde{\bPhi}_C
    \right).
\end{align}

The capacitors for each nonreciprocal oscillator contribute with
\begin{align}
    \omega_{2B}^{G,k}
    =
    -\frac{1}{2}
    \left(
        \dbQ_{G,k}
        +
        \dbQ_{GT,k}
    \right)^T
    \wedge
    \dbPhi_{G,k},
\end{align}
where the capacitor-branch currents (arrows without names in Fig.~\ref{fig:App_Zcircuit}) are expressed in terms of the gyrator and transformer branch currents ($-\left(
        \dbQ_{G,k}
        +
        \dbQ_{GT,k}
    \right)=\dbQ_{C,k}$). Using the right-transformer constraint in \cref{Eq:AppRightTransformerConstraint},
\begin{align}
    \dbQ_{GT,k}
    =
    -\msT_k\dbQ_P,
    \label{Eq:TransformerCurrentConstraint}
\end{align}
the nonreciprocal contribution becomes
\begin{align}
\label{eq:w_Gk_Z_init}
\begin{aligned}
    \omega^{G,k}
    ={}&\frac{1}{2}
    \dbQ_P^T
    \wedge
    \left(
        \msT_k^T
        \dbPhi_{G,k}
    \right)    -\frac{1}{2}
    \dbQ_{G,k}^T
    \wedge
    \dbPhi_{G,k}
    .
\end{aligned}
\end{align}

As shown in \cref{AppSec:FirstOrderReview}, the gyrator constraint for oscillator $k$ reads
\begin{align}
    \dbQ_{G,k}
    =
    G_k
    \msf{J}
    \dbPhi_{G,k},
    \label{Eq:GyratorConstraint}
\end{align}
where $G_k$ denotes its gyration parameter, and $\msJ$ is the standard
two-dimensional symplectic matrix, i.e.,
\begin{align}
    \mathrm{d}Q_{G,k}^{a}
    &=
    G_k
    \mathrm{d}\Phi_{G,k}^{b},
    \nonumber\\
    \mathrm{d}Q_{G,k}^{b}
    &=
    -G_k
    \mathrm{d}\Phi_{G,k}^{a}.
\end{align}
The gyrator constraint reduces the four variables
$(\Phi_{G,k}^{a},\Phi_{G,k}^{b},
Q_{G,k}^{a},Q_{G,k}^{b})$
to one canonical pair, which, after trivial integration, can be re-expressed as
\begin{align}
    X_k
    &\equiv
    \Phi_{G,k}^{a}
    =
    -\frac{Q_{G,k}^{b}}{G_k},
    \nonumber\\
    P_k
    &\equiv
    -G_k\Phi_{G,k}^{b}
    =
    -Q_{G,k}^{a}.
    \label{Eq:GyratorCanonicalVariables}
\end{align}
Thus, we can collect the gyrator variable vectors in
\begin{align}
    \bQ_{G,k}
    &=
    -
    \begin{pmatrix}
        P_k\\
        G_kX_k
    \end{pmatrix},
    \quad
    \bPhi_{G,k}
    =
    \begin{pmatrix}
        X_k\\
        -P_k/G_k
    \end{pmatrix}
    \label{Eq:GyratorChargeCanonicalVariables}
\end{align}
to reduce \cref{eq:w_Gk_Z_init} to
\begin{align}
    \omega^{G,k}
    =
    \mathrm{d}P_k
    \wedge
    \mathrm{d}X_k
    +
    \frac{1}{2}
    \dbQ_P^T
    \wedge
    \left(
        \msT_k^T
        \dbPhi_{G,k}
    \right).
\end{align}

Finally, the reduced nonlinear ports contribute (see details in
\cref{AppSubsec:ParallelNonlinearInductors})
\begin{align}
    \omega^P = \frac{1}{2} \dbQ_P^T \wedge \dbPhi_P.
\end{align}
Note that the current direction at the port is opposite to that of the (potentially parallel) inductor elements, hence the sign in the previous equation~\cite{ParraRodriguez:2024}. Using the voltage constraints for the transformer
\begin{align}
    \dbPhi_P
    =
    \msT_C^T
    \mathrm{d}\tilde{\bPhi}_C
    +
    \sum_k
    \msT_k^T
    \dbPhi_{G,k},
\end{align}
the total two-form reduces directly to a canonical form
\begin{align}
    \omega
    =
    \dbQ_P^T
    \wedge
    \dbPhi_P
    +
    \sum_k
    \mathrm{d}P_k
    \wedge
    \mathrm{d}X_k.
\end{align}

For full-rank $\msC_{\msZ,0}^{-1}$, we have identified independent canonical pairs with Poisson brackets $\{\bPhi_P,\bQ_P^T\}=\msId$ and $\{X_k,P_{k'}\}=\delta_{kk'}$. In the finite-mode galvanic case $\msC_{\msZ,0}^{-1}=0$, however, $\msT_C=0$ and the voltage constraint reduces to
\begin{align}
    \mathrm d\bPhi_P = \sum_k \msT_k^T\,\mathrm d\bPhi_{G,k}.
\end{align}
The port and mode variables are therefore not all independent. The two-form and Hamiltonian must be restricted to this constraint manifold, with any remaining null directions reduced before quantization. More generally, if $\msC_{\msZ,0}^{-1}$ is rank deficient, the
same constrained reduction applies to
\begin{align*}
    \boldsymbol{\lambda}^{T}
    \left(
        \mathrm d\bPhi_P
        -\sum_k\msT_k^T\,\mathrm d\bPhi_{G,k}
    \right)=0
\end{align*}
for every $\boldsymbol{\lambda}\in\ker\msC_{\msZ,0}^{-1}$.

\subsubsection{Hamiltonian construction}
The canonical circuit representation for a given pole is not unique, and different physical parameters give the same response~\cite{Newcomb:1966}. As mentioned in the previous subsubsection, without loss of generality, one can assume the internal capacitors to be normalized to unity ($C_{k}^a=C_{k}^b=1$), with their physical values absorbed into their corresponding transformer matrices ($\msT_k$). Using this convention, the energy function summing contributions from each capacitor and (nonlinear) inductor of the circuit is
\begin{align}
    H
    &=
    U(\bPhi_P,\bPhi_{e,P})
    +
    \frac{1}{2}
    \tilde{\bQ}_C^T
    \tilde{\bQ}_C
    \\
    &\quad+
    \frac{1}{2}
    \sum_k
    \left(
        \bQ_{G,k}
        +
        \bQ_{GT,k}
    \right)^T
    \left(
        \bQ_{G,k}
        +
        \bQ_{GT,k}
    \right).\nonumber
\end{align}
Inserting the constraints derived in \cref{Sec:TwoForm}, this becomes
\begin{align}
    H = H_P^\msZ + H_B + H_{PB},
    \label{Eq:FluxChargeHamiltonian}
\end{align}
with
\begin{align}
\begin{aligned}
    H_P^\msZ
    &=
    U(\bPhi_P,\bPhi_{e,P})
    +
    \frac{1}{2}
    \bQ_P^T
    \left(
        \msC_{\msZ,0}^{-1}
        +
        \msC_{\msZ,\infty}^{-1}
    \right)
    \bQ_P,
    \\
    H_B
    &=
    \frac{1}{2}
    \sum_k
    \left(
        P_k^2
        +
        G_k^2X_k^2
    \right),
    \\
    H_{PB}
    &=
    \sum_k
    \bQ_P^T
    \left(
        \msT_{k,a}^T P_k
        +
        G_k
        \msT_{k,b}^T X_k
    \right).
\end{aligned}
\label{Eq:DiscreteHamiltonianZ}
\end{align}
Here, the unit-capacitance normalization has been used explicitly, such that the zero-pole capacitive matrix is decomposed as
\begin{align}
    \msT_C^T\msId\msT_C
    =
    \msC_{\msZ,0}^{-1}.
\end{align}
Similarly, we have defined the high-frequency inverse-capacitance dressing as
\begin{align}
    \msC_{\msZ,\infty}^{-1}
    \equiv
    \sum_k
    \msT_k^T\msId\msT_k
    =
    \sum_\alpha
    \msA_\alpha
    =
    \lim_{\eta\rightarrow\infty}
    \eta\msZ^+(\eta).
    \label{Eq:ImpedanceHighFreqCapacitance}
\end{align}

\subsubsection{Quantization and internal modes in the Fock space}
\label{AppSubsubsec:Quantization_Fock_Z}

Before quantizing the internal modes, we briefly recall the
quantization of the dressed port degrees of freedom discussed in the
main text. Since $\bQ_P\equiv\bQ_C$ in the impedance representation,
extended canonical pairs are promoted according to
\begin{align}
    \left[
        \hat{\Phi}_p,
        \hat{Q}_{Cp'}
    \right]
    =
    i\hbar\delta_{pp'},
\end{align}
whereas compact flux directions are described by the globally defined
operators $\hat n_{Cp}$ and $e^{i\hat{\varphi}_p}$, with
\begin{align}
    Q_{Cp}
    &\overset{q.}{\longrightarrow}
    2e
    \left(
        \hat n_{Cp}-n_{g,p}
    \right),
    \nonumber\\
    e^{i(2\pi/\Phi_0)\Phi_p}
    &\overset{q.}{\longrightarrow}
    e^{i\hat{\varphi}_p}.
\end{align}
Thus, for both compact and extended directions, the port-charge
variables entering the quantum Hamiltonian are replaced according to
\begin{align}
    \bQ_P
    \equiv
    \bQ_C
    \overset{q.}{\longrightarrow}
    \hat{\bQ}_C-\bQ_{g,P},
    \label{Eq:PortChargeQuantizationZ}
\end{align}
where the components of $\bQ_{g,P}$ associated with extended flux
coordinates are unitarily removable. The identification of compact
and extended directions, and the corresponding quantization of the
port potential $U$, are discussed in the main text (Subsec.~\ref{Sec:DiscreteQuantizedHamiltonian}).

For the canonical realization chosen in \cref{Eq:GaugeChoiceZ},
the gyration parameters are fixed by the corresponding pole
frequencies ($G_k=\omega_k$). We can therefore introduce the bosonic
operators through
\begin{align}
    X_k
    &=
    \sqrt{\frac{\hbar}{2\omega_k}}
    \left(
        a_k+a_k^\dag
    \right),
    \nonumber\\
    P_k
    &=
    -i\sqrt{\frac{\hbar\omega_k}{2}}
    \left(
        a_k-a_k^\dag
    \right).
\end{align}
The bath Hamiltonian is then simply given by
\begin{align}
    H_B
    =
    \sum_k
    \hbar\omega_k
    \left(
        a_k^\dag
        a_k
        +
        \frac{1}{2}
    \right).
\end{align}
Using \cref{Eq:PortChargeQuantizationZ}, the interaction becomes
\begin{align}
\begin{aligned}
    H_{PB}
    =&
    \sum_k
    \left(
        \hat{\bQ}_C-\bQ_{g,P}
    \right)^T
    \\&\times
    \sqrt{
        \frac{\hbar\omega_k}{2}
    }
    \bigg[
        \left(
            \msT_{k,b}^T
            -
            i\msT_{k,a}^T
        \right)
        a_k+
        \left(
            \msT_{k,b}^T
            +
            i\msT_{k,a}^T
        \right)
        a_k^\dag
    \bigg]
    \\
    =&
    \sum_\alpha
    \left[
        \left(
            \hat{\bQ}_C-\bQ_{g,P}
        \right)^T
        \msG_\alpha\ba_\alpha
        +
        \mathrm{h.c.}
    \right].
\end{aligned}
\end{align}
Here, we have collected again the $r_\alpha$ polarizations associated with each pole $\alpha$, and the vector of modes per frequency
$\ba_\alpha=(a_{\alpha,1},\ldots,a_{\alpha,r_\alpha})^T$,
and introduced the coupling matrix
\begin{align}
    \msG_\alpha
    =
    \sqrt{
        \frac{\hbar\omega_\alpha}{2}
    }
    \left(
        \msT_{\alpha,b}^T
        -
        i\msT_{\alpha,a}^T
    \right).
    \label{Eq:NRCoupling}
\end{align}
Using \cref{Eq:DiscretePoleMapping,Eq:NonReciprocalTransformerImpedanceRelation}, the coupling matrices are related to the residues at the pole through
\begin{align}
    \msG_\alpha\msG_\alpha^\dag
    =
    \hbar\omega_\alpha\msR_\alpha.
\end{align}

\subsection{Continuum limit}
We finally take the same Riemann limit introduced
in \cref{Sec:CauerDecomposition}. With the normalization chosen above,
\begin{align}
\begin{aligned}
    \msT_\alpha
    &=
    \msT(\omega_\alpha)\sqrt{\Delta\omega},
    \\
    \msG_\alpha
    &=
    \msG(\omega_\alpha)\sqrt{\Delta\omega},
    \\
    \ba_\alpha
    &=
    \ba(\omega_\alpha)\sqrt{\Delta\omega}.
\end{aligned}
\end{align}
The continuum operators obey
\begin{align}
    \left[
        a_\epsilon(\omega),
        a_{\epsilon'}^\dagger(\omega')
    \right]
    =
    \delta_{\epsilon\epsilon'}
    \delta(\omega-\omega').
\end{align}

Dropping the bath zero-point constant and applying the port
quantization \cref{Eq:PortChargeQuantizationZ}, the Hamiltonian becomes
\begin{align}
\begin{aligned}
    &H_P^\msZ
    =U(\hat{\bPhi}_P,\bPhi_{e,P})
    \\
    &+
    \frac{1}{2}
    \left(
        \hat{\bQ}_C-\bQ_{g,P}
    \right)^T
    \left(
        \msC_{\msZ,0}^{-1}
        +
        \msC_{\msZ,\infty}^{-1}
    \right)
    \left(
        \hat{\bQ}_C-\bQ_{g,P}
    \right),
    \\
    &H_B
    =
    \int_{\mathcal B_{\msZ}}
    \mathrm{d}\omega\,
    \hbar\omega\,
    \ba^\dagger(\omega)
    \ba(\omega),
    \\
    &H_{PB}
    =
    \int_{\mathcal B_{\msZ}}
    \mathrm{d}\omega\,
    \left[
        \left(
            \hat{\bQ}_C-\bQ_{g,P}
        \right)^T
        \msG(\omega)
        \ba(\omega)
        +
        \mathrm{h.c.}
    \right].
\end{aligned}\nonumber
\end{align}
Here, $U$ denotes the corresponding quantum port-potential operator,
with its compact and extended directions quantized as described in
the main text.

The high-frequency capacitive dressing entering $H_P^\msZ$ takes the
continuum form
\begin{align}
    \msC_{\msZ,\infty}^{-1}
    =
    \int_0^\infty
    \mathrm{d}\omega\,
    \msA(\omega)
    =
    \frac{2}{\pi}
    \int_0^\infty
    \mathrm{d}\omega\,
    \Re\{\msZ^+[\omega]\},
    \label{Eq:DressedQubitHamiltonian}
\end{align}
where we have used \cref{Eq:ContinuousPoleMapping}, whenever the
corresponding high-frequency dressing is finite.

The residue factorization above has a direct continuum counterpart.
Combining \cref{Eq:FullPoleResidue,Eq:RiemannPoleMapping,Eq:ContinuousPoleMapping}
gives
\begin{align}
    \msR_\alpha
    =
    \frac{\Delta\omega}{\pi}
    \msZH[\omega_\alpha],
    \qquad
    \msZH[\omega]
    \equiv
    \frac{
        \msZ[\omega]
        +
        \msZ^\dag[\omega]
    }{2}.
    \label{Eq:ContinuumResidueMapping}
\end{align}
If its frequency-local eigendecomposition is written as
\begin{align}
    \msZH[\omega]
    =
    \msU^\dag(\omega)
    \msZD(\omega)
    \msU(\omega),
\end{align}
the continuum limit of \cref{Eq:TransformerEigenGauge} gives directly
\begin{align}
    \msT_b^T(\omega)
    -
    i\msT_a^T(\omega)
    =
    \sqrt{
        \frac{2}{\pi}
    }
    \msU^\dag(\omega)
    \msZD^{1/2}(\omega).
    \label{Eq:ContinuumTransformerGauge}
\end{align}
Using the continuum version of \cref{Eq:NRCoupling}, the coupling
matrix is therefore
\begin{align}
\begin{aligned}
    \msG(\omega)
    &=
    \sqrt{
        \frac{\hbar\omega}{2}
    }
    \left[
        \msT_b^T(\omega)
        -
        i\msT_a^T(\omega)
    \right]
    \\
    &=
    \sqrt{
        \frac{\hbar\omega}{\pi}
    }
    \msU^\dag(\omega)
    \msZD^{1/2}(\omega),
\end{aligned}
\end{align}
which satisfies
\begin{align}
    \msG(\omega)
    \msG^\dag(\omega)
    =
    \frac{\hbar\omega}{\pi}
    \msZH[\omega].
    \label{Eq:ZCouplingSpectralIdentities}
\end{align}
Thus, although the transformer realization provides the underlying
canonical circuit construction, its explicit synthesis is not
required to obtain the Hamiltonian couplings: a frequency-local
eigendecomposition of $\msZH[\omega]$ directly determines an
equivalent coupling matrix $\msG(\omega)$.

\section{Details on the quantization of the nonlinear port coupled to a
$\msY$-environment}
\label{AppSec:Y_construction_and_quantization}

\subsection{Cauer decompositions of discrete and continuous
$\msY$ environments}
\label{Sec:CauerDecompositionY}

In this Appendix, we provide the admittance counterpart of the construction developed in
\cref{AppSec:Z_construction_and_quantization}. We first consider
the canonical Cauer decomposition of a discrete $\msY$-environment
and then take the same continuum limit introduced there. The exact
Hamiltonian is obtained by applying the first-order construction
summarized in \cref{AppSec:FirstOrderReview}. Both the discrete
and continuous Hamiltonians are written in the flux gauge.

\begin{figure}[t]
    \centering
    \includegraphics[width=\linewidth]
    {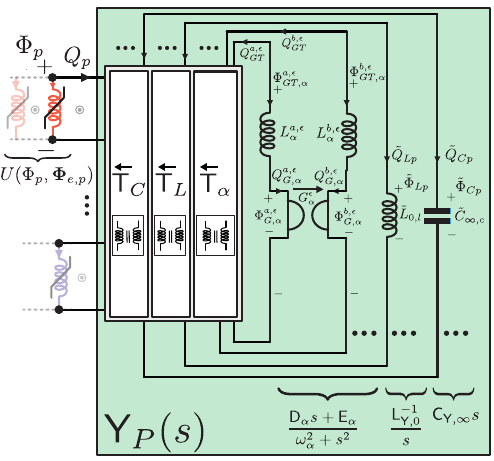}
    \caption{Schematic of the nonlinear inductive ports coupled to the canonical admittance representation of \cref{Eq:AdmittanceSplit}, with dual nonreciprocal harmonic oscillator modes and reversed transformers. All static flux offsets are assigned to the external port potential $U(\bPhi_P,\bPhi_{e,P})$. Parallel port capacitances are included in $\msC_{\msY,\infty}$.}
    \label{fig:SchemaY}
\end{figure}

We consider an $N$-port admittance matrix of the form
\begin{align}
    \msY_P(s)
    =
    s\msC_{\msY,\infty}
    +
    \frac{\msL_{\msY,0}^{-1}}{s}
    +
    \sum_{\alpha=1}^{M}
    \msY_\alpha(s),
    \label{Eq:AdmittanceSplit}
\end{align}
where $\msC_{\msY,\infty}$ is the capacitive pole at infinity,
$\msL_{\msY,0}^{-1}$ is the inductive pole at zero, and
$\msY_\alpha(s)$ denotes a finite-frequency pole, see Fig.~\ref{fig:SchemaY}. Observe that generic passivity allows for
an extra constant nonreciprocal contribution but for simplicity here, we have set it to zero~\cite{Newcomb:1966,Labarca:2024}.

The finite-frequency poles are written as
\begin{align}
    \msY_\alpha(s)
    =
    \frac{
        \msD_\alpha s+\msE_\alpha
    }{
        \omega_\alpha^2+s^2
    },
    \label{Eq:YDiscretePoles}
\end{align}
where $\msD_\alpha$ are symmetric positive-semidefinite matrices,
whereas $\msE_\alpha$ are antisymmetric matrices. We define the
finite-frequency admittance $\msY(s) = \msY_P(s) - s\msC_{\msY,\infty} -\msL_{\msY,0}^{-1}/s$.

The residue and boundary-value constructions are identical to those
given explicitly for the impedance in
\cref{AppSec:Z_construction_and_quantization}, under the
replacements
$\msZ\rightarrow\msY$,
$\msA\rightarrow\msD$, and
$\msB\rightarrow\msE$.
In particular, the full residue at $s=-i\omega_\alpha$ is
\begin{align}
    \msR_\alpha^\msY
    &\equiv
    \underset{s=-i\omega_\alpha}{\operatorname{Res}}
    \msY(s)=
    \frac{1}{2}
    \left(
        \msD_\alpha
        +
        \frac{i}{\omega_\alpha}
        \msE_\alpha
    \right)
    \succeq0,
    \label{Eq:YFullPoleResidue}
\end{align}
and we denote $r_\alpha \equiv\operatorname{rank}\msR_\alpha^\msY$.

For a continuous admittance, the corresponding positive-frequency
densities are
\begin{align}
    \msD(\omega)
    &=
    \frac{2}{\pi}
    \Re\{\msY^+[\omega]\}, \quad
    \msE(\omega)
    =
    \frac{2\omega}{\pi}
    \Im\{\msY^-[\omega]\},
    \label{Eq:YContinuousPoleMapping}
\end{align}
where $\msY^\pm=(\msY\pm\msY^T)/2$. For bounded or disconnected continua, these spectral densities vanish outside $\mathcal B_{\msY}$, so the following integrals may still be written over $0<\omega<\infty$. A uniform discretization gives $\msD_\alpha =\msD(\omega_\alpha)\Delta\omega, \msE_\alpha=   \msE(\omega_\alpha)\Delta\omega$. Accordingly,
\begin{align}
    \msY(s)
    =
    \msY^+(s)+\msY^-(s)
    =
    \int_0^\infty
    \mathrm{d}\omega\,
    \frac{
        s\msD(\omega)+\msE(\omega)
    }{
        s^2+\omega^2
    }.
    \label{Eq:CauerContinuumY}
\end{align}

The singular terms are synthesized using the left-transformer
orientation introduced in \cref{AppSec:FirstOrderReview}.
As in the impedance construction, we normalize the internal energetic
branches to the identity and absorb their physical values into the
corresponding transformer matrices,
\begin{align}
    \msC_{\msY,\infty}
    &=
    \msT_C\msT_C^T,
    \nonumber\\
    \msL_{\msY,0}^{-1}
    &=
    \msT_L\msT_L^T.
    \label{Eq:YInfZeroSynthesis}
\end{align}

For the finite-frequency sections, it is convenient to use the
impedance representation of the fundamental gyrator. From
\cref{Eq:AppGyratorConstraint},
\begin{align}
    \dbPhi_{G,k}
    =
    \msZ_{G,k}\dbQ_{G,k},
    \label{Eq:YGyratorImpedance}
\end{align}
with $\msZ_{G,k} =-R_k\msf{J},$ and $ R_k\equiv G_k^{-1}$. For the sake of simplicity and without loss of generality, we choose unit internal inductances and
\begin{align}
    R_\alpha  = \omega_\alpha
    \label{Eq:GaugeChoiceY}
\end{align}
for all polarizations associated with pole $\alpha$.

As in the impedance construction, the equalities
$\msL_\alpha=\msId$ and $R_\alpha=\omega_\alpha$ are understood in
normalized internal units. With a slight abuse of notation, the
dimensional factors associated with this normalization are absorbed
into the transformer matrices, or equivalently into the normalization
of the internal branch variables.

For each polarization
$\epsilon\in\{1,\ldots,r_\alpha\}$, we define the left transformer
\begin{align}
    \msT_\alpha^\epsilon
    =
    \begin{pmatrix}
        \msT_{\alpha,a}^\epsilon
        &
        \msT_{\alpha,b}^\epsilon
    \end{pmatrix},
    \qquad
    \msT_{\alpha,a}^\epsilon,
    \msT_{\alpha,b}^\epsilon
    \in
    \mathbb{R}^{N\times1}.
    \label{Eq:YDiscreteTransformerDefinition}
\end{align}
Collecting the polarization-resolved columns,
\begin{align}
    \msT_{\alpha,a}
    &=
    \begin{pmatrix}
        \msT_{\alpha,a}^{1}
        &
        \cdots
        &
        \msT_{\alpha,a}^{r_\alpha}
    \end{pmatrix},
    \nonumber\\
    \msT_{\alpha,b}
    &=
    \begin{pmatrix}
        \msT_{\alpha,b}^{1}
        &
        \cdots
        &
        \msT_{\alpha,b}^{r_\alpha}
    \end{pmatrix},
\end{align}
with
$\msT_{\alpha,a},\msT_{\alpha,b}
\in\mathbb{R}^{N\times r_\alpha}$.

The corresponding pole matrices are then related to the transformer matrices through
\begin{align}
\begin{aligned}
    \msD_\alpha
    &=
    \msT_{\alpha,a}
    \msT_{\alpha,a}^T
    +
    \msT_{\alpha,b}
    \msT_{\alpha,b}^T,
    \\
    \msE_\alpha
    &=
    \omega_\alpha
    \left(
        \msT_{\alpha,a}
        \msT_{\alpha,b}^T
        -
        \msT_{\alpha,b}
        \msT_{\alpha,a}^T
    \right).
\end{aligned}
\label{Eq:NonReciprocalTransformerAdmittanceRelation}
\end{align}
Equivalently, each pole is realized as
\begin{align}
    \msY_\alpha(s)
    =
    \sum_{\epsilon=1}^{r_\alpha}
    \msT_\alpha^\epsilon
    \frac{1}{s^2+\omega_\alpha^2}
    \begin{pmatrix}
        s & \omega_\alpha\\
        -\omega_\alpha & s
    \end{pmatrix}
    \left(
        \msT_\alpha^\epsilon
    \right)^T.
    \label{Eq:YCanonicalPoleRealization}
\end{align}
The two relations above combine into
\begin{align}
\begin{aligned}
    &
    \left(
        \msT_{\alpha,a}
        -
        i\msT_{\alpha,b}
    \right)
    \left(
        \msT_{\alpha,a}
        -
        i\msT_{\alpha,b}
    \right)^\dag
    \\
    &\qquad=
    \msD_\alpha
    +
    \frac{i}{\omega_\alpha}
    \msE_\alpha
    =
    2\msR_\alpha^\msY.
\end{aligned}
\label{Eq:YTransformerResidueFactorization}
\end{align}
Thus, the same residue eigendecomposition used in
\cref{AppSec:Z_construction_and_quantization} directly determines
the transformer columns, up to unitary rotations among the
$r_\alpha$ polarization channels.

\subsection{Derivation of the exact circuit Hamiltonian}
\label{Sec:CanonicalHamiltonianY}
We now apply the first-order construction summarized in
\cref{AppSec:FirstOrderReview} to the canonical circuit in
\cref{fig:SchemaY}. The external ports are treated as in \cref{Sec:CanonicalHamiltonian}, i.e., if there are several parallel nonlinear inductors at a port, they can be grouped together and described by a single pair of variables.

Denoting the charge of the normalized internal capacitive branch by
$\tilde{\bQ}_C$, we define
\begin{align}
    \bQ_C
    \equiv
    \msT_C\tilde{\bQ}_C.
    \label{Eq:QCwidetildeDefinition}
\end{align}
The pair $(\bPhi_P,\bQ_C)$ will emerge below as the external
canonical pair.

As in the previous Appendix, we combine the pole and polarization
indices into
$ k\equiv(\alpha,\epsilon)$, $\sum_k  \equiv \sum_{\alpha=1}^{M}
    \sum_{\epsilon=1}^{r_\alpha}$, $ \omega_k \equiv \omega_\alpha$.
For each $k$,
\begin{align}
    \msT_k
    =
    \begin{pmatrix}
        \msT_{k,a}
        &
        \msT_{k,b}
    \end{pmatrix}
    \in
    \mathbb{R}^{N\times2},
\end{align}
while $\bPhi_{G,k}$, $\bQ_{G,k}$, and $\bPhi_{GT,k}$ are
two-component vectors in the $a,b$ space of the corresponding
nonreciprocal oscillator.

\subsubsection{Canonical two-form}
\label{Sec:TwoFormY}

Following \cref{AppSec:FirstOrderReview}, we impose Kirchhoff's
constraints together with the left-transformer and gyrator relations. The nonlinear branch charges at each port have already been reduced to the port charge $Q_p$ as described in
\cref{AppSubsec:ParallelNonlinearInductors}. With the orientations
shown in \cref{fig:SchemaY}, the current constraint therefore reads
\begin{align}
    -\dbQ_P
    =
    -\msT_C
    \mathrm{d}\tilde{\bQ}_C
    -
    \msT_L
    \dbQ_L
    +
    \sum_k
    \msT_k
    \dbQ_{GT,k}.
    \label{Eq:YCurrentConstraint}
\end{align}
The corresponding differential left-transformer flux constraints are
\begin{align}
    \mathrm{d}\tilde{\bPhi}_C
    &=
    \msT_C^T
    \dbPhi_P,
    \quad
    \dbPhi_L
    =
    \msT_L^T
    \dbPhi_P,
    \nonumber\\
    \dbPhi_{GT,k}
    &=
    \msT_k^T
    \dbPhi_P.
    \label{Eq:YFluxConstraints}
\end{align}

We assign the static flux offsets to the external potential
$U(\bPhi_P,\bPhi_{e,P})$, following
\cref{AppSubsec:ParallelNonlinearInductors}. With the flux origins of
the passive realization chosen consistently with this allocation, the
integrated linear constraints are

\begin{align}
    \tilde{\bPhi}_C
    &=
    \msT_C^T\bPhi_P,\quad
    \bPhi_L
    =
    \msT_L^T\bPhi_P,\quad
    \bPhi_{GT,k}
    =
    \msT_k^T\bPhi_P.
    \label{Eq:YIntegratedFluxConstraints}
\end{align}
The offsets are constant, so this choice of reference does not
alter the differential constraints or the reduced two-form. Using the energetic-element rules of \cref{AppSec:FirstOrderReview}, the pullback of the precanonical two-form is
\begin{align}
    \omega
    =
    \dbQ_C^T
    \wedge
    \dbPhi_P
    +
    \frac{1}{2}
    \sum_k
    \dbQ_{G,k}^T
    \wedge
    \msZ_{G,k}
    \dbQ_{G,k}.
    \label{Eq:TwoFormFinalYShort}
\end{align}

Using \cref{Eq:YGyratorImpedance}, we introduce the canonical pair
\begin{align}
    X_k
    &\equiv
    Q_{G,k}^{b},\qquad
    P_k
    \equiv
    -R_kQ_{G,k}^{a}.
    \label{Eq:YGyratorCanonicalVariables}
\end{align}
The corresponding gyrator variables are
\begin{align}
    \bQ_{G,k}
    &=
    \begin{pmatrix}
        -P_k/R_k\\
        X_k
    \end{pmatrix},
    \qquad
    \bPhi_{G,k}
    =
    -
    \begin{pmatrix}
        R_kX_k\\
        P_k
    \end{pmatrix}.
    \label{Eq:YGyratorCanonicalVectors}
\end{align}
It follows directly that
\begin{align}
    \frac{1}{2}
    \dbQ_{G,k}^T
    \wedge
    \msZ_{G,k}
    \dbQ_{G,k}
    =
    \mathrm{d}P_k
    \wedge
    \mathrm{d}X_k.
\end{align}
Hence,
\begin{align}
    \omega
    =
    \dbQ_C^T
    \wedge
    \dbPhi_P
    +
    \sum_k
    \mathrm{d}P_k
    \wedge
    \mathrm{d}X_k,
\end{align}
with
$\{\Phi_p,Q_{Cp'}\}=\delta_{pp'}$ and
$\{X_k,P_{k'}\}=\delta_{kk'}$.

\subsubsection{Hamiltonian construction}

With the normalized internal energetic elements introduced above, the
energy function is
\begin{align}
\begin{aligned}
    H
    ={}&
    U(\bPhi_P,\bPhi_{e,P})
    +
    \frac{1}{2}
    \tilde{\bQ}_C^T
    \tilde{\bQ}_C
    +
    \frac{1}{2}
    \bPhi_L^T
    \bPhi_L
    \\
    &+
    \frac{1}{2}
    \sum_k
    \left(
        \bPhi_{GT,k}
        -
        \bPhi_{G,k}
    \right)^T
    \left(
        \bPhi_{GT,k}
        -
        \bPhi_{G,k}
    \right).
\end{aligned}
\label{Eq:YOriginalEnergy}
\end{align}
Using \cref{Eq:QCwidetildeDefinition,Eq:YIntegratedFluxConstraints,Eq:YGyratorCanonicalVectors}, this becomes
\begin{align}
    H
    =
    H_P^\msY
    +
    H_B
    +
    H_{PB},
\end{align}
with

\begin{align}
\begin{aligned}
    H_P^\msY
    ={}&
    U(\bPhi_P,\bPhi_{e,P})+
    \frac{1}{2}
    \bPhi_P^T
    (\msL_{\msY,0}^{-1}+\msL_{\msY,\infty}^{-1})
    \bPhi_P
    \\
    &
    +
    \frac{1}{2}
    \bQ_C^T
    \msC_{\msY,\infty}^{-1}
    \bQ_C,
    \\
    H_B
    ={}&
    \frac{1}{2}
    \sum_k
    \left(
        P_k^2
        +
        R_k^2X_k^2
    \right),
    \\
    H_{PB}
    ={}&
    \sum_k
    \bPhi_P^T
    \left(
        \msT_{k,b}P_k
        +
        R_k
        \msT_{k,a}X_k
    \right).
\end{aligned}
\label{Eq:YExternalHamiltonianDiscrete}
\end{align}

Here, we have defined the finite-frequency inductive dressing
\begin{align}
    \msL_{\msY,\infty}^{-1}
    &\equiv
    \sum_k
    \msT_k\msT_k^T
    =
    \sum_\alpha
    \msD_\alpha
    =
    \lim_{\eta\rightarrow\infty}
    \eta\msY^+(\eta).
    \label{Eq:YHighFreqInductance}
\end{align}
Similarly, under the full-rank assumption for
$\msC_{\msY,\infty}$,
$\tilde{\bQ}_C^T\tilde{\bQ}_C
=
\bQ_C^T\msC_{\msY,\infty}^{-1}\bQ_C$
via $\msT_C$. The zero-frequency term follows from
$\msL_{\msY,0}^{-1}=\msT_L\msT_L^T$ and
\cref{Eq:YIntegratedFluxConstraints}.

\subsubsection{Quantization and internal modes in the Fock space}
\label{AppSubsubsec:Quantization_Fock_Y}

We briefly recall the port quantization discussed in the main text. We restrict here to extended port canonical pairs, as discussed in the main text. Full rank of $\msL_{\msY,0}^{-1}$ is sufficient, but explicit inductive terms in $U$ can also enforce this property. The static branch offsets are classical parameters and do not modify the canonical algebra, so that $\left[\hat{\Phi}_p, \hat{Q}_{Cp'} \right] =    i\hbar\delta_{pp'}$.

When $\msL_{\msY,0}^{-1}$ is not full rank and $U$ retains
nontrivial periodic directions, compact coordinates may instead
correspond to collective port--environment symmetries. Their systematic treatment and the additional subtleties arising in gapless environments are left for future work, as discussed in the main text (Subsec.~\ref{Sec:DiscreteQuantizedHamiltonian}).

For the canonical realization chosen in \cref{Eq:GaugeChoiceY},
the gyration resistances are fixed by the corresponding pole
frequencies ($R_k=\omega_k$). We can therefore introduce the bosonic
operators through

\begin{align}
    X_k
    &=
    \sqrt{
        \frac{\hbar}{2\omega_k}
    }
    \left(
        a_k+a_k^\dag
    \right),
    \quad
    P_k
    =
    -i
    \sqrt{
        \frac{\hbar\omega_k}{2}
    }
    \left(
        a_k-a_k^\dag
    \right),
\end{align}
and $H_B = \sum_k\hbar\omega_k\left( a_k^\dag a_k  + \frac{1}{2}    \right)$. The interaction then becomes
\begin{align}
\begin{aligned}
    H_{PB}
    ={}&
    \sum_k
    \hat{\bPhi}_P^T
    \sqrt{
        \frac{\hbar\omega_k}{2}
    }
    \bigg[
        \left(
            \msT_{k,a}
            -
            i\msT_{k,b}
        \right)
        a_k
    \\
    &\qquad\qquad+
        \left(
            \msT_{k,a}
            +
            i\msT_{k,b}
        \right)
        a_k^\dag
    \bigg]
    \\
    ={}&
    \sum_\alpha
    \left[
        \hat{\bPhi}_P^T
        \msG_\alpha
        \ba_\alpha
        +
        \mathrm{h.c.}
    \right],
\end{aligned}
\end{align}
where again
$\ba_\alpha=
(a_{\alpha,1},\ldots,a_{\alpha,r_\alpha})^T$, and
\begin{align}
    \msG_\alpha
    =
    \sqrt{
        \frac{\hbar\omega_\alpha}{2}
    }
    \left(
        \msT_{\alpha,a}
        -
        i\msT_{\alpha,b}
    \right).
    \label{Eq:YNRCoupling}
\end{align}
Using \cref{Eq:YFullPoleResidue,Eq:YTransformerResidueFactorization}, the coupling matrices satisfy
\begin{align}
    \msG_\alpha
    \msG_\alpha^\dag
    =
    \hbar\omega_\alpha
    \msR_\alpha^\msY.
    \label{Eq:YDiscreteCouplingResidue}
\end{align}

\subsection{Continuum limit}

We finally take the same Riemann limit introduced in
\cref{AppSec:Z_construction_and_quantization}, with
\begin{align}
    \msG_\alpha
    &=
    \msG(\omega_\alpha)
    \sqrt{\Delta\omega},
    &
    \ba_\alpha
    &=
    \ba(\omega_\alpha)
    \sqrt{\Delta\omega}.
\end{align}
The continuum operators obey
\begin{align}
    \left[
        a_\epsilon(\omega),
        a_{\epsilon'}^\dagger(\omega')
    \right]
    =
    \delta_{\epsilon\epsilon'}
    \delta(\omega-\omega').
\end{align}
Dropping the bath zero-point constant, the quantum Hamiltonian becomes

\begin{align}
\begin{aligned}
    H_B
    ={}&
    \int_{\mathcal B_{\msY}}
    \mathrm{d}\omega\,
    \hbar\omega\,
    \ba^\dag(\omega)
    \ba(\omega),
    \\
    H_{PB}
    ={}&
    \int_{\mathcal B_{\msY}}
    \mathrm{d}\omega\,
    \left[
        \hat{\bPhi}_P^T
        \msG(\omega)
        \ba(\omega)
        +
        \mathrm{h.c.}
    \right],
\end{aligned}
\label{Eq:YContinuumHamiltonian}
\end{align}
with the same $H_P^{\msY}$ as in \cref{Eq:YExternalHamiltonianDiscrete}. The finite-frequency inductive dressing entering $H_P^\msY$ takes the
continuum form
\begin{align}
    \msL_{\msY,\infty}^{-1}
    =
    \int_0^\infty
    \mathrm{d}\omega\,
    \msD(\omega)
    =
    \frac{2}{\pi}
    \int_0^\infty
    \mathrm{d}\omega\,
    \Re\{\msY^+[\omega]\},
    \label{Eq:YContinuumInductiveDressing}
\end{align}
whenever the corresponding dressing is finite.

Finally, defining
\begin{align}
    \msYH[\omega]
    \equiv
    \frac{
        \msY[\omega]
        +
        \msY^\dag[\omega]
    }{2},
\end{align}
and writing its frequency-local eigendecomposition as
\begin{align}
    \msYH[\omega]
    =
    \msU^\dag(\omega)
    \msD_{\msY}(\omega)
    \msU(\omega),
\end{align}
the continuum counterpart of \cref{Eq:YDiscreteCouplingResidue} gives
directly
\begin{align}
    \msG(\omega)
    =
    \sqrt{
        \frac{\hbar\omega}{\pi}
    }\,
    \msU^\dag(\omega)
    \msD_{\msY}^{1/2}(\omega),
\end{align}
and therefore satisfies
\begin{align}
    \msG(\omega)
    \msG^\dag(\omega)
    =
    \frac{\hbar\omega}{\pi}
    \msYH[\omega].
    \label{Eq:YCouplingSpectralIdentities}
\end{align}
Thus, as in the impedance representation, the explicit continuum transformer synthesis is not required to obtain the Hamiltonian couplings.

\section{Time-dependent fluxes and parallel capacitances at nonlinear ports}
\label{AppSec:DrivenNonlinearPorts}

\begin{figure}[t]
    \centering
    \includegraphics[width=\linewidth]{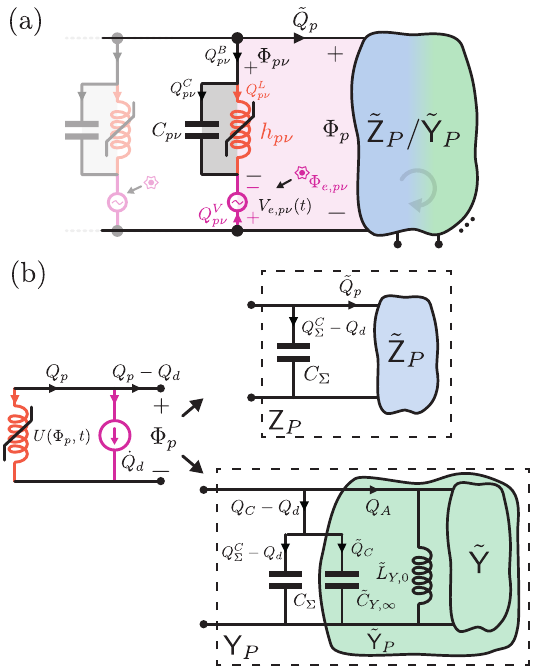}
    \caption[Time-dependent fluxes at nonlinear ports]{    Time-dependent fluxes at nonlinear ports and their equivalent
    circuit representations.
    (a) Each nonlinear branch $h_{p\nu}\parallel C_{p\nu}$ is placed
    in series with $V_{e,p\nu}(t)=\dot{\Phi}_{e,p\nu}(t)$.
    No independent time-dependent external fluxes thread loops
    inside the bare box. The bare box and its port charge are
    denoted by $\tilde{\msZ}_P/\tilde{\msY}_P$ and $\widetilde Q_p$.
    (b) Equivalent reduced circuit with nonlinear potential $U(\Phi_p,t)$,
    Norton current source $\dot Q_d$, and passive environment.
    The explicit branch capacitances are incorporated into the
    complete impedance $\msZ_P$ or admittance $\msY_P$, shown by
    the dashed enclosures.
    }
    \label{fig:DrivenPorts}
\end{figure}

We extend the nonlinear-port reduction to branches with parallel capacitances and prescribed time-dependent external fluxes. As shown in \cref{fig:DrivenPorts}, the latter are represented by ideal series voltage sources
$V_{e,p\nu}(t)=\dot\Phi_{e,p\nu}(t)$, with any static biases included in $\Phi_{e,p\nu}$. Each pair $h_{p\nu}\parallel C_{p\nu}$ is treated as a lumped
two-element model with no external flux threading its local loop. The quantities $\Phi_{e,p\nu}(t)$ are the branch flux offsets introduced in \cref{AppSubsec:ParallelNonlinearInductors}, now
allowed to depend on time. Their signs follow
\cref{Eq:DrivenOuterKirchhoff}. We assume the source allocation of \cref{fig:DrivenPorts}, with no additional drives acting inside the bare environment. The drive responses must be determined separately, for example, by electromagnetic simulation or experimental calibration.
They are not specified by the passive matrices $\widetilde{\msZ}_P$ and $\tilde{\msY}_P$ alone.
Static flux offsets must also be specified, and applied fluxes exclude self-flux already accounted for by the circuit inductances.

We first reduce the external branches independently of the linear environment and then incorporate their passive capacitances into the corresponding impedance or admittance.

\subsection{Reduction of the external branches}
\label{AppSubsec:DrivenExternalBranches}

An ideal voltage source with branch variables $(Q_v,\Phi_v)$
contributes~\cite{ParraRodriguez:2024}
\begin{align}
    \omega_{2B}^{V}
    &=
    \frac{1}{2}
    \mathrm dQ_v
    \wedge
    \mathrm d\Phi_v,
    \qquad
    H_d^{V}(t)
    =
    Q_vV_v(t).
    \label{Eq:AppVoltageSourceRule}
\end{align}

At port $p$, branch $\nu$ contains an inductive element with energy
$h_{p\nu}(\Phi_{p\nu})$, a parallel capacitance $C_{p\nu}$, and an
ideal series voltage source (see \cref{fig:DrivenPorts}). We define
$Q_{p\nu}^{B}=Q_{p\nu}^{L}+Q_{p\nu}^{C}$.

Collecting the contributions of these external elements before imposing the source fluxes gives
\begin{align}
\begin{aligned}
    \omega_1
    ={}&
    \frac{1}{2}\sum_{p,\nu}
    \Bigl[
        \mathrm d\Phi_{p\nu}\wedge\mathrm dQ_{p\nu}^{L}
        +
        \mathrm dQ_{p\nu}^{C}\wedge\mathrm d\Phi_{p\nu}
        \\
        &\qquad
        +
        \mathrm dQ_{p\nu}^{V}\wedge\mathrm d\Phi_{e,p\nu}
    \Bigr],
    \\
    H_1
    ={}&
    \sum_{p,\nu}
    \left[
        h_{p\nu}(\Phi_{p\nu})
        +
        \frac{(Q_{p\nu}^{C})^2}{2C_{p\nu}}
        +
        Q_{p\nu}^{V}V_{e,p\nu}(t)
    \right].
\end{aligned}
\end{align}

We first impose the Kirchhoff constraints, retaining the source
fluxes as branch variables until the prescribed voltage relations
are imposed.
With the orientations of \cref{fig:DrivenPorts}, these constraints read
\begin{align}
\begin{aligned}
    \Phi_{p\nu}
    &=
    \Phi_p+\Phi_{e,p\nu},
    \\
    Q_{p\nu}^{V}
    &=
    -Q_{p\nu}^{B},
    \qquad
    \tilde{Q}_p
    =
    -\sum_\nu Q_{p\nu}^{B}.
\end{aligned}
\label{Eq:DrivenOuterKirchhoff}
\end{align}
Here $\tilde{Q}_p$ is the integrated current entering the bare
linear environment. Additive charge constants are set to zero
in this classical reduction.

Introducing
$Q_{\Sigma,p}^{C}\equiv\sum_\nu Q_{p\nu}^{C}$ and
$\bQ_\Sigma^C=(Q_{\Sigma,1}^C,\ldots,Q_{\Sigma,N}^C)^T$,
we obtain
\begin{align}
\begin{aligned}
    \omega_1
    ={}&
    \left[
        \frac{1}{2}\mathrm d\tilde{\bQ}_P
        +
        \mathrm d\bQ_\Sigma^C
    \right]^T
    \wedge
    \mathrm d\bPhi_P
    -
    \sum_{p,\nu}
    \mathrm dQ_{p\nu}^{L}
    \wedge
    \mathrm d\Phi_{e,p\nu},
    \\
    H_1
    ={}&
    U(\bPhi_P,t)
    +\sum_{p,\nu}
    \left[
        \frac{(Q_{p\nu}^{C})^2}{2C_{p\nu}}
        -
        Q_{p\nu}^{B}V_{e,p\nu}(t)
    \right],
\end{aligned}
\end{align}
where the nonlinear potential is defined as
\begin{align}
    U(\bPhi_P,t)
    =
    \sum_{p,\nu}
    h_{p\nu}
    \left[
        \Phi_p+\Phi_{e,p\nu}(t)
    \right].
    \label{Eq:DrivenPortPotential}
\end{align}
Up to a total time derivative, the corresponding first-order Lagrangian is
\begin{align}
\begin{aligned}
    L_1
    ={}&
    \left(
        \frac{1}{2}\tilde{\bQ}_P+\bQ_\Sigma^C
    \right)^T
    \dot{\bPhi}_P
    -
    \sum_{p,\nu}Q_{p\nu}^{L}\dot\Phi_{e,p\nu}
    -
    H_1.
\end{aligned}
\end{align}

Using $\dot\Phi_{e,p\nu}=V_{e,p\nu}(t)$ and
$Q_{p\nu}^{B}-Q_{p\nu}^{L}=Q_{p\nu}^{C}$, the source terms in $L_1$
combine and we obtain
\begin{align}
\begin{aligned}
    L_1
    ={}&
    \left(
        \frac12\tilde{\bQ}_P+\bQ_\Sigma^C
    \right)^T
    \dot{\bPhi}_P
    -
    U(\bPhi_P,t)
    \\
    &-
    \sum_{p,\nu}
    \left[
        \frac{(Q_{p\nu}^{C})^2}{2C_{p\nu}}
        -
        Q_{p\nu}^{C}V_{e,p\nu}(t)
    \right].
\end{aligned}
\label{Eq:DrivenSourceFirstOrder}
\end{align}

Varying $Q_{p\nu}^{C}$ at fixed $\tilde{Q}_p$ gives the capacitor relation
$Q_{p\nu}^{C}=C_{p\nu}[\dot\Phi_p+V_{e,p\nu}(t)]$.
With $C_{\Sigma,p}\equiv\sum_\nu C_{p\nu}$ and
\begin{align}
    Q_{d,p}(t)
    \equiv
    \sum_\nu C_{p\nu}V_{e,p\nu}(t),
    \label{Eq:DrivenCharge}
\end{align}
we obtain
$Q_{\Sigma,p}^{C}-Q_{d,p}=C_{\Sigma,p}\dot\Phi_p$.
The individual capacitor charges are therefore
\begin{align}
    Q_{p\nu}^{C}
    =
    \frac{C_{p\nu}}{C_{\Sigma,p}}
    \left[
        Q_{\Sigma,p}^{C}-Q_{d,p}(t)
    \right]
    +
    C_{p\nu}V_{e,p\nu}(t).
    \label{Eq:CapacitorCharge}
\end{align}

This eliminates the relative capacitor charges while retaining
$Q_{\Sigma,p}^{C}$, on which the kinetic term already depends.
Substituting into the capacitor and source terms gives
\begin{align}
\begin{aligned}
    H_{C,p}(t)
    &=
    \sum_\nu
    \left[
        \frac{(Q_{p\nu}^{C})^2}{2C_{p\nu}}
        -
        Q_{p\nu}^{C}V_{e,p\nu}(t)
    \right]
    \\
    &=
    \frac{
        [Q_{\Sigma,p}^{C}-Q_{d,p}(t)]^2
    }{
        2C_{\Sigma,p}
    }
    -
    \frac12\sum_\nu C_{p\nu}V_{e,p\nu}^{2}(t).
\end{aligned}
\label{Eq:DrivenReducedCapacitor}
\end{align}
The last term depends only on time and can be omitted.

Writing
$\msC_\Sigma=\operatorname{diag}
(C_{\Sigma,1},\ldots,C_{\Sigma,N})$ and
$\bQ_d=(Q_{d,1},\ldots,Q_{d,N})^T$,
we obtain the reduced external Lagrangian where we eliminated the $Q_{p\nu}^C$ variables
\begin{align}
\begin{aligned}
    L_1
    ={}&
    \left(
        \frac{1}{2}\tilde{\bQ}_P
        +
        \bQ_\Sigma^C
    \right)^T
    \dot{\bPhi}_P
    -
    U(\bPhi_P,t)
    \\
    &-
    \frac{1}{2}
    \left(
        \bQ_\Sigma^C-\bQ_d
    \right)^T
    \msC_\Sigma^{-1}
    \left(
        \bQ_\Sigma^C-\bQ_d
    \right).
\end{aligned}
\label{Eq:DrivenL1}
\end{align}

The external KCL also identifies the collective nonlinear-port
charge,
\begin{align}
    \bQ_P
    \equiv
    \tilde{\bQ}_P+\bQ_\Sigma^C
    =
    -
    \begin{pmatrix}
        \sum_\nu Q_{1\nu}^{L}\\
        \vdots\\
        \sum_\nu Q_{N\nu}^{L}
    \end{pmatrix}.
    \label{Eq:DrivenPortCharge}
\end{align}

This reduction implements the Norton equivalence of the driven capacitive branches, as shown in \cref{fig:DrivenPorts}(b). The passive capacitances $\msC_\Sigma$ carry charge 
$\bQ_\Sigma^C-\bQ_d$, with parallel current sources $\dot{\bQ}_d$ oriented as indicated.
The nonlinear branches are represented by $U(\bPhi_P,t)$, and the total charge entering the passive environment is $\bQ_P-\bQ_d$.

The factor $1/2$ in \cref{Eq:DrivenL1} reflects that $L_1$
contains only the external branches.
As shown below, the bare-environment contribution
$L_2^{\msX}$, with $\msX=\msZ$ or $\msY$, supplies the boundary
term $\frac{1}{2}\tilde{\bQ}_P^T\dot{\bPhi}_P$.
These boundary contributions combine as
\begin{align}
    \left(
        \frac{1}{2}\tilde{\bQ}_P+\bQ_\Sigma^C
    \right)^T\dot{\bPhi}_P
    +
    \frac{1}{2}\tilde{\bQ}_P^T\dot{\bPhi}_P
    =
    \bQ_P^T\dot{\bPhi}_P.
\end{align}
We can therefore add \cref{Eq:DrivenL1} to either representation
of the bare environment and complete the passive-network reduction.

\subsection{Impedance representation}
\label{AppSubsec:DrivenZ}

Let $\tilde{\msZ}_P$ denote the bare impedance before including
$\msC_\Sigma$. If it admits the charge-gauge realization of
Appendix~\ref{AppSec:Z_construction_and_quantization}, its first-order
contribution after imposing the internal constraints can be written as
\begin{align}
\begin{aligned}
    L_2^{\msZ}
    ={}&
    \frac{1}{2}
    \tilde{\bQ}_P^T
    \dot{\bPhi}_P
    +
    \sum_k
    \widetilde P_k
    \dot{\widetilde X}_k
    \\
    &-
    H_{\tilde{\msZ}}^{\rm lin}
    \left(
        \tilde{\bQ}_P,
        \{\widetilde X_k,\widetilde P_k\}
    \right).
\end{aligned}
\label{Eq:DrivenL2Z}
\end{align}

Using $\bQ_\Sigma^C=\bQ_P-\tilde{\bQ}_P$,
\cref{Eq:DrivenL1,Eq:DrivenL2Z} give
\begin{align}
\begin{aligned}
    L_1+L_2^{\msZ}
    ={}&
    \bQ_P^T\dot{\bPhi}_P
    +
    \sum_k
    \widetilde P_k\dot{\widetilde X}_k
    -
    U(\bPhi_P,t)
    \\
    &-
    H_{\tilde{\msZ}}^{\rm lin}
    \left(
        \tilde{\bQ}_P,
        \{\widetilde X_k,\widetilde P_k\}
    \right)
    \\
    &-
    \frac{1}{2}
    \left[
        \bQ_P-\tilde{\bQ}_P-\bQ_d
    \right]^T
    \msC_\Sigma^{-1}
    \\
    &\quad\times
    \left[
        \bQ_P-\tilde{\bQ}_P-\bQ_d
    \right].
\end{aligned}
\label{Eq:DrivenCombinedZ}
\end{align}

The charge $\tilde{\bQ}_P$ does not enter the symplectic term
in \cref{Eq:DrivenCombinedZ} and is therefore an auxiliary
variable. To eliminate it, we use its algebraic Euler--Lagrange
equation,
\begin{align}
    \frac{\partial H_{\tilde{\msZ}}^{\rm lin}}
    {\partial \tilde{\bQ}_P}
    &=
    \msC_\Sigma^{-1}
    \left(
        \bQ_P-\tilde{\bQ}_P-\bQ_d
    \right).
    \label{Eq:EulerLagrange}
\end{align}
Varying with respect to $\bQ_P$ and using
\cref{Eq:DrivenPortCharge} gives
\begin{align}
    \bQ_P-\bQ_d
    =
    \tilde{\bQ}_P
    +
    \msC_\Sigma\dot{\bPhi}_P.
    \label{Eq:ChargeConstraint}
\end{align}
Combining \cref{Eq:EulerLagrange,Eq:ChargeConstraint} then yields
\begin{align}
    \frac{\partial H_{\tilde{\msZ}}^{\rm lin}}
    {\partial \tilde{\bQ}_P}
    =
    \dot{\bPhi}_P.
\end{align}
The left-hand side is the port voltage of the bare environment~\footnote{
Using Hamilton's equations for the
internal environment variables gives
$\mathrm{d}H_{\tilde{\msZ}}^{\rm lin}/\mathrm{d}t
=
(\partial H_{\tilde{\msZ}}^{\rm lin}/
\partial\tilde{\bQ}_P)^T\dot{\tilde{\bQ}}_P$.
Since $\dot{\tilde{\bQ}}_P$ is the current entering the bare
environment, comparison with the electrical input power
identifies $\partial H_{\tilde{\msZ}}^{\rm lin}/
\partial\tilde{\bQ}_P$ as its port voltage.}.
Hence, the bare environment and the added passive capacitances
$\msC_\Sigma$ share the same voltage $\dot{\bPhi}_P$.
Furthermore, \cref{Eq:ChargeConstraint} shows that the passive
port charge $\bQ_P-\bQ_d$ is the sum of the charge entering the
bare environment and the charge stored in the added passive
capacitances.

The resulting passive environment is therefore the parallel
combination of the capacitance network $\msC_\Sigma$ and the
bare environment with impedance $\tilde{\msZ}_P$, with total
port current $\dot{\bQ}_P-\dot{\bQ}_d$.
Its impedance is
\begin{align}
    \msZ_P(s)
    =
    \left[
        \tilde{\msZ}_P^{-1}(s)
        +
        s\msC_\Sigma
    \right]^{-1}.
    \label{Eq:DrivenFinalZ}
\end{align}

We now apply the construction of
\cref{AppSec:Z_construction_and_quantization} to the modified
impedance $\msZ_P(s)$, with passive port charge
$\bQ_P-\bQ_d(t)$. All capacitance matrices and mode parameters
below refer to this modified impedance. This gives
\begin{align}
\begin{aligned}
    H_P^{\msZ}
    ={}&
    U(\bPhi_P,t)
    +
    \frac{1}{2}
    \left(
        \bQ_P-\bQ_d
    \right)^T
    \\
    &\times
    \left(
        \msC_{\msZ,0}^{-1}
        +
        \msC_{\msZ,\infty}^{-1}
    \right)
    \left(
        \bQ_P-\bQ_d
    \right),
    \\
    H_B
    ={}&
    \frac{1}{2}
    \sum_k
    \left(
        P_k^2+G_k^2X_k^2
    \right),
    \\
    H_{PB}^{\msZ}
    ={}&
    \sum_k
    \left(
        \bQ_P-\bQ_d
    \right)^T
    \left(
        \msT_{k,a}^TP_k
        +
        G_k\msT_{k,b}^TX_k
    \right).
\end{aligned}
\label{Eq:DrivenFinalHZ}
\end{align}
Thus the final result applies whenever $\msZ_P$ satisfies the class and
convergence assumptions of Appendix~\ref{AppSec:Z_construction_and_quantization}.
All coefficients in \cref{Eq:DrivenFinalHZ} are obtained from this complete
response, so the branch capacitances are included only once.

\subsection{Admittance representation}
\label{AppSubsec:DrivenY}

Let $\tilde{\msY}_P$ denote instead the bare admittance. Before
completing the port KCL, its first-order contribution can be written as
\begin{align}
    L_2^{\msY}
    =
    \frac12
    \tilde{\bQ}_P^T
    \dot{\bPhi}_P
    +
    L_{\tilde{\msY},{\rm int}},
    \label{Eq:DrivenL2Y}
\end{align}
where $L_{\tilde{\msY},{\rm int}}$ contains the remaining
first-order terms and energies of its Cauer realization.
Combining this with \cref{Eq:DrivenL1} and using
$\bQ_\Sigma^C=\bQ_P-\tilde{\bQ}_P$ gives
\begin{align}
\begin{aligned}
    L_1+L_2^{\msY}
    ={}&
    \bQ_P^T\dot{\bPhi}_P
    -
    U(\bPhi_P,t)
    +
    L_{\tilde{\msY},{\rm int}}
    \\
    &-
    \frac12
    \left[
        \bQ_P-\tilde{\bQ}_P-\bQ_d
    \right]^T
    \msC_\Sigma^{-1}
    \\
    &\quad\times
    \left[
        \bQ_P-\tilde{\bQ}_P-\bQ_d
    \right].
\end{aligned}
\label{Eq:DrivenCombinedY}
\end{align}

The bare port KCL expresses $\tilde{\bQ}_P$ in terms of the
internal branch charges and can therefore again be used to eliminate it from \cref{Eq:DrivenCombinedY}.
Since $\bQ_\Sigma^C-\bQ_d=\msC_\Sigma\dot{\bPhi}_P$,
the added capacitances contribute $s\msC_\Sigma$ to the passive
admittance. The complete passive network has injected charge
$\bQ_P-\bQ_d$ and admittance
\begin{align}
    \msY_P(s)
    =
    \tilde{\msY}_P(s)
    +
    s\msC_\Sigma.
    \label{Eq:DrivenFinalY}
\end{align}
In particular,
$\msC_{\msY,\infty}
=\tilde{\msC}_{\msY,\infty}+\msC_\Sigma$, while the
zero-frequency inductive sector and all finite-frequency poles and
residues are unchanged.

Let $\bQ_A$ denote the integrated current entering the bare
admittance branches other than its shunt capacitance. The bare
port KCL gives
\begin{align}
    \tilde{\bQ}_P
    =
    \tilde{\msC}_{\msY,\infty}\dot{\bPhi}_P
    +\bQ_A.
\end{align}
Here $\tilde{\bQ}_C\equiv
\tilde{\msC}_{\msY,\infty}\dot{\bPhi}_P$ is the charge of the
bare shunt capacitance shown in \cref{fig:DrivenPorts}(b). Here the tilde indicates the bare environment. This charge corresponds to $\bQ_C$ in \cref{Eq:QCwidetildeDefinition}, evaluated for the bare admittance. The internal reduction of
Appendix~\ref{AppSec:Y_construction_and_quantization} supplies the kinetic contribution
$-\bQ_A^T\dot{\bPhi}_P$. Thus the port kinetic term in \cref{Eq:DrivenCombinedY}
becomes $(\bQ_P-\bQ_A)^T\dot{\bPhi}_P$, identifying the canonical capacitive charge as
\begin{align}
\begin{aligned}
    \bQ_C
    &\equiv \bQ_P-\bQ_A
    \\
    &=\tilde{\bQ}_C+\bQ_\Sigma^C=\msC_{\msY,\infty}\dot{\bPhi}_P+\bQ_d.
\end{aligned}
\end{align}

Thus the passive charging energy depends on $\bQ_C-\bQ_d(t)$,
while the canonical kinetic term is $\bQ_C^T\dot{\bPhi}_P$.
The bath modes and transformer matrices of the bare admittance
realization can be retained, giving
\begin{align}
\begin{aligned}
    L^{\msY}
    ={}&
    \bQ_C^T\dot{\bPhi}_P
    +
    \sum_kP_k\dot X_k
    -
    U(\bPhi_P,t)
    \\
    &-
    H_{\msY}^{\rm lin}
    \left(
        \bQ_C-\bQ_d(t),
        \bPhi_P,
        \{X_k,P_k\}
    \right),
\end{aligned}
\label{Eq:DrivenFinalLY}
\end{align}
with the two-form
$\omega^{\msY}
=\mathrm d\bQ_C^T\wedge\mathrm d\bPhi_P
+\sum_k\mathrm dP_k\wedge\mathrm dX_k$.

The Hamiltonian terms are therefore
\begin{align}
\begin{aligned}
    H_P^{\msY}
    ={}&
    U(\bPhi_P,t)
    +
    \frac12
    \left(
        \bQ_C-\bQ_d
    \right)^T
    \msC_{\msY,\infty}^{-1}
    \left(
        \bQ_C-\bQ_d
    \right)
    \\
    &+
    \frac12
    \bPhi_P^T
    \left(
        \msL_{\msY,0}^{-1}
        +
        \msL_{\msY,\infty}^{-1}
    \right)
    \bPhi_P,
    \\
    H_B
    ={}&
    \frac12
    \sum_k
    \left(
        P_k^2+\omega_k^2X_k^2
    \right),
    \\
    H_{PB}^{\msY}
    ={}&
    \sum_k
    \bPhi_P^T
    \left(
        \msT_{k,b}P_k
        +
        \omega_k\msT_{k,a}X_k
    \right).
\end{aligned}
\label{Eq:DrivenFinalHY}
\end{align}
Here $\omega_k=R_k$, as in
Appendix~\ref{AppSec:Y_construction_and_quantization}.
All static biases are included in the prescribed branch fluxes
$\Phi_{e,p\nu}(t)$ entering $U(\bPhi_P,t)$.
The final result applies whenever $\msY_P$ satisfies the class
and convergence assumptions of
Appendix~\ref{AppSec:Y_construction_and_quantization}.
All coefficients in \cref{Eq:DrivenFinalHY} are obtained from this
complete response, so the branch capacitances are included only once.

\subsection*{Summary}

For a given bare environment and set of nonlinear branch energies,
the extension requires only the prescribed branch fluxes
$\Phi_{e,p\nu}(t)$, the charge-drive vector $\bQ_d(t)$ of
\cref{Eq:DrivenCharge}, and the shunt-capacitance matrix $\msC_\Sigma$.
One first forms the complete passive responses
\begin{align}
\begin{aligned}
    \msZ_P(s)
    &=
    \left[\tilde{\msZ}_P^{-1}(s)+s\msC_\Sigma\right]^{-1},
    \\
    \msY_P(s)
    &=
    \tilde{\msY}_P(s)+s\msC_\Sigma,
\end{aligned}
\end{align}
and extracts the Hamiltonian coefficients from the chosen response.
The constructions of Appendices~\ref{AppSec:Z_construction_and_quantization} and~\ref{AppSec:Y_construction_and_quantization} then apply with the
flux-shifted potential $U(\bPhi_P,t)$ of
\cref{Eq:DrivenPortPotential} and the replacements
$\bQ_P\mapsto\bQ_P-\bQ_d(t)$ or
$\bQ_C\mapsto\bQ_C-\bQ_d(t)$, respectively, throughout the linear
Hamiltonian. This prescription assumes the source allocation stated
at the beginning of this appendix.

\subsection{One-port examples}
\label{AppSubsec:DrivenExamples}

We illustrate the construction with capacitive and inductive coupling
to a parallel LC resonator. In both cases, the nonlinear port is a SQUID formed by two branches $h_1\parallel C_{J1}$ and $h_2\parallel C_{J2}$, with the local Josephson energy $h_i(\Phi_i)=-E_{Ji}\cos(2\pi\Phi_i/\Phi_0)$. We consider a flux $\Phi_o(t)$ through the outer loop closed through branch~1, the coupling element, and $L_r$, and a SQUID-loop flux $\Phi_s(t)$. No flux threads the local junction-capacitor or resonator LC loops. The source fluxes are
$\Phi_{e,1}=\Phi_o$ and $\Phi_{e,2}=\Phi_o+\Phi_s$, giving
$V_{e,1}=V_o=\dot\Phi_o$ and
$V_{e,2}=V_o+V_s=\dot\Phi_o+\dot\Phi_s$.
Here $C_\Sigma=C_{J1}+C_{J2}$ includes only the SQUID branch capacitances.

Denoting the dimensional port flux by $\Phi$, the nonlinear potential is
\begin{align}
    U(\Phi,t)
    ={}&
    h_1\left[\Phi+\Phi_o(t)\right]+
    h_2\left[\Phi+\Phi_o(t)+\Phi_s(t)\right],
\end{align}
and $Q_d(t)=C_\Sigma V_o(t)+C_{J2}V_s(t)$.

\begin{figure}[t]
    \centering
    \includegraphics[width=\linewidth]{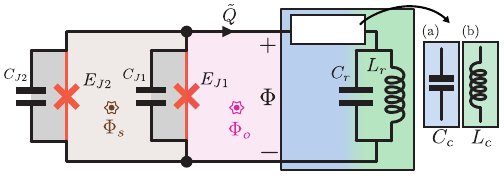}
    \caption{Driven one-port examples. A SQUID with branch capacitances $C_{J1}$ and $C_{J2}$ is coupled to the parallel resonator $L_r\parallel C_r$ through either (a) $C_c$ or (b) $L_c$. The loop fluxes correspond to the source allocation $V_{e,1}=\dot\Phi_o$ and $V_{e,2}=\dot\Phi_o+\dot\Phi_s$.
    }
    \label{fig:DrivenExamples}
\end{figure}

\subsubsection{Capacitive coupling}

For the circuit in \cref{fig:DrivenExamples}(a), the bare impedance,
before including the SQUID capacitances, is
\begin{align}
    \tilde{Z}_P(s)
    =
    \frac{1}{sC_c}
    +
    \frac{sL_r}{1+s^2L_rC_r}.
    \label{Eq:DrivenExampleBareZ}
\end{align}
The complete passive impedance is
$Z_P=[\tilde{Z}_P^{-1}+sC_\Sigma]^{-1}$.
Defining $D_Z=C_c+C_\Sigma+C_cC_\Sigma/C_r$, it can be written as
\begin{align}
    Z_P(s)
    =
    \frac{1}{s(C_c+C_\Sigma)}
    +
    g_Z^2\frac{s}{s^2+\Omega_Z^2},
    \label{Eq:DrivenExampleZ}
\end{align}
where
\begin{align}
    \Omega_Z^2
    =
    \frac{C_c+C_\Sigma}{L_rC_rD_Z},
    \qquad
    g_Z^2
    =
    \frac{C_c^2}{C_r(C_c+C_\Sigma)D_Z}.
    \label{Eq:DrivenExampleZParameters}
\end{align}

We define the charging capacitance $C_Z$ through
\begin{align}
    C_Z^{-1}
    \equiv
    C_{\msZ,0}^{-1}+C_{\msZ,\infty}^{-1}
    =
    \left(
        C_\Sigma+\frac{C_cC_r}{C_c+C_r}
    \right)^{-1}.
\end{align}
The canonical port pair is $(\Phi,Q)$, with the port
quantization specified in
Appendix~\ref{AppSec:Z_construction_and_quantization}.
Writing the internal mode in terms of
$[a_Z,a_Z^\dag]=1$ and omitting its zero-point energy gives
\begin{align}
\begin{aligned}
    H^{\msZ}
    ={}&
    U(\Phi,t)
    +
    \frac{[Q-Q_d(t)]^2}{2C_Z}
    +
    \hbar\Omega_Z a_Z^\dag a_Z
    \\
    &+
    i g_Z\sqrt{\frac{\hbar\Omega_Z}{2}}\,
    [Q-Q_d(t)]
    (a_Z^\dag-a_Z).
\end{aligned}
\label{Eq:DrivenExampleHZ}
\end{align}
Here the oscillator phase convention is that of Appendix~\ref{AppSec:Z_construction_and_quantization}, so its momentum quadrature is proportional to
$i(a_Z^\dag-a_Z)$.

\subsubsection{Inductive coupling}

For the circuit in \cref{fig:DrivenExamples}(b), the corresponding bare
admittance is
\begin{align}
    \tilde{Y}_P(s)
    =
    \left[
        sL_c
        +
        \frac{sL_r}{1+s^2L_rC_r}
    \right]^{-1}.
    \label{Eq:DrivenExampleBareY}
\end{align}
It can be decomposed as
\begin{align}
    \tilde{Y}_P(s)
    =
    \frac{1}{s(L_c+L_r)}
    +
    g_Y^2\frac{s}{s^2+\Omega_Y^2},
    \label{Eq:DrivenExampleBareYDecomp}
\end{align}
with
\begin{align}
    \Omega_Y^2
    =
    \frac{L_c+L_r}{L_cL_rC_r},
    \qquad
    g_Y^2
    =
    \frac{L_r}{L_c(L_c+L_r)}.
    \label{Eq:DrivenExampleYParameters}
\end{align}
Adding the SQUID capacitances only modifies the pole at infinity,
\begin{align}
    Y_P(s)
    =
    \tilde{Y}_P(s)
    +
    sC_\Sigma.
    \label{Eq:DrivenExampleFinalY}
\end{align}

The canonical port pair has commutator
$[\Phi,Q_C]=i\hbar$, whereas the internal mode satisfies
$[a_Y,a_Y^\dag]=1$. Using $1/(L_c+L_r)+g_Y^2=1/L_c$ and
$\Omega_Yg_Y=1/(L_c\sqrt{C_r})$, the exact Hamiltonian is
\begin{align}
\begin{aligned}
    H^{\msY}
    ={}&
    U(\Phi,t)
    +
    \frac{[Q_C-Q_d(t)]^2}{2C_\Sigma}
    +
    \frac{\Phi^2}{2L_c}
    \\
    &+
    \hbar\Omega_Y a_Y^\dag a_Y
    -
    g_Y\sqrt{\frac{\hbar\Omega_Y}{2}}\,
    \Phi(a_Y+a_Y^\dag),
\end{aligned}
\label{Eq:DrivenExampleHY}
\end{align}
where the oscillator phase is chosen to give the negative
interaction shown, and its zero-point energy has been omitted.

A direct node-flux construction and Legendre transformation, followed by canonical normalization and quantization of the internal oscillator, reproduce \cref{Eq:DrivenExampleHZ,Eq:DrivenExampleHY} up to functions of time only. We have kept the specified voltage-source allocation throughout. A global irrotational transformation~\cite{You:2019} would depend on the complete linear realization and can mix the port and internal coordinates. Once an exact Hamiltonian has been obtained, unitary transformations can be used to put it in a more convenient form.

\section{Details for the dispersive models: Schrieffer--Wolff transformation}
\label{Sec:AppendixDiscreteSW}
In this appendix, we derive the effective dispersive Hamiltonian for discrete environments. To keep the discussion general for both the $\msZ$ and $\msY$ immittances, we use the generic notation $\msX\in\{\msZ,\msY\}$ and introduce the generic port coupling operator $O_p^\msX$, which, for the two specific cases, reads
\begin{align}
    O_{p}^{\msZ}=\hat Q_{Cp}-Q_{g,p},
    \qquad
    O_{p}^{\msY}=\hat\Phi_p.
\end{align}
We further diagonalize the local port Hamiltonians $h_p$, while assuming weak direct coupling between external ports (see \cref{Sec:MainDirectCoupling}).

\subsection{Generator and second-order Hamiltonian}
Omitting the explicit $\msX$ label of the Hamiltonian, we start from the partition

\begin{align}
    H
    =
    H_0
    +
    H_{PB}
    +
    H_{\mathrm{int}},
\label{Eq:AppendixSWStartingHamiltonian}
\end{align}
with
\begin{align}
    H_0=\sum_p h_p+H_B,
    \qquad
    H_{\mathrm{int}}
    =
    H_{\mathrm{int}}^{(0)}
    +
    H_{\mathrm{int}}^{\mathrm{CT}}.
\end{align}
We assume the perturbative ordering $H_{PB}=O(\lambda)$ and, for the port-local organization used here,
$H_{\mathrm{int}}=O(\lambda^2)$.

Expanding the physical port operators in the eigenbasis of $h_p$, the port-bath interaction reads
\begin{align}
\begin{aligned}
    H_{PB}
    ={}&
    \sum_{\alpha,\epsilon,p}
    \sum_{i,j}
    O_{p,ij}^{\msX}
    \left[
        (\msG_\alpha)_{p,\epsilon}
        a_{\alpha,\epsilon}
        +
        (\msG_\alpha)_{p,\epsilon}^*
        a_{\alpha,\epsilon}^\dagger
    \right]
    \sigma_{ji}^{(p)}.
\end{aligned}
\label{Eq:AppendixSWInteractionExpanded}
\end{align}
The elementary eigenoperator relations are
\begin{align}
\begin{aligned}
    [H_0,
    a_{\alpha,\epsilon}
    \sigma_{ji}^{(p)}]
    &=
    \hbar
    \left(
        \omega_{ij}^{(p)}
        -
        \omega_\alpha
    \right)
    a_{\alpha,\epsilon}
    \sigma_{ji}^{(p)},
    \\
    [H_0,
    a_{\alpha,\epsilon}^\dagger
    \sigma_{ji}^{(p)}]
    &=
    \hbar
    \left(
        \omega_{ij}^{(p)}
        +
        \omega_\alpha
    \right)
    a_{\alpha,\epsilon}^\dagger
    \sigma_{ji}^{(p)}.
\end{aligned}
\end{align}

The generator $S$ that eliminates the port-bath coupling to first order must satisfy the usual condition $[S,H_0]=-H_{PB}$ and is given by
\begin{align}
\begin{aligned}
    S
    ={}&
    \sum_{\alpha,\epsilon,p}
    \sum_{i,j}
    \frac{
        (\msG_\alpha)_{p,\epsilon}
        O_{p,ij}^{\msX}
    }{
        \hbar
        \left(
            \omega_{ij}^{(p)}
            -
            \omega_\alpha
        \right)
    }
    a_{\alpha,\epsilon}
    \sigma_{ji}^{(p)}
    \\
    &+
    \sum_{\alpha,\epsilon,p}
    \sum_{i,j}
    \frac{
        (\msG_\alpha)_{p,\epsilon}^*
        O_{p,ij}^{\msX}
    }{
        \hbar
        \left(
            \omega_{ij}^{(p)}
            +
            \omega_\alpha
        \right)
    }
    a_{\alpha,\epsilon}^\dagger
    \sigma_{ji}^{(p)}.
\end{aligned}
\label{Eq:AppendixSWGenerator}
\end{align}
Hermiticity of $O_p^{\msX}$ and
$\omega_{ji}^{(p)}=-\omega_{ij}^{(p)}$ guarantee $S^\dagger=-S$.

Applying the transformation to the full Hamiltonian $\tilde{H}=e^SHe^{-S}$, and retaining terms up to second order in $\lambda$, we obtain 
\begin{align}
    \tilde{H} = H_0 + H_{\mathrm{int}}  + \frac{1}{2}[S,H_{PB}] +
    O(\lambda^3).
\label{Eq:AppendixSWSecondOrderHamiltonian}
\end{align}
Note that due to the weak port-port coupling assumption $H_{\mathrm{int}}=O(\lambda^2)$, the commutator of $S$ with the interaction Hamiltonian only yields corrections at order $O(\lambda^3)$.

\subsection{Effective dispersive model in the vacuum subspace}

We now project onto the vacuum of all internal modes using the projector $P_0$.
Using
\begin{align}
\begin{aligned}
    P_0[
        a_{\alpha,\epsilon}A,
        a_{\beta,\epsilon'}^\dagger B
    ]P_0
    &=
    \delta_{\alpha\beta}
    \delta_{\epsilon\epsilon'}
    AB,
    \\
    P_0[
        a_{\alpha,\epsilon}^\dagger A,
        a_{\beta,\epsilon'}B
    ]P_0
    &=
    -
    \delta_{\alpha\beta}
    \delta_{\epsilon\epsilon'}
    BA,
\end{aligned}
\end{align}
we obtain
\begin{align}
    &\frac{1}{2}
    P_0[S,H_{PB}]P_0
    =
    \frac{1}{2\hbar}
    \sum_{\alpha,\epsilon}
    \sum_{p,p'}
    \sum_{i,j,k,l}
    O_{p,ij}^{\msX}
    O_{p',kl}^{\msX}\label{Eq:AppendixSWProjectedCommutator}
    \\
    &\times
    \left[
        \frac{
            (\msG_\alpha)_{p,\epsilon}
            (\msG_\alpha)_{p',\epsilon}^*
        }{
            \omega_{ij}^{(p)}
            -
            \omega_\alpha
        }
        \sigma_{ji}^{(p)}
        \sigma_{lk}^{(p')}
        -
        \frac{
            (\msG_\alpha)_{p,\epsilon}^*
            (\msG_\alpha)_{p',\epsilon}
        }{
            \omega_{ij}^{(p)}
            +
            \omega_\alpha
        }
        \sigma_{lk}^{(p')}
        \sigma_{ji}^{(p)}
    \right].\nonumber
\end{align}

The corresponding second-order correction can be related to the residue of $\msX$ through the relation

\begin{align}
    \sum_\epsilon
    (\msG_\alpha)_{p,\epsilon}
    (\msG_\alpha)_{p',\epsilon}^*
    =
    \hbar\omega_\alpha
    (\msR_\alpha)_{p,p'}.
\label{Eq:AppendixSWResidueIdentity}
\end{align}

We separate \cref{Eq:AppendixSWProjectedCommutator} into its local
($p=p'$) and cross-port ($p\neq p'$) contributions,
\begin{align}
    \frac{1}{2}P_0[S,H_{PB}]P_0
    =
    H_{\rm LS}^{(2)}
    +
    \tilde{H}_{\rm int}^{(2)}.
\label{Eq:AppendixSWLocalCrossSplit}
\end{align}
The effective Hamiltonian in the environmental vacuum subspace can therefore be written as
\begin{align}
    H_{\rm SW}
    =
    \sum_p h_p
    +
    H_{\mathrm{int}}^{(0)}
    +
    H_{\rm LS}^{(2)}
    +
    H_{\rm int}^{(2)},
\label{Eq:AppendixSWEffectiveHamiltonian}
\end{align}
where we define the total second-order cross-port interaction as the sum of the SW-generated term and the high-frequency counterterm

\begin{align}
    H_{\rm int}^{(2)}
    \equiv
    \tilde{H}_{\rm int}^{(2)}
    +
    H_{\mathrm{int}}^{\mathrm{CT}}.
\label{Eq:AppendixSWTotalInteraction}
\end{align}

For the local terms, using
\begin{align}
    \sigma_{ji}^{(p)}
    \sigma_{lk}^{(p)}
    =
    \delta_{j,k}
    \ket{i}_p\bra{l}_p,
\end{align}
and combining the two operator orderings gives
\begin{align}
\begin{aligned}
    H_{\rm LS}^{(2)}
    ={}&
    -\frac{1}{2}
    \sum_p
    \sum_{i,l,j}
    O_{p,ij}^{\msX}
    O_{p,jl}^{\msX}
    \\
    &\times
    \left[
        \left(
            \msI_{\msX}^{\rm d}
            \left[
                \omega_{ji}^{(p)}
            \right]
        \right)_{p,p}
        +
        \left(
            \msI_{\msX}^{\rm d}
            \left[
                \omega_{jl}^{(p)}
            \right]
        \right)_{p,p}
    \right]
    \ket{i}_p\bra{l}_p,
\end{aligned}
\label{Eq:AppendixSWLocalOperator}
\end{align}
where
\begin{align}
    \left(
        \msI_{\msX}^{\rm d}[\omega]
    \right)_{p,p}
    \equiv
    \sum_\alpha
    \frac{
        \omega_\alpha
        \left(
            \msR_\alpha
        \right)_{p,p}
    }{
        \omega_\alpha+\omega
    }=\sum_{\alpha}\left(\msR_\alpha
        \right)_{p,p}\left[1-\frac{\omega}{\omega_\alpha+\omega}\right].
% =    
%     \frac{\left(\msM_{\msX,\infty}\right)_{p,p}}{2}-\sum_{\alpha}\frac{
%         \omega
%         \left(
%             \msR_\alpha
%         \right)_{p,p}
%     }{
%         \omega_\alpha+\omega
%     }.
\label{Eq:DiscreteIKernel}
\end{align}
Using the previous definition $\sum_{\alpha}(\msR_\alpha)_{pp}=(\msM_{\msX,\infty})_{pp}/2$, we obtain the decomposition \cref{Eq:DiscreteStaticLambShiftCoefficientMain,Eq:DiscreteLambShiftCoefficientMain} in the main text.

Equation~\eqref{Eq:AppendixSWLocalOperator} retains the complete local
second-order correction, including the matrix elements with $i\neq l$;
no local secular approximation has been performed.
We next consider $p\neq p'$. Since operators belonging to different ports
commute, collecting the two orderings in
\cref{Eq:AppendixSWProjectedCommutator} gives
\begin{align}
    \tilde{H}_{\rm int}^{(2)}/\hbar
    =
    \frac{1}{2}
    \sum_{p'>p}
    \sum_{i,j,k,l}
    \left(
        \tilde{g}_{ij,kl}^{p,p'}
        \sigma_{ji}^{(p)}
        \sigma_{lk}^{(p')}
        +
        \hc
    \right),
\label{Eq:DiscreteInteractionSW}
\end{align}
with
\begin{align}
    \tilde{g}_{ij,kl}^{p,p'}
    ={}&
    \frac{
        O_{p,ij}^{\msX}
        O_{p',kl}^{\msX}
    }{2\hbar}
    \sum_\alpha
    \omega_\alpha\label{Eq:AppendixSWMediatedResidueCoefficient}
    \\
    &\times
    \Bigg\{
        (\msR_\alpha)_{p,p'}
        \left(
            \frac{1}{
                \omega_{ij}^{(p)}
                -
                \omega_\alpha
            }
            -
            \frac{1}{
                \omega_{kl}^{(p')}
                +
                \omega_\alpha
            }
        \right)\nonumber
        \\
        &\hspace{7mm}
        +
        (\msR_\alpha)_{p',p}
        \left(
            \frac{1}{
                \omega_{kl}^{(p')}
                -
                \omega_\alpha
            }
            -
            \frac{1}{
                \omega_{ij}^{(p)}
                +
                \omega_\alpha
            }
        \right)
    \Bigg\}.\nonumber
\end{align}

For frequencies away from the discrete poles,
\begin{align}
    \msXA[\omega]
    =
    i\sum_\alpha
    \left[
        \frac{
            \msR_\alpha
        }{
            \omega-\omega_\alpha
        }
        +
        \frac{
            \msR_\alpha^T
        }{
            \omega+\omega_\alpha
        }
    \right],
\label{Eq:AppendixSWXADiscrete}
\end{align}
and
\begin{align}
    \msM_{\msX,\infty}
    =
    \sum_\alpha
    \left(
        \msR_\alpha+\msR_\alpha^T
    \right).
\label{Eq:AppendixSWAinftyResidueSum}
\end{align}
It follows that
\begin{align}
\begin{aligned}
    \sum_\alpha
    \omega_\alpha
    \left[
        \frac{
            \msR_\alpha
        }{
            \omega-\omega_\alpha
        }
        -
        \frac{
            \msR_\alpha^T
        }{
            \omega+\omega_\alpha
        }
    \right]
    =
    -i\omega\msXA[\omega]
    -
    \msM_{\msX,\infty}.
\end{aligned}
\label{Eq:AppendixSWResponseIdentity}
\end{align}
Using the symmetry of $\msM_{\msX,\infty}$,
\cref{Eq:AppendixSWMediatedResidueCoefficient} becomes
\begin{align}
\label{Eq:AppendixSWMediatedImmittanceCoefficient}
    \tilde{g}_{ij,kl}^{p,p'}
    =&
    \frac{
        O_{p,ij}^{\msX}
        O_{p',kl}^{\msX}
    }{2\hbar}
    \Bigg\{
        -i\omega_{ij}^{(p)}
        \left(
            \msXA[
                \omega_{ij}^{(p)}
            ]
        \right)_{p,p'}
        \\
        &
        -
        i\omega_{kl}^{(p')}
        \left(
            \msXA[
                \omega_{kl}^{(p')}
            ]
        \right)_{p',p}
        -
        2
        \left(
            \msM_{\msX,\infty}
        \right)_{p,p'}
    \Bigg\}.\nonumber
\end{align}
The last term combines with
$H_{\mathrm{int}}^{\mathrm{CT}}$ in \cref{Eq:AppendixSWTotalInteraction},
yielding \cref{Eq:DiscreteGeneralCouplingMain}. Thus, the second-order vacuum-projected Hamiltonian is given by \cref{Eq:AppendixSWEffectiveHamiltonian}, with the complete local correction in \cref{Eq:AppendixSWLocalOperator} and the cross-port interaction in \cref{Eq:AppendixSWTotalInteraction}. No secular or rotating-wave approximation has been made in obtaining these expressions.

\subsection{Validity of the dispersive SW transformation}

The perturbative SW transformation requires
\begin{align}
    \left|
        \frac{
            (\msG_\alpha)_{p,\epsilon}
            O_{p,ij}^{\msX}
        }{
            \hbar
            \left(
                \omega_{ij}^{(p)}
                \mp
                \omega_\alpha
            \right)
        }
    \right|
    \ll1
\label{Eq:AppendixSWValidity}
\end{align}
for all retained system transitions and eliminated environmental modes. For countably infinite discrete environments, the corresponding sums defining the generator and the second-order corrections must additionally converge.

We have further assumed that the residual cross-port terms in $H_{\mathrm{int}}=H_{\mathrm{int}}^{(0)}+
H_{\mathrm{int}}^{\mathrm{CT}}$ enter at second order in the chosen bookkeeping. If some of them are strong,
they must instead be included in the unperturbed collective system Hamiltonian, which is diagonalized before constructing the SW generator.

Since the complete local matrix in \cref{Eq:AppendixSWLocalOperator} is
retained, no additional local secular approximation is required. Exact or near
degeneracies among retained system levels can be treated by retaining and
diagonalizing the corresponding blocks of the effective Hamiltonian. The
dispersive condition \cref{Eq:AppendixSWValidity} requires that the retained system transitions remain sufficiently off resonance with the environmental modes that are eliminated.

\section{Adiabatic elimination of continuous modes}
\label{Sec:AdiabaticElimination}

After finding the canonical circuit representations of the impedance and
admittance matrices, we proceed to the elimination of the internal modes.
In direct analogy with \cref{Sec:AppendixDiscreteSW}, we write
\begin{align}
    H
    =
    H_0
    +
    H_{PB}
    +
    H_{\mathrm{int}},
\end{align}
with
\begin{align}
    H_0
    =
    h_P+H_B,
    \qquad
    h_P
    \equiv
    \sum_p h_p.
\end{align}
The diagonal port dressings are included in the local Hamiltonians $h_p$,
while the residual terms
$H_{\mathrm{int}}=H_{\mathrm{int}}^{(0)}+
H_{\mathrm{int}}^{\mathrm{CT}}$
are retained to second order in the weak-coupling bookkeeping. If some
direct couplings are strong, they must instead be included in a collective
system Hamiltonian before the interaction-picture decomposition. As in the
main text, we use the eigen-decomposition in
\cref{Eq:DiagonalizedPortHamiltonian}.

\subsection{Born--Markov equation before secularization}

In the interaction picture generated by $H_0$, the port-bath coupling is
\begin{align}
\begin{aligned}
    H_{PB}(t)
    ={}&
    \sum_{p,i,j,\epsilon}
    \int_0^\infty\!\mathrm d\omega\,
    O_{p,ij}^{\msX}
    e^{i\omega_{ij}^{(p)}t}
    \sigma_{ji}^{(p)}
    \\
    &\times
    \left[
        (\msG(\omega))_{p,\epsilon}
        a_\epsilon(\omega)e^{-i\omega t}
        +\hc
    \right].
\end{aligned}
\label{Eq:AEInteractionPicture}
\end{align}

Under the standard second-order Born--Markov approximation at zero
temperature~\cite{BreuerPetruccione:2002},
\begin{align}
\begin{aligned}
    \frac{\mathrm d}{\mathrm dt}\rho_P(t)
    ={}&
    -\frac{1}{\hbar^2}
    \int_0^\infty\!\mathrm d\tau\,
    \operatorname{Tr}_B
    \left[
        H_{PB}(t),\right.\\
        &\left.
        \left[
            H_{PB}(t-\tau),
            \rho_P(t)\otimes\rho_B
        \right]
    \right].
\end{aligned}
\label{Eq:BornMarkovME}
\end{align}
The one-sided coefficients appearing after evaluating the bath correlator
are
\begin{align}
\begin{aligned}
    \Gamma_{ij,kl}^{p,p'}
    ={}&
    \frac{
        O_{p,ij}^{\msX}
        O_{p',kl}^{\msX}
    }{\hbar^2}
    \int_0^\infty\!\mathrm d\omega\,
    \left(
        \msG(\omega)\msG^\dagger(\omega)
    \right)_{p,p'}
    \\
    &\times
    \int_0^\infty\!\mathrm d\tau\,
    e^{-i(\omega+\omega_{kl}^{(p')})\tau}.
\end{aligned}
\label{Eq:AECoefs}
\end{align}
Using the Sokhotski--Plemelj formula $\int_0^\infty\!\mathrm d\tau\,
    e^{-ix\tau} =\pi\delta(x) - i\mathcal P\frac{1}{x}$ and $\msG(\omega)\msG^\dagger(\omega)=\hbar\omega\msXH[\omega]/\pi$ from \cref{Eq:ZCouplingSpectralIdentities,Eq:YCouplingSpectralIdentities}, we define
\begin{align}
    \msKX[\omega]
    &\equiv
    |\omega|\msXH[\omega],
    \\
    \msIX[\omega]
    &\equiv
    \frac{1}{\pi}
    \mathcal P\!\int_0^\infty
    \frac{
        \msKX[\omega']
    }{
        \omega'+\omega
    }
    \,\mathrm d\omega',
\end{align}
and obtain
\begin{align}
\begin{aligned}
    \Gamma_{ij,kl}^{p,p'}
    =
    \frac{
        O_{p,ij}^{\msX}
        O_{p',kl}^{\msX}
    }{\hbar}
    \Big[
        &\Theta(-\omega_{kl}^{(p')})
        \left(
            \msKX[-\omega_{kl}^{(p')}]
        \right)_{p,p'}
        \\
        &-
        i
        \left(
            \msIX[\omega_{kl}^{(p')}]
        \right)_{p,p'}
    \Big],
\end{aligned}
\label{Eq:AEGammaCompact}
\end{align}
where $\Theta(x)$ denotes the Heaviside step function. For later convenience, we separate the hermitian and anti-hermitian
combinations
\begin{align}
\begin{aligned}
    \gamma_{ij,kl}^{p,p'}
    &\equiv
    \Gamma_{ij,kl}^{p,p'}
    +
    \left(
        \Gamma_{lk,ji}^{p',p}
    \right)^*,
    \\
    J_{ij,kl}^{p,p'}
    &\equiv
    \frac{1}{2i}
    \left[
        \Gamma_{ij,kl}^{p,p'}
        -
        \left(
            \Gamma_{lk,ji}^{p',p}
        \right)^*
    \right].
\end{aligned}
\label{Eq:EffCoefsME}
\end{align}
Before a secular or coarse-graining step, the resulting Born--Markov generator is a Redfield-type equation and is not, in general, of GKSL form; the matrix built from the coefficients
$\gamma_{ij,kl}^{p,p'}$ need not be positive
semidefinite~\cite{Schaller:2008,Trushechkin:2021}.

In terms of the generalized dissipator
\begin{align}
    \mathcal D[A,B]\rho
    \equiv
    A\rho B
    -
    \frac{1}{2}
    \left\{
        BA,\rho
    \right\},
\end{align}
the interaction-picture equation may be written as
\begin{align}
\begin{aligned}
    \mathcal L_R^{(2)}\rho_P(t)
    ={}&
    -i
    \sum_{p,p'}
    \sum_{i,j,k,l}
    J_{ij,kl}^{p,p'}
    \left[
        \sigma_{ji}^{(p)}(t)
        \sigma_{lk}^{(p')}(t),
        \rho_P(t)
    \right]
    \\
    &+
    \sum_{p,p'}
    \sum_{i,j,k,l}
    \gamma_{ij,kl}^{p,p'}
    \mathcal D\!\left[
        \sigma_{lk}^{(p')}(t),
        \sigma_{ji}^{(p)}(t)
    \right]
    \rho_P(t).
\end{aligned}
\label{Eq:RedfieldCompact}
\end{align}

Returning to the Schr\"odinger picture and restoring the residual direct
terms gives
\begin{align}
\begin{aligned}
    \mathcal L\rho_P
    ={}&
    -\frac{i}{\hbar}
    \left[
        h_P
        +H_{\mathrm{int}}^{(0)}
        +H_{\mathrm{int}}^{\mathrm{CT}},
        \rho_P
    \right]
    \\
    &-
    \frac{i}{\hbar}
    \left[
        H_{\rm LS}^{(2)}
        +\tilde{H}_{\rm int}^{(2)},
        \rho_P
    \right]
    +
    \mathcal L_P^{(2)}\rho_P
    +
    \mathcal L_{\rm int}^{(2)}\rho_P.
\end{aligned}
\label{Eq:FullME}
\end{align}
Here,
\begin{align}
\begin{aligned}
    H_{\rm LS}^{(2)}/\hbar
    &=
    \sum_p\sum_{i,l}
    \left(
        \sum_j J_{ij,jl}^{p,p}
    \right)
    \ket{i}_p\bra{l}_p,
    \\
    \tilde{H}_{\rm int}^{(2)}/\hbar
    &=
    \frac{1}{2}
    \sum_{p\neq p'}
    \sum_{i,j,k,l}
    \left(
        J_{ij,kl}^{p,p'}
        \sigma_{ji}^{(p)}
        \sigma_{lk}^{(p')}
        +\hc
    \right),
\end{aligned}
\end{align}
while $\mathcal L_P^{(2)}$ and
$\mathcal L_{\rm int}^{(2)}$ denote, respectively, the $p=p'$ and
$p\neq p'$ parts of the second-order dissipative contribution in
\cref{Eq:RedfieldCompact}.

Using the Kramers--Kronig relations and the associated high-frequency integral rule, one can extract the following useful decomposition for the diagonal integral
\begin{align}
\begin{aligned}
    \left(
        \msIX[\omega]
    \right)_{p,p}
    ={}&
    \frac{1}{2}
    \left(
        \msM_{\msX,\infty}
    \right)_{p,p}
    \\
    &-
    \frac{\omega}{\pi}
    \mathcal P\!\int_0^\infty
    \frac{
        \left(
            \msXH[\omega']
        \right)_{p,p}
    }{
        \omega'+\omega
    }
    \,\mathrm d\omega',
\end{aligned}
\label{Eq:IntegralDecomposition}
\end{align}
whereas the off-diagonal symmetrized integral satisfies
\begin{align}
\begin{aligned}
    \left(
        \msIX[-\omega]
        +
        \msIX[\omega]^T
    \right)_{p,p'}
    =&
    \left(
        \msM_{\msX,\infty}
    \right)_{p,p'}+
    i\omega
    \left(
        \msXA[\omega]
    \right)_{p,p'}.
\end{aligned}
\label{Eq:AppPVCoherentCouplingResult}
\end{align}

\subsection{Secular and partial-secular transition blocks}

The secular approximation acts on \emph{Bohr frequencies}, not on
individual transitions. Hence all transitions with the same Bohr frequency
must remain in the same block. For each port, define the exact
positive-frequency block
\begin{align}
    \mathcal T_p(\Omega)
    \equiv
    \left\{
        (i,j):
        i>j,\;
        \omega_{ij}^{(p)}=\Omega
    \right\},
    \qquad
    \Omega>0,
\end{align}
and the corresponding lowering component of the physical port operator,
\begin{align}
    O_p^{\msX}(\Omega)
    \equiv
    \sum_{(i,j)\in\mathcal T_p(\Omega)}
    O_{p,ji}^{\msX}
    \sigma_{ij}^{(p)}.
\label{Eq:TransitionBlockOperator}
\end{align}

The zero-temperature second-order dissipator obtained after exact
secularization is
\begin{align}
\begin{aligned}
    \mathcal L_D^{(2)}\rho_P
    ={}&
    \frac{2}{\hbar}
    \sum_{\Omega>0}
    \sum_{p,p'}
    \left(
        \msKX[\Omega]
    \right)_{p,p'}
    \mathcal D\!\left[
        O_{p'}^{\msX}(\Omega),
        {O_p^{\msX}(\Omega)}^\dagger
    \right]\rho_P.
\end{aligned}
\label{Eq:SecularDissipatorBlock}
\end{align}
Because
$\msKX[\Omega]=\Omega\msXH[\Omega]\succeq0$ for a passive environment, every exactly degenerate frequency block is positive semidefinite and \cref{Eq:SecularDissipatorBlock} is of GKSL form.

Expanding \cref{Eq:SecularDissipatorBlock} gives, for
$(i,j)\in\mathcal T_p(\Omega)$ and
$(k,l)\in\mathcal T_{p'}(\Omega)$,
\begin{align}
    \gamma_{ij,lk}^{p,p'}
    =
    \frac{
        2
        O_{p,ij}^{\msX}
        O_{p',lk}^{\msX}
    }{\hbar}
    \left(
        \msKX[\Omega]
    \right)_{p,p'}.
\label{Eq:SecularTransitionBlockRate}
\end{align}
For $p=p'$ and a one-dimensional transition block this reduces to
\begin{align}
    \gamma_{ij}^{p}
    =
    \frac{
        2|O_{p,ij}^{\msX}|^2
    }{\hbar}
    \left(
        \msKX[
            \omega_{ij}^{(p)}
        ]
    \right)_{p,p},
\end{align}
which is \cref{Eq:LocalDecay}.
Equation~\eqref{Eq:SecularTransitionBlockRate} also contains the
correlated decay coefficients between different ports.

If the transition frequencies are not exactly but only nearly degenerate, a partial-secular approximation needs to be performed. To do so, one clusters together all nearly degenerate transitions and uses a representative frequency $\bar\Omega$ within each cluster, i.e.,
\begin{align}
    \msKX[
        \omega_{ij}^{(p)}
    ]
    \longrightarrow
    \msKX[\bar\Omega],
\end{align}
with the analogous replacement in the principal-value coefficients.
This is justified when the bath response is smooth across the cluster and
preserves the positive-semidefinite block
structure~\cite{Schaller:2008,Cattaneo:2019,Trushechkin:2021}. Exact degeneracy is the special case in which no such replacement is
required.

For the coherent local contribution, secularization retains matrix elements
within the same exact or unresolved energy block. 
\begin{align}
    H_{\rm LS}^{(2)}/\hbar
    =
    \sum_p
    \sum_{i,l:\,\omega_i^{(p)}\simeq\omega_l^{(p)}}
    \left(
        \sum_j
        J_{ij,jl}^{p,p}
    \right)
    \ket{i}_p\bra{l}_p.
\label{Eq:SecularLocalLambBlock}
\end{align}
For nondegenerate local energy eigenspaces this reduces to
\begin{align}
    H_{\rm LS}^{(2)}
    =
    \sum_p\sum_i
    \hbar\overline\Lambda_i^{(p)}
    \ket{i}_p\bra{i}_p,
\end{align}
with
\begin{align}
    \overline\Lambda_i^{(p)}
    =
    -
    \sum_j
    \frac{
        |O_{p,ij}^{\msX}|^2
    }{\hbar}
    \left(
        \msIX[
            \omega_{ji}^{(p)}
        ]
    \right)_{p,p}.
\end{align}
Using \cref{Eq:IntegralDecomposition} and completeness yields the two
contributions in \cref{Eq:LambShift,Eq:LambRenormalization}.
If a local energy eigenspace is degenerate, the block in
\cref{Eq:SecularLocalLambBlock} must instead be diagonalized within that
eigenspace.

For the coherent interaction between distinct ports, let
$\mathcal R_{ij}^{p,p'}$ denote the transitions belonging to the frequency
cluster opposite to $(i,j,p)$. The second-order interaction generated by the bath elimination is
\begin{align}
\begin{aligned}
    \tilde{H}_{\rm int}^{(2)}/\hbar
    ={}&
    \sum_{p'>p}
    \sum_{i>j}
    \sum_{(k,l)\in\mathcal R_{ij}^{p,p'}}
    \left(
        \tilde{g}_{ij,kl}^{p,p'}
        \sigma_{ji}^{(p)}
        \sigma_{lk}^{(p')}
        +\hc
    \right),
\end{aligned}
\label{Eq:AECoherentCoupling}
\end{align}
where
\begin{align}
\begin{aligned}
    \tilde{g}_{ij,kl}^{p,p'}
    ={}&
    -
    \frac{
        O_{p,ij}^{\msX}
        O_{p',kl}^{\msX}
    }{\hbar}
    \\
    &\times
    \left[
        \left(
            \msIX[
                \omega_{kl}^{(p')}
            ]
        \right)_{p,p'}
        +
        \left(
            \msIX[
                \omega_{ij}^{(p)}
            ]
        \right)_{p',p}
    \right].
\end{aligned}
\end{align}

For an exactly resonant block,
$\omega_{kl}^{(p')}=-\omega_{ij}^{(p)}$, or after replacing a
near-degenerate cluster by its common representative $\bar\Omega$, the
Kramers--Kronig identity \cref{Eq:AppPVCoherentCouplingResult} shows that
the $\msM_{\msX,\infty}$ contribution from the principal-value integral is
canceled by the high-frequency counterterm. We therefore group the two
contributions together and define
\begin{align}
    H_{\rm int}^{(2)}
    \equiv
    \tilde{H}_{\rm int}^{(2)}
    +
    H_{\mathrm{int}}^{\mathrm{CT}}.
\label{Eq:AppendixTotalInteraction}
\end{align}
Here, $\tilde{H}_{\rm int}^{(2)}$ is generated by the second-order Born--Markov elimination, whereas
$H_{\mathrm{int}}^{\mathrm{CT}}$ is already present in the exact
Hamiltonian. Their combination gives the finite immittance expression in
\cref{Eq:CoherentCoupling}.

Collecting the secular or partial-secular contributions, the reduced
generator can finally be written as
\begin{align}
\begin{aligned}
    \mathcal L\rho_P
    ={}&
    -\frac{i}{\hbar}
    \left[
        h_P
        +
        H_{\mathrm{int}}^{(0)},
        \rho_P
    \right]
    \\
    &-
    \frac{i}{\hbar}
    \left[
        H_{\rm LS}^{(2)}
        +
        H_{\rm int}^{(2)},
        \rho_P
    \right]
    +
    \mathcal L_P^{(2)}\rho_P
    +
    \mathcal L_{\rm int}^{(2)}\rho_P.
\end{aligned}
\label{Eq:EffectiveMEAppendix}
\end{align}
Here,
$\mathcal L_P^{(2)}$
contains the $p=p'$ terms of the secular or partial-secular dissipator,
including all transitions belonging to a common local Bohr-frequency
block, whereas
$\mathcal L_{\rm int}^{(2)}$
contains the corresponding $p\neq p'$ correlated decay terms.

\section{Details on Examples}
\label{AppSec:DetailsExamples}
\subsection{Transmission-line impedance matrices for TEM propagation}
\label{AppSec:TransmissionLineImpedances}

\begin{figure}[t]
    \centering
    \includegraphics[width=\linewidth]{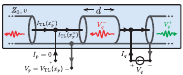}
    \caption{Voltage and current conventions for an infinite lossless TEM transmission line. Injecting a current $I_q$ at port $q$ with $I_p=0$ gives the transfer impedance $Z_{pq}=V_p/I_q$. The excitation generates outgoing left- and right-propagating waves $V_q^-$ and $V_q^+$.}
    \label{fig:App_TL_impedance_conventions}
\end{figure}

For completeness, we first collect the TL impedance matrices used throughout the examples for ideal TEM propagation, see further details in standard microwave engineering literature~\cite{Pozar:2009}. We consider lossless TLs with characteristic impedance $Z_0$ and propagation velocity $v$, with the voltage and current conventions illustrated in \cref{fig:App_TL_impedance_conventions}.

For later use, we first recall the general Laplace-domain traveling-wave solution. Taking the distributed transmission-line current
$I_{\rm TL}(x,s)$ to be positive toward increasing $x$, the voltage and
current are
\begin{align}
\begin{aligned}
    V_{\rm TL}(x,s)
    &=
    V^+(s)e^{-sx/v}
    +
    V^-(s)e^{sx/v},
    \\
    I_{\rm TL}(x,s)
    &=
    \frac{1}{Z_0}
    \left[
        V^+(s)e^{-sx/v}
        -
        V^-(s)e^{sx/v}
    \right],
\end{aligned}
    \label{Eq:TLTravelingWaveSolution}
\end{align}
where $V^+$ and $V^-$ denote the corresponding right- and left-propagating wave amplitudes within each source-free segment. Thus, a pure right- (left-)propagating wave satisfies $V_{\rm TL}=Z_0I_{\rm TL}$
($V_{\rm TL}=-Z_0I_{\rm TL}$).

\subsubsection{Semi-infinite TL}

For a semi-infinite transmission line extending toward increasing $x$, a current $I(s)$ injected at its end excites only the right-propagating solution in \cref{Eq:TLTravelingWaveSolution}. Hence $V(s)=Z_0I(s)$, and the impedance seen from the injection point is simply
\begin{align}
    Z_{\rm TL}^{\rm semi}(s)=Z_0.
    \label{Eq:SemiInfiniteTLImpedance}
\end{align}

\subsubsection{(Reciprocal) infinite TL}

We next consider an infinite transmission line with $N$ injection points at
ordered positions
\begin{align}
    x_1<x_2<\cdots<x_N.
\end{align}
To obtain column $q$ of the impedance matrix, we inject a current $I_q(s)$
at $x_q$ while setting all other port currents to zero, as illustrated in
\cref{fig:App_TL_impedance_conventions}. The excitation launches outgoing
waves in both directions. Since the two semi-infinite branches have equal
impedance $Z_0$, the injected current splits equally between them. With
$I_{\rm TL}$ defined positive toward increasing $x$,
\begin{align}
    I_{\rm TL}(x_q^+,s)
    =
    \frac{I_q(s)}{2},
    \qquad
    I_{\rm TL}(x_q^-,s)
    =
    -\frac{I_q(s)}{2}.
\end{align}
Using \cref{Eq:TLTravelingWaveSolution}, both branches therefore produce the
same local voltage,
\begin{align}
    V_q(s)=\frac{Z_0}{2}I_q(s).
\end{align}
The outgoing wave amplitudes in
\cref{fig:App_TL_impedance_conventions} therefore satisfy
$V_q^+=V_q^-=V_q$. The corresponding voltage waves propagate in both directions, acquiring the
factor $e^{-s|x_p-x_q|/v}$ between ports $q$ and $p$. By superposition,
\begin{align}
    V_p(s)
    =
    \frac{Z_0}{2}
    \sum_{q=1}^{N}
    e^{-s|x_p-x_q|/v}
    I_q(s),
\end{align}
and hence
\begin{align}
    \left[\msZTL^{\rm R}(s)\right]_{p,q}
    = \frac{Z_0}{2} e^{-s|x_p-x_q|/v}.
    \label{Eq:InfiniteTLNPort}
\end{align}

For equally spaced injection points, $x_{p+1}-x_p=d$, defining the propagation factor
\begin{align}
    r(s)\equiv e^{-sd/v},
    \label{Eq:PropagationFactorTL}
\end{align}
we obtain
\begin{align}
    \msZTL^{\rm R}(s)
    =
    \frac{Z_0}{2}
    \begin{pmatrix}
        1&r&\cdots&r^{N-1}\\
        r&1&\cdots&r^{N-2}\\
        \vdots&\vdots&\ddots&\vdots\\
        r^{N-1}&r^{N-2}&\cdots&1
    \end{pmatrix}.
    \label{Eq:InfiniteTLNPortEquallySpaced}
\end{align}
On the boundary $s=-i\omega+0^+$, the propagation factor becomes
$r=e^{i\omega d/v}$.

\subsubsection{Infinite TL with circulators between injection points}

We next consider $N$ equally spaced injection points, $x_{p+1}-x_p=d$, with an ideal matched three-port circulator placed halfway between every pair of neighboring injection points. The $N-1$ circulators are oriented in the same direction, such that propagation from smaller to larger port indices is transmitted, whereas propagation in the opposite direction is redirected into their matched third ports.

A current $I_q$ injected at port $q$ locally sees two matched branches of impedance $Z_0$ in parallel, and therefore produces the voltage
\begin{align}
    V_q(s)=\frac{Z_0}{2}I_q(s).
\end{align}
The wave propagating toward increasing $x$ is transmitted through all
subsequent circulators and acquires only the usual propagation phase.
Consequently, at a downstream injection point $p>q$,
\begin{align}
    V_p(s)
    =
    \frac{Z_0}{2}
    e^{-s(x_p-x_q)/v}
    I_q(s).
\end{align}
The wave propagating toward decreasing $x$ is redirected into the matched
third port of the first circulator it encounters, so that it cannot reach
any port $p<q$. By superposition, the $N$-port impedance matrix is therefore
\begin{align}
    \left[
        \msZTL^{\rm IC}(s)
    \right]_{p,q}
    =
    \begin{cases}
        0,
        & p<q,
        \\[1mm]
        \dfrac{Z_0}{2}
        e^{-s(x_p-x_q)/v},
        & p\geq q.
    \end{cases}
    \label{Eq:InfiniteTLNPortIntermediateCirculators}
\end{align}
For equally spaced injection points, using
$r(s)=e^{-sd/v}$, this becomes
\begin{align}
    \msZTL^{\rm IC}(s)
    =
    \frac{Z_0}{2}
    \begin{pmatrix}
        1&0&\cdots&0\\
        r&1&\cdots&0\\
        \vdots&\vdots&\ddots&\vdots\\
        r^{N-1}&r^{N-2}&\cdots&1
    \end{pmatrix}.
    \label{Eq:InfiniteTLNPortIntermediateCirculatorsEquallySpaced}
\end{align}
Reversing the orientation of all circulators gives the transpose of this
matrix. The two-port configuration considered in
\cref{fig:Examples_waveguide_mediated_qudits}(c) follows directly by
setting $N=2$.

\subsubsection{Infinite TL with circulators at the injection points}

We finally place an ideal matched three-port circulator at each injection
point, oriented such that signals propagate from smaller to larger port
indices. Consider first a current injected only at port $q$. Since the circulator directs the injected signal into a single matched semi-infinite
channel, the local voltage is
\begin{align}
    V_q=Z_0I_q.
\end{align}
The corresponding outgoing voltage wave propagates only downstream, so that the wave arriving at a port $p>q$ has amplitude
\begin{align}
    V_{p}^{\rm in}
    =
    Z_0
    e^{-s(x_p-x_q)/v}
    I_q.
\end{align}
When computing the transfer impedance, no current is injected at port $p$.
The condition $I_p=0$ implies equal incoming and outgoing voltage-wave
amplitudes at the accessible port, and hence
\begin{align}
    V_p
    =
    2V_p^{\rm in}
    =
    2Z_0
    e^{-s(x_p-x_q)/v}
    I_q,
    \qquad
    p>q.
\end{align}
No signal propagates upstream. By superposition, the $N$-port impedance is
therefore
\begin{align}
    \left[\msZTL^{\rightarrow}(s)\right]_{p,q}
    =
    \begin{cases}
        0,
        & p<q,
        \\[1mm]
        Z_0,
        & p=q,
        \\[1mm]
        2Z_0e^{-s(x_p-x_q)/v},
        & p>q.
    \end{cases}
    \label{Eq:InfiniteTLNPortCirculators}
\end{align}
For equally spaced injection points,
\begin{align}
    \msZTL^{\rightarrow}(s)
    =
    Z_0
    \begin{pmatrix}
        1&0&\cdots&0\\
        2r&1&\cdots&0\\
        \vdots&\vdots&\ddots&\vdots\\
        2r^{N-1}&2r^{N-2}&\cdots&1
    \end{pmatrix}.
    \label{Eq:InfiniteTLNPortCirculatorsEquallySpaced}
\end{align}
Reversing the orientation of all circulators simply reverses the propagation
direction,
\begin{align}
    \msZTL^{\leftarrow}(s)
    =
    \left[
        \msZTL^{\rightarrow}(s)
    \right]^T.
\end{align}

\subsection{Charge qudit capacitively coupled to a (semi-)infinite TL}
\label{Sec:ImpedanceSemiInf}

The semi-infinite and infinite transmission lines considered in \cref{fig:1CQ_Cc_TLs}(a,c) are locally equivalent to frequency-independent resistances
\begin{align}
    R=
    \begin{cases}
        Z_0, & \text{semi-infinite \gls{TL}},\\[1mm]
        Z_0/2, & \text{infinite \gls{TL}},
    \end{cases}
\end{align}
as follows directly from
\cref{Eq:SemiInfiniteTLImpedance,Eq:InfiniteTLNPort}. We therefore treat
both cases simultaneously.

The total impedance seen from the Josephson junction is obtained by
combining $R$ in series with the coupling capacitor and the resulting
branch in parallel with the junction capacitance,
\begin{align}
\begin{aligned}
    Z_P(s)
    &=
    \left[
        sC_J
        +
        \left(1/(s C_c)+R
        \right)^{-1}
    \right]^{-1}
    \\
    &=
    \frac{1}{sC_\Sigma}
    +
    \frac{C_c^2}{C_\Sigma^2}
    \frac{R}{1+s/\wcut},
\end{aligned}
\end{align}
where $C_\Sigma=C_J+C_c$, and $\wcut
    = \left( RC_cC_J/C_\Sigma\right)^{-1}$.

The residue of the zero-frequency pole is
$C_{Z,0}^{-1}=C_\Sigma^{-1}$, while the remaining impedance has the
high-frequency contribution $C_{Z,\infty}^{-1}  = C_c/(C_JC_\Sigma)$, which is independent of $R$. The external Hamiltonian
\cref{Eq:ExternalHamiltonianZ} thus simply reads
\begin{align}
    H_P^\msZ
    =
    4E_C(\hat n-n_g)^2
    -
    E_J\cos(\hat\varphi),
\end{align}
where the capacitance felt by the charge qudit is the bare junction
capacitance, $E_C=e^2/(2C_J)$.

The semi-infinite and infinite cases are obtained by setting
$R=Z_0$ and $R=Z_0/2$, respectively. In particular, the cutoff frequency
of the infinite-line case is twice that of the semi-infinite-line case.

\subsection{Transmon coupled to a low-$Q$ mode}
\label{App:LowQEnvironment}
The circuit shown in \cref{fig:Low-Q-resonator}(a) consists of a Josephson junction capacitively coupled to an LC resonator, itself capacitively coupled to a semi-infinite transmission line. The \gls{TL} contributes the boundary impedance $Z_0$, which is filtered by the resonator before reaching the qubit.

\subsubsection{Impedance}

We model the readout resonator as a parallel $LC$ circuit with capacitance
$C_r$ and inductance $L_r$, coupled to the semi-infinite transmission line
through $C_\kappa$. The total impedance seen from the Josephson junction is
\begin{align}
    Z_P(s)
    =
    \frac{1}{sC_\Sigma}
    +
    \frac{C_c^2}{C_\Sigma^2}
    \frac{Z_r(s)}{1+sC_\ast Z_r(s)},
\end{align}
where
\begin{align}
    Z_r(s) =\left[      sC_r
        +
        \frac{1}{sL_r}
        +
        \frac{sC_\kappa}{1+sC_\kappa Z_0}
    \right]^{-1},
\end{align}
$C_\Sigma=C_J+C_c$ and $C_\ast=C_JC_c/C_\Sigma$. The high-frequency contribution is $C_{Z,\infty}=(C_r+C_*)C_\Sigma^2/C_c^2$, yielding the total effective capacitance 
\begin{align}
    C_{\mathrm{eff}}\equiv(C_{Z,0}^{-1}+C_{Z,\infty}^{-1})^{-1}=C_J+\frac{C_cC_r}{C_c+C_r}.
\end{align}

\subsubsection{Poles of the linearized circuit}

To obtain a nonperturbative linear benchmark, we replace the Josephson
junction by its linearized inductance $L_J
    = \hbar^2/[(2e)^2E_J]$. The circuit then consists of two dynamical nodes, the junction node with flux $\Phi_J$ and the resonator node with flux $\Phi_r$. The nodal admittance matrix of the complete linearized circuit is
\begin{align}
\begin{aligned}
    &\msY_{\mathrm{full}}(s)
    \\
    &=
    \begin{pmatrix}
        \frac{1}{sL_J}
        +sC_\Sigma
        &
        -sC_c
        \\
        -sC_c
        &
        \frac{1}{sL_r}
        +sC_{\Sigma,R}
        +Y_{\mathrm{out}}(s)
    \end{pmatrix},
    \end{aligned}
    \label{Eq:FullLinearizedAdmittance}
\end{align}
with $C_{\Sigma,R}=C_r+C_c$, and $Y_{\rm out}(s)=sC_\kappa/(1+sC_\kappa Z_0)$.

The solutions $s_q$ of
$\det\msY_{\mathrm{full}}(s_q)=0$ are the complex poles of the linearized
open circuit~\cite{Nigg:2012,Hassler:2019}. For a stable mode,
$s_q=-\gamma_q^{\mathrm{pole}}/2-i\omega_q$, so that
$-\Im\{s_q\}$ gives its oscillation frequency and
\begin{align}
    \gamma_q^{\mathrm{pole}}
    =
    -2\Re(s_q)
\end{align}
its energy-decay rate.

\subsection{Linearized Josephson-junction-array metamaterial}
\label{AppSec:JJAMetamaterial}

\subsubsection{Chain impedance, spectrum, and port coupling}

For the linearized array in Fig.~\ref{fig:Metamaterials}, let
$Y(s)=sC+1/(sL)$ be the admittance of each intersite branch and define
$Z_b(s)=Y^{-1}(s)$ and $Y_g(s)=sC_g$. The self-similarity of the
semi-infinite chain gives the fixed-point relation
\begin{align}
Z_{\rm ch}(s)
=
\left[
Y_g+\frac{1}{Z_b+Z_{\rm ch}(s)}
\right]^{-1}.
\end{align}
Solving this quadratic equation gives
\begin{align}
Z_{\rm ch}(s)
=
\frac{Z_b}{2}
\left[
\sqrt{1+\frac{4}{Y_gZ_b}}-1
\right],
\end{align}
which yields \cref{Eq:JJAChainImpedance} upon using
$Z_b=Y^{-1}$ and $Y_g=sC_g$. The square-root branch is fixed by
analyticity and passivity for $\Re \{s\}>0$.

The same unit cell, comprising, from the input side, a shunt admittance $Y_g$
followed by a series impedance $Z_b$, has the transfer matrix
\begin{align}
\mathsf W(s)
=
\begin{pmatrix}
1 & Z_b\\
Y_g & 1+Y_gZ_b
\end{pmatrix}.
\label{Eq:JJATransferMatrix}
\end{align}
Since $\det\{\mathsf W\}=1$, propagating solutions of the infinite periodic
array have eigenvalues $\lambda_\pm=e^{\pm ik}$. Hence,
$2\cos(k)=\Tr\{\mathsf W\}=2+Y_gZ_b$, which on the passive
boundary $s=-i\omega+0^+$ gives
\begin{align}
\omega_k
=
\frac{2\sin(k/2)}
{\sqrt{L[C_g+4C\sin^2(k/2)]}}.
\label{Eq:JJADispersion}
\end{align}
For $0<k<\pi$, this defines the continuous spectral domain
$\mathcal B_Z=(0,\omega_{\rm UV})$, with
$\omega_{\rm UV}=2/\sqrt{L(C_g+4C)}$.

Using $C_\Sigma=C_d+C_{gd}$,
$C_*=C_dC_{gd}/C_\Sigma$, and
$\eta=C_{gd}/C_\Sigma$, subtraction of the zero-frequency pole in
\cref{Eq:JJARegularZ} gives
\begin{align}
Z(s)
=
\eta^2
\frac{Z_{\rm ch}(s)}
{1+sC_*Z_{\rm ch}(s)}.
\label{Eq:JJARegularZApp}
\end{align}
At low frequency,
$Z_{\rm ch}(s)\rightarrow\sqrt{L/C_g}$, and therefore
$C_{Z,0}^{-1}=C_\Sigma^{-1}$. At high frequency,
$Z_{\rm ch}(s)\sim1/(sC_b)$, with
$C_b=[C_g+\sqrt{C_g^2+4CC_g}]/2$, giving
$C_{Z,\infty}^{-1}=\eta^2/(C_b+C_*)$. Combining the two contributions yields $C_{\rm eff}\equiv(C_{Z,0}^{-1}+C_{Z,\infty}^{-1})^{-1}
=C_d+C_{gd}C_b/(C_{gd}+C_b)$.

Within the propagation band, the passive boundary value of the chain
impedance can be written as
\begin{align}
Z_{\rm ch}[\omega_k]
=
\frac{2\sin(k/2)}{\omega_k C_g}e^{ik/2}.
\end{align}
It follows from \cref{Eq:JJARegularZApp} that
\begin{align}
\Re\{Z[\omega_k]\}
=
\frac{\eta^2\sin(k)}
{\omega_k C_g
\left[
1+4r(1+r)\sin^2(k/2)
\right]},
\end{align}
where $r=C_*/C_g$. Eliminating $k$ with
\cref{Eq:JJADispersion} and using
$G^2(\omega)=\hbar\omega\Re \{Z[\omega]\}/\pi$ gives the continuum coupling
in \cref{Eq:JJAG}. The resulting frequency dependence is consistent with the continuum description of Josephson-junction arrays in Ref.~\cite{Snyman:2015}.

\subsection{Two braided giant atoms capacitively coupled to an infinite TL}
\label{App:BraidedGiantAtomsImpedance}

\begin{figure}[t]
    \centering
    \includegraphics[width=\linewidth]{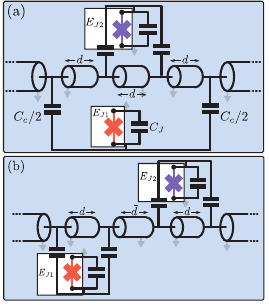}
    \caption{Two superconducting giant atoms coupled at four points of the same infinite TL in (a) nested and (b) separate configurations.}
    \label{fig:Examples_TLs_Giant_atoms_Appendix}
\end{figure}

The braided configuration shown in
\cref{fig:Examples_TLs_Giant_atoms}(a) consists of two Josephson elements capacitively coupled to four points of the same reciprocal infinite transmission line. The ordering of the coupling points is
\begin{align}
    x_1:a,\qquad
    x_2:b,\qquad
    x_3:a,\qquad
    x_4:b,
\end{align}
with equal spacing
\begin{align}
    x_{\mu+1}-x_\mu=d,
    \qquad
    \mu=1,2,3.
\end{align}
We denote by $C_c$ the total coupling capacitance of one giant atom, so each
physical coupling leg has capacitance $C_c/2$. The four-port transmission-line
impedance is the $N=4$ specialization of
$\msZTL^{\rm R}(s)$ in \cref{Eq:InfiniteTLNPortEquallySpaced}.

To map the two injection points of each giant atom onto the corresponding
junction node, we define the incidence matrix
\begin{align}
    \msP_{\rm Braided}
    = \begin{pmatrix}
        1&0\\
        0&1\\
        1&0\\
        0&1
    \end{pmatrix},
    \label{Eq:BraidedIncidenceMatrix}
\end{align}
and write $\msP\equiv\msP_{\rm Braided}$ in the following (and in the main
text). The junction-voltage vector is
\begin{align}
    \mbf V_J
    =
    \begin{pmatrix}
        V_a\\
        V_b
    \end{pmatrix}.
\end{align}
The vector $\msP\mbf V_J$ gives the voltages at the junction-side terminals of the four coupling capacitors.

Note that different configurations of the injection points (see \cref{fig:Examples_TLs_Giant_atoms_Appendix}) can be obtained by considering different incidence matrices. For the nested and separate configurations, these read
\begin{align}
    \msP_{\rm Nested}
    &=
    \begin{pmatrix}
        1&0\\
        0&1\\
        0&1\\
        1&0
    \end{pmatrix},
    \quad
    \msP_{\rm Separate}
    =
    \begin{pmatrix}
        1&0\\
        1&0\\
        0&1\\
        0&1
    \end{pmatrix}.
\end{align}

\subsubsection{Exact admittance}
The exact environmental admittance seen from the two junction nodes is
\begin{align}
    \msYnode(s)
    =
    \msP^T
    \left[
        \left(\frac{2}{sC_c}\right)\msId_4
        +
        \msZTL^{\rm R}(s)
    \right]^{-1}
    \msP.
    \label{Eq:YenvBraidedExact}
\end{align}
Including the junction capacitances in parallel gives the full impedance
seen by the two Josephson branches,
\begin{align}
    \msZ_P(s)
    =
    \left[
        sC_J\msId_2
        +
        \msYnode(s)
    \right]^{-1}.
    \label{Eq:ZwidetildeBraidedExact}
\end{align}
For unequal junction or coupling capacitances, $C_J\msId_2$ and $2/(C_c) \msId_4$ should be replaced by the appropriate diagonal capacitance matrices.

The total impedance can be computed analytically, and one can extract the
pole at zero to obtain
\begin{align}
    \msZ_P(s)
    =
    \frac{1}{sC_\Sigma}\msId_2
    +
    \msZ(s),
    \qquad
    C_\Sigma=C_J+C_c.
    \label{Eq:BraidedZeroPole}
\end{align}

\subsubsection{Perturbative impedance in
the weak-capacitive-loading regime}

To obtain the weak-coupling limit, we expand the exact impedance in the
coupling capacitance. Starting from \cref{Eq:YenvBraidedExact} and assuming
the weak-capacitive-loading condition \cref{Eq:ValidityPerturbativeImpedance}
with $\msZTL=\msZ^{\mathrm R}_{\mathrm{TL}}$, we use perturbative matrix inversion $\left(
\msId+\msX\right)^{-1}  =  \msId-\msX+\msX^2-\dots $. The admittance in \cref{Eq:YenvBraidedExact} can then be perturbatively
computed as
\begin{align}
    \msYnode(s)
    &=
    sC_c\msId_2
    -
    \frac{s^2C_c^2}{4}
    \msP^T\msZTL^{\rm R}(s)\msP
    +
    O(C_c^3),
    \label{Eq:ApproxImpedanceEnv}
\end{align}
where we used $\msP^T\msP=2\msId_2$. Applying perturbative matrix inversion for a second time, we obtain
\begin{align}
    \msZ_P(s)
    \approx
    \frac{1}{sC_\Sigma}\msId_2
    + \left(
        \frac{C_c}{2C_\Sigma}
    \right)^2
    \msP^T\msZTL^{\rm R}(s)\msP.
\end{align}
For the braided ordering considered in the main text, this leads to
\begin{align}
    \msZ(s)
    \simeq
    \frac{Z_0}{4}
    \left(
        \frac{C_c}{C_\Sigma}
    \right)^2
    \begin{pmatrix}
        1+r^2
        &
        \dfrac{3r+r^3}{2}
        \\
        \dfrac{3r+r^3}{2}
        &
        1+r^2
    \end{pmatrix}.
    \label{Eq:ZregBraidedWeakMatrix}
\end{align}

\subsubsection{Perturbative rates and coherent couplings}

For simplicity, we assume that the two giant atoms are identical ($E_{J,i}=E_J$) and restrict the dynamics to their qubit subspaces with transition frequency $\omega_{10}$. Defining the delay time $\tau=d/v$, the associated phase $\theta=\omega_{10}\tau$, and the decay-rate scale associated with a single physical connection point,
\begin{align}
    \gamma_0
    =
    \frac{|Q_{10}|^2}{\hbar}
    \omega_{10} Z_0
    \left(
        \frac{C_c}{2C_\Sigma}
    \right)^2,
    \label{Eq:SinglePointRateBraided}
\end{align}
the diagonal elements of \cref{Eq:ZregBraidedWeakMatrix} give
\begin{align}
    \gamma_a=\gamma_b
    =
    2\gamma_0
    \left(
        1+\cos(2\theta)
    \right).
    \label{Eq:BraidedLocalDecay}
\end{align}
On the other hand, the off-diagonal element gives the collective decay rate
\begin{align}
    \gamma_{12}
    =
    \gamma_0
    \left(
        3\cos(\theta)+\cos(3\theta)
    \right).
    \label{Eq:BraidedCollectiveDecay}
\end{align}
Within the same weak-capacitive-loading and Markov approximations, the coherent exchange coupling is given by
\begin{align}
    g
    =
    \frac{\gamma_0}{2}
    \left(
        3\sin(\theta)+\sin(3\theta)
    \right).
    \label{Eq:BraidedExchange}
\end{align}
These expressions reproduce the standard braided-giant-atom decoherence-free interaction condition~\cite{Kockum:2018,Soro:2022}. At $\theta=\pi/2$, the local and collective decay rates vanish, whereas the exchange interaction remains finite, i.e., $g=\gamma_0$. These approximate results are naturally accompanied by divergent local Lamb shifts, which have been typically absorbed into fitted transition frequencies.

\subsection{Two giant atoms coupled to an infinite TL via circulators at injection points}
\label{App:BraidedGACirculators}
We now consider the same four equally spaced injection points with ideal, matched circulators at each, enforcing propagation from smaller to larger port indices. The corresponding four-port
transmission-line impedance is the $N=4$ specialization of $\msZTL^{\rightarrow}(s)$ in \cref{Eq:InfiniteTLNPortCirculatorsEquallySpaced}. Assuming again that each giant atom has total coupling capacitance $C_c$ split into two equal branches, the same incidence matrices as in the previous example allow us to project the two legs onto the corresponding junction nodes.

\subsubsection{Exact admittance}

For the braided configuration shown in
\cref{fig:Examples_TLs_Giant_atoms}(b), we use the incidence matrix in
\cref{Eq:BraidedIncidenceMatrix}. The exact environmental admittance seen
from the two Josephson nodes is slightly modified by
\begin{align}
    \msYnode(s)
    =
    \msP^T
    \left[
        \left(\frac{2}{sC_c}\right)\msId_4
        +
        \msZTL^{\rightarrow}(s)
    \right]^{-1}
    \msP.
\end{align}
Combining this with the parallel junction capacitances gives again a port impedance matrix $\msZ_P(s)  = \left[ sC_J\msId_2  + \msYnode(s)   \right]^{-1}$. The inversion can again be performed analytically, yielding $\msZ_P(s) =  1/(sC_\Sigma) \msId_2 + \msZ(s)$, and $C_\Sigma=C_J+C_c$.

\subsubsection{Perturbative impedance in the weak-capacitive-loading regime}
\label{sec:weakZ}
In the weak-capacitive-loading regime (\cref{Eq:ValidityPerturbativeImpedance}
with $\msZTL=\msZTL^{\rightarrow}$), proceeding as in the reciprocal case gives the approximate impedance
\begin{align}
    \msZ(s)
    &\simeq
    \left(
        \frac{C_c}{2C_\Sigma}
    \right)^2
    \msP^T
    \msZTL^{\rightarrow}(s)
    \msP
    \\
    &=
    Z_0
    \left(
        \frac{C_c}{2C_\Sigma}
    \right)^2
    \begin{pmatrix}
        2(1+r^2)
        &
        2r
        \\
        2r(r^2+2)
        &
        2(1+r^2)
    \end{pmatrix}.
    \label{eq:Zweak_general}
\end{align}

\subsubsection{Perturbative rates and coherent couplings}

We now extract the weak-coupling decay rates and coherent exchange
couplings. As in the reciprocal braided-atom example, we restrict the
dynamics to the qubit subspace of the two giant atoms and use the same
definitions of $\gamma_0$, $\tau$, and $\theta$. Inserting
\cref{eq:Zweak_general} into the immittance formulas gives
\begin{align}
    \gamma_a=\gamma_b
    =
    4\gamma_0
    \left(
        1+\cos(2\theta)
    \right).
\end{align}
The collective decay coefficient is
\begin{align}
    \gamma_{12}
    =
    2\gamma_0
    \left(
        e^{i\theta}
        +
        2e^{-i\theta}
        +
        e^{-3i\theta}
    \right).
\end{align}
The coherent exchange coupling is
\begin{align}
    g
    =
    i\gamma_0
    \left(
        2e^{-i\theta}
        +
        e^{-3i\theta}
        -
        e^{i\theta}
    \right).
\end{align}
At $\theta=\pi/2$, the local and collective decay rates vanish, whereas the
coherent coupling remains finite, $g=2\gamma_0$. Thus, as in the reciprocal
braided giant-atom geometry, the braided ordering supports a finite
waveguide-mediated exchange interaction while all decay channels vanish.
\makeatletter
\renewcommand{\bibsection}{%
  \par
  \begingroup
    \baselineskip26\p@
    \bib@device{\hsize}{72\p@}%
  \endgroup
  \nobreak\@nobreaktrue
  \addvspace{19\p@}%
}
\makeatother

\bibliography{bibliography}

\end{document}